\documentclass[prx,aps,twocolumn,footinbib,superscriptaddress,notitlepage]{revtex4-2}
\usepackage[T1]{fontenc}
\usepackage{amsmath}
\usepackage{amssymb}
\usepackage{graphicx}
\usepackage{dcolumn}
\usepackage{enumerate}
\usepackage{bm}
\usepackage{xcolor}
\usepackage{comment}
\usepackage{subfigure}
\usepackage{tikz}
\DeclareRobustCommand*{\circled}[1]{%
  \tikz[baseline=(char.base)]{%
    \node[shape=circle,draw,inner sep=1pt] (char) {#1};%
  }%
}
\usepackage{breqn}
\usepackage{hyperref}
\hypersetup{colorlinks=true, citecolor=blue, urlcolor=blue, linkcolor=blue}

\makeatletter
\let\cat@comma@active\@empty
\makeatother

\begin{document}
\title{Simulation of topological superconductors and their competing orders using photon-mediated interactions}
\author{Anjun Chu}
\email{anjunchu@uchicago.edu}
\affiliation{JILA, NIST and Department of Physics, University of Colorado, Boulder, Colorado 80309, USA}
\affiliation{Center for Theory of Quantum Matter, University of Colorado, Boulder, Colorado 80309, USA}
\affiliation{Pritzker School of Molecular Engineering, University of Chicago, Chicago, Illinois 60637, USA}
\author{Joyce Kwan}
\affiliation{JILA, NIST and Department of Physics, University of Colorado, Boulder, Colorado 80309, USA}
\author{Eric Yilun Song}
\affiliation{JILA, NIST and Department of Physics, University of Colorado, Boulder, Colorado 80309, USA}
\author{Seth Hew Peng Chew}
\affiliation{JILA, NIST and Department of Physics, University of Colorado, Boulder, Colorado 80309, USA}
\author{\\James K. Thompson}
\affiliation{JILA, NIST and Department of Physics, University of Colorado, Boulder, Colorado 80309, USA}
\author{Ana Maria Rey}
\affiliation{JILA, NIST and Department of Physics, University of Colorado, Boulder, Colorado 80309, USA}
\affiliation{Center for Theory of Quantum Matter, University of Colorado, Boulder, Colorado 80309, USA}
\date{\today}

\begin{abstract}
Realizing and controlling the unconventional pairing featured by topological superconductors remains a central challenge. 
We introduce a cavity QED quantum simulator that engineers competing chiral $p_x+ip_y$ and $d_{x^2-y^2}+id_{xy}$ orders by tailoring cavity-mediated couplings between atomic pseudospins that emulate momentum-dependent pairing channels. 
The desired spatially inhomogeneous cavity-mediated couplings can be engineered in a 2D optical lattice using incommensurate cavity-lattice wavelengths naturally occurring in cavity QED systems.
This minimal and fully tunable platform enables controlled state preparation and continuous measurement of superconducting order parameters. 
To probe this competition, the platform enables selective excitation and direct time-domain detection of Bardasis-Schrieffer modes---collective excitations of the subdominant pairing channel. The same competing channel can have an even stronger impact out of equilibrium, where a sudden quench can drive its growth and abruptly reshape the dominant superconducting order, despite the stability of the dominant order in equilibrium. Together, these results establish cavity QED as a controlled setting for revealing hidden competing orders and exploring nonlinear dynamics of topological superconductivity that remain elusive in solid-state and ultracold-atom systems.
\end{abstract}

\maketitle

\section{Introduction}
Topological superconductors and superfluids intertwine nontrivial topology with many-body physics \cite{Qi2011,Sato2017}, hosting protected Majorana zero modes \cite{Chiu2016} with prospects for fault-tolerant quantum computation \cite{Nayak2008}. 
Realizing such phases remains difficult because they require unconventional Cooper pairing with nonzero orbital angular momentum (e.g. $p_x \pm i p_y$, $d_{x^2-y^2} \pm i d_{xy}$), which is typically much weaker than conventional $s$-wave pairing \cite{Sato2017}. 
Establishing these exotic pairing symmetries as the dominant mechanism in a stable and controllable platform therefore remains an outstanding challenge.

Progress toward $p$-wave topological superconductivity in solid-state and engineered nanowire systems has been encouraging \cite{Kallin2016,Sato2017}, but disorder and ambiguous signatures continue to obscure definitive observation. Superfluid $^{3}$He provides the only confirmed example of a topological superfluid \cite{Volovik2003,Pautti2016}, and even in this setting Majorana modes remain unverified. Although theoretical proposals exist for $d$-wave topological superconductors \cite{Laughlin1998,Black-Schaffer2012,Nandkishore2012,Kiesel2012,Kiesel2013,Fischer2014}, experimental confirmation has proven challenging. Ultracold Fermi gases offer controllability for realizing topological superfluidity \cite{Gurarie2007}, yet $p$- and $d$-wave interactions are intrinsically weak and their enhancement via Feshbach resonances induces detrimental losses \cite{Regal2003,Schunck2005,Gunter2005,Gaebler2007}. Despite several promising proposals \cite{Zhang2008,Han2009,Cooper2009,Fedorov2016,Zhang2015}, accessing robust topological phases remains elusive.

\begin{figure*}[t]
    \centering
    \includegraphics[width=0.8\textwidth]{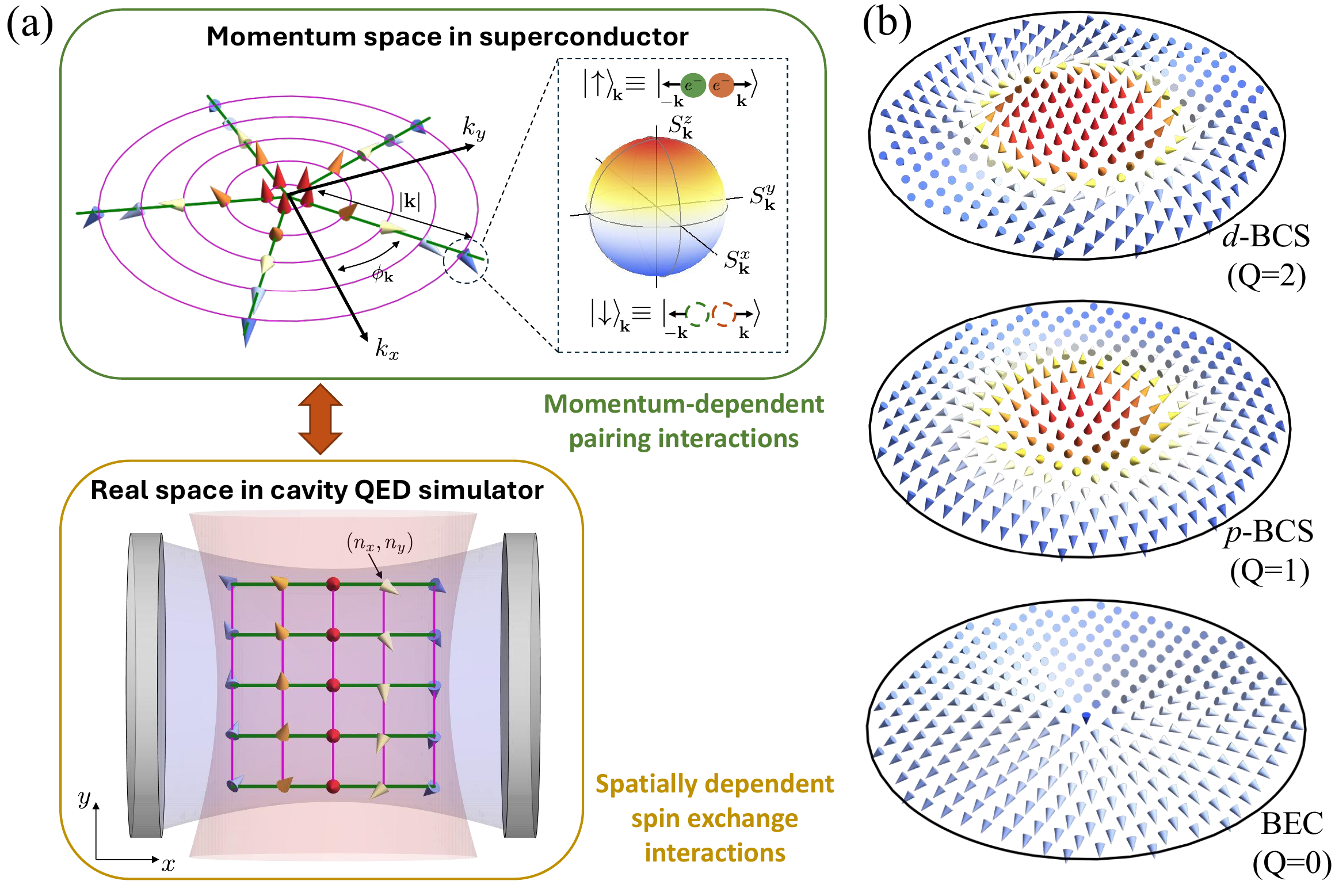}
    \caption{\textbf{Mapping topological BCS superconductors to cavity QED simulators.} 
    (a) We consider Anderson pseudospin mapping between the presence/absence of a Cooper pair and pseudospin states $|\uparrow\rangle_{\mathbf{k}}$/$|\downarrow\rangle_{\mathbf{k}}$, and then relate the Anderson pseudospins in 2D momentum space (top panel) to the atomic spins pinned in 2D lattice sites inside a cavity QED simulator (bottom panel).
    We can thus implement the momentum dependence of pairing interactions via the spatial inhomogeneity of spin exchange interactions, which is generated by the standing-wave structure of the cavity mode (blue) and the running-wave structure of the laser drive (red). The $x,y$ directions of the lattice are mapped to magnitude (indicated by green color) and phase (indicated by magenta color) of momentum respectively.
    (b) The pseudospin textures for different topological phases (characterized by Chern number $Q$) in topological BCS superconductors: $d$-BCS phase ($Q=2$), $p$-BCS phase ($Q=1$), and BEC phase ($Q=0$). See Sec.~\ref{sec:topology}-B for the definition of Chern number $Q$.}
    \label{fig:cartoon}
\end{figure*}

Beyond realizing an individual topological pairing channel, a central challenge is to understand the competition among distinct unconventional pairing tendencies, which underlies many unresolved phenomena in quantum
materials~\cite{Fradkin2015,Proust2019,Kim2018Science,Cao2021Science,Comin2015NatMat,Jang2022SciAdv,Oike2018SciAdv,Sun2020PRX,Li2021PRX,Zhang2024PRX}. 
In many candidate materials, the microscopic interactions can favor several pairing symmetries with comparable energies, so that small changes in band structure, density, pressure, or external fields may select different superconducting states or stabilize phases in which multiple orders coexist~\cite{Fradkin2015,Proust2019}. Identifying only the symmetry of the condensed order is therefore insufficient to determine the underlying pairing mechanism, as it leaves potentially strong subdominant channels largely hidden.
Probing these latent pairing tendencies and determining how close they are to condensation are thus key steps toward understanding and controlling unconventional superconductivity.

Bardasis-Schrieffer (BS) modes provide a powerful means of probing these hidden pairing tendencies and of constraining microscopic theories of unconventional superconductivity~\cite{Bardasis1961,Kretzschmar2013,Boehm2014,Maiti2015,Maiti2016}.
They are collective excitations in which Cooper pairs coherently fluctuate from the dominant superconducting channel into an attractive but uncondensed pairing channel. 
In particular, the symmetry of a BS mode identifies the subdominant pairing channel, while its energy provides a measure of the relative strength of that channel and softens as the dominant superconducting order in the system approaches an instability.  
Evidence for these modes has been reported through Raman spectroscopy in iron-based superconductors~\cite{Kretzschmar2013,Boehm2014}.
In quantum materials, however, their identification is indirect and can be complicated by multiband structure, quasiparticle continua, disorder, and other collective excitations.
Moreover, the competing interactions are fixed by the material and cannot generally be tuned independently. 
A controlled setting in which the pairing channels can be tuned separately, selectively excited, and directly monitored would transform BS modes from indirect spectroscopic signatures into dynamical probes of competing superconducting orders.

Here we introduce such a controlled setting: a cavity QED quantum simulator that realizes and probes competing chiral $p_x+ip_y$ and $d_{x^2-y^2}+id_{xy}$ pairing orders (Sec.~\ref{sec:impl}). 
As illustrated in Fig.~\ref{fig:cartoon}, our scheme exploits the Anderson pseudospin mapping~\cite{Anderson1958} between the presence or absence of a Cooper pair and the internal states of atoms pinned in a two-dimensional optical lattice.
Spatially structured cavity-mediated interactions then emulate the momentum-dependent form factors of the desired pairing channels. 
These couplings are generated using the incommensurate wavelengths of the cavity modes and optical lattice that arise naturally in cavity QED systems~\cite{Hu2015,Chu2021}.
Building on recent cavity-QED realizations of $s$-wave superconducting dynamics~\cite{Lewis-Swan2021,Kelly2022,Young2024,Young2025} and proposals for simulating chiral $p$-wave pairing with trapped ions~\cite{Shankar2022}, our platform enables multiple unconventional pairing channels to be engineered simultaneously and their interaction strengths to be tuned independently.

We develop experimentally realistic protocols for controlled state preparation and continuous, nondestructive measurement of the complex order parameters, providing access to both equilibrium and nonequilibrium phenomena (Sec.~\ref{sec:impl}). 
For a pure $p_x+ip_y$ interaction, we show that the simulator can resolve the full dynamical phase diagram and its associated topological transitions (Sec.~\ref{sec:topology}). 
In the presence of both pairing channels, we identify regimes in which one order dominates and regimes in which both $p_x+ip_y$ and
$d_{x^2-y^2}+id_{xy}$ orders are locally stable (Sec.~\ref{sec:compete}-A).
Near the corresponding stability boundary, the dominant superconducting channel maintains the equilibrium value, while the subdominant channel supports collective Bardasis-Schrieffer modes~\cite{Bardasis1961,Kretzschmar2013,Boehm2014,Maiti2015,Maiti2016}.
We show that they can be directly detected through the relative phase of the two complex order parameters, and directly excited via an
angular-momentum-selective protocol, even in the presence of realistic cavity dissipation (Sec.~\ref{sec:compete}-B and \ref{sec:compete}-C). 
Interestingly, we find that a sudden quench  can drive the growth of even a weak competing pairing channel, abruptly changing the amplitude and oscillation frequency of the dominant superconducting order, despite the stability of the dominant order in equilibrium (Sec.~\ref{sec:compete}-D).

Taken together, our results introduce a powerful and versatile route to the quantum simulation of competing topological superconducting phases.
The combination of programmability, real-time measurement, and the ability to realize multiple pairing channels in a clean environment opens new opportunities for addressing open questions in the physics of unconventional superconductivity, including the microscopic mechanisms by which competing orders emerge, coexist, or suppress one another.

\section{Topological BCS superconductors and the pseudospin picture}
\label{sec:bcs}
Topological superconductors provide a unique setting where unconventional Cooper pairing not only drives superconductivity but also endows the bulk with nontrivial topology. The Bardeen-Cooper-Schrieffer (BCS) framework offers the minimal and universal description of these phases \cite{Bardeen1957,Read2000}. In its pseudospin formulation, the BCS Hamiltonian of a pairing channel $\alpha$ can be written as a collective spin model with global interactions of the form,
\begin{equation}
    \hat{H}_{\rm BCS}^{\alpha} 
    = \sum_{\mathbf{k}} 2\varepsilon_{\mathbf{k}} \hat{S}^{z}_{\mathbf{k}}
    - U_{\alpha}
    \sum_{\mathbf{k},\mathbf{q}} 
    f^{\alpha *}_{\mathbf{k}} f^{\alpha}_{\mathbf{q}}
    \hat{S}^{+}_{\mathbf{k}} \hat{S}^{-}_{\mathbf{q}},
    \label{eq:hamilspin}
\end{equation} 
where $\varepsilon_{\mathbf{k}}=\hbar^2\mathbf{k}^2/2M$ is the fermion kinetic energy in 2D, with $\mathbf{k}=(k_x,k_y)=|\mathbf{k}|(\cos\phi_{\mathbf{k}},\sin\phi_{\mathbf{k}})$, $M$ is the fermion mass, and $U_{\alpha}$ is the pairing interaction strength ($U_\alpha>0$) with $\alpha$ labeling the pairing channel. 
The form factor ${f}_{\mathbf{k}}^\alpha$ encodes the orbital structure of the Cooper pairs \cite{Read2000}:
\begin{equation}
f_{\mathbf{k}}^{s}=1,\qquad
f_{\mathbf{k}}^{p}=|\mathbf{k}|e^{i\phi_{\mathbf{k}}},\qquad
f_{\mathbf{k}}^{d}=|\mathbf{k}|^2 e^{2i\phi_{\mathbf{k}}},
\label{eq:channel}
\end{equation}
corresponding to $\ell=0,1,2$ orbital angular momentum,  associated to $s$, $p_x+ip_y$ and $d_{x^2-y^2}+id_{xy}$ pairing channels. 
Topological superconductivity exists for $\ell>0$ pairing channels.

The spin operators in Eq.~(\ref{eq:hamilspin}) are derived from the Anderson pseudospin mapping \cite{Anderson1958},
\begin{eqnarray}
\hat{S}_{\mathbf{k}}^{-}
    &=& \hat{c}_{-\mathbf{k},\Downarrow}\hat{c}_{\mathbf{k},\Uparrow},\nonumber\\
2\hat{S}_{\mathbf{k}}^{z}+1 
    &=& \hat{c}^{\dagger}_{\mathbf{k},\Uparrow}\hat{c}_{\mathbf{k},\Uparrow}
     + \hat{c}^{\dagger}_{-\mathbf{k},\Downarrow}\hat{c}_{-\mathbf{k},\Downarrow}, 
     \label{eq:and}
\end{eqnarray}
where $\hat{c}_{\mathbf{k},\sigma}$ is a field operator that annihilates a fermion with momentum $\mathbf{k}$ and spin $\sigma=\{\Uparrow,\Downarrow\}$.
Eq.~(\ref{eq:and}) maps the occupation of a pair of fermions (Cooper pair) onto a pseudospin aligned up ($|\uparrow\rangle$) or down ($|\downarrow\rangle$), as shown in Fig.~\ref{fig:cartoon}(a). Note that the Cooper pair defined in Eq.~(\ref{eq:and}) is a linear combination of spin singlet and triplet, which interacts via pairing channels with even and odd $\ell$ respectively. 
Therefore, for interactions between opposite-spin fermions, the same pseudospin representation can describe both even and odd pairing channels (see App.~\ref{sec:origin}).

In this work, we focus on the interplay between chiral $p_x+ip_y$ and $d_{x^2-y^2}+id_{xy}$ pairing channels, described by 
\begin{equation}
    \begin{aligned}
    \hat{H}_{\rm BCS}^{pd}
    &= \sum_{\mathbf{k}} 2\varepsilon_{\mathbf{k}} \hat{S}^{z}_{\mathbf{k}}
    - 
    U_p\sum_{\mathbf{k},\mathbf{q}} 
   |\mathbf{k}|\,|\mathbf{q}|\,e^{-i\phi_{\mathbf{k}}}e^{i\phi_{\mathbf{q}}}
    \hat{S}^{+}_{\mathbf{k}} \hat{S}^{-}_{\mathbf{q}}\\
    &- U_d\sum_{\mathbf{k},\mathbf{q}} 
   |\mathbf{k}|^2\,|\mathbf{q}|^2\,e^{-2i\phi_{\mathbf{k}}}e^{2i\phi_{\mathbf{q}}}
    \hat{S}^{+}_{\mathbf{k}} \hat{S}^{-}_{\mathbf{q}}.
    \end{aligned}
    \label{eq:hamilspinpd}
\end{equation} 
This Hamiltonian contains the essential ingredients for topological superconductivity: chiral momentum-resolved orbital pairing channels with nontrivial Chern numbers, and a built-in mechanism for competition and coexistence between topological superconducting orders.
In Fig.~\ref{fig:cartoon}(b), we show the pseudospin textures of different topological phases that can be realized in this minimal model. 
As we show below, this minimal model can be engineered in a cavity QED platform with full control over the effective couplings and real-time access to the pseudospin dynamics. 
It thus opens the door to experimentally probing  topological transitions and competing orders beyond the reach of conventional materials.

\section{cavity QED implementation}
\label{sec:impl}

\subsection{Engineering competing topological pairing interactions}

\begin{figure}[t]
    \centering
    \includegraphics[width=1.0\columnwidth]{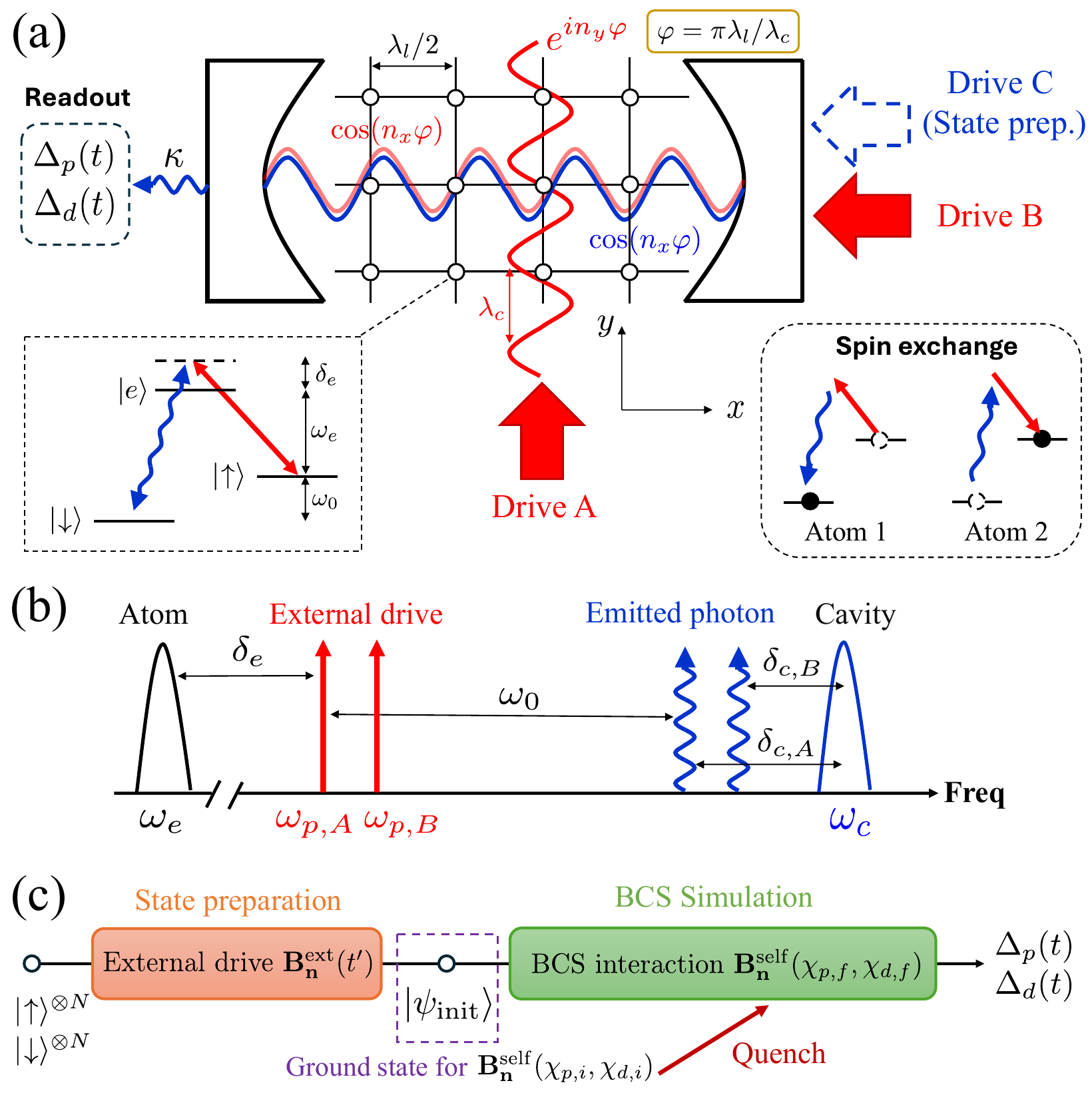}
    \caption{\textbf{Cavity QED setup for the BCS model with both $p_x+ip_y$ and $d_{x^2-y^2}+id_{xy}$ pairings.} (a) We consider atomic spins pinned in 2D lattice sites inside a standing-wave optical cavity, and apply external laser drives (red color, drive A from the side and drive B along the cavity) to couple $|\uparrow\rangle$ to $|e\rangle$ (see the left inset). The standing wave cavity mode $b$ (blue color) couples $|\downarrow\rangle$ to $|e\rangle$ and mediates spin exchange interactions between atoms (see the right inset). Another cavity mode $r$ (light red color) supports external drive B and its induced interatomic interactions is negligible. The dynamics of superconducting order parameters $\Delta_p(t)$ and $\Delta_d(t)$ can be observed by continuously tracking the light leaking out of the cavity. For state preparation, we engineer Raman transitions via an additional two-tone drive $C$ coupling $|\downarrow\rangle$ to $|e\rangle$. (b) Frequency diagram for engineering interaction Hamiltonian $\hat{H}_{\rm cav}$, including atomic transitions ($\omega_e$), external laser drives $A$ and $B$ ($\omega_{p,A}$, $\omega_{p,B}$), cavity modes ($\omega_c$) and emitted photons (see text). For clarity, external drive $C$ only used in state preparation is not shown in this diagram. (c) Proposed experimental sequence for probing sudden quench dynamics of topological superconductors. We prepare the mean-field ground state aligned to self-consistent field $\mathbf{B}^{\rm self}_{\mathbf{n}}(\chi_{p,i},\chi_{d,i})$ (see Eq.~(\ref{eq:bself})) by ramping external field $\mathbf{B}^{\rm ext}_{\mathbf{n}}(t')$ (see Eq.~(\ref{eq:drive1})). We then let the system evolve under self-consistent field $\mathbf{B}^{\rm self}_{\mathbf{n}}(\chi_{p,f},\chi_{d,f})$.}
    \label{fig:schematic}
\end{figure}

We consider an ensemble of $N$ atoms that have two relevant long-lived ground levels ($|\uparrow\rangle, |\downarrow\rangle$) separated by a frequency $\omega_0$, and an optical excited state $|e\rangle$ with frequency $\omega_e$.
Here we are focusing on a general scheme that can be realized in  different types of  atomic internal structures. Detailed implementations  on specific atomic species are discussed in App.~\ref{sec:experiment}.
We assume atoms are trapped in a 2D square optical lattice generated by beams with wavelength $\lambda_l$ in the $x$-$y$ plane (see Fig.~\ref{fig:cartoon}(a) and Fig.~\ref{fig:schematic}(a)).
The lattice is assumed to be deep enough to suppress tunneling between lattice sites, which we index as $\mathbf{n}=(n_x,n_y)$.

The 2D lattice is placed inside a standing-wave optical cavity aligned along the $x$ direction of the lattice. 
Atoms are coupled to two degenerate cavity modes with frequency $\omega_c$  and two different polarizations (labeled by $r$ and $b$). 
The cavity mode $b$ excites the  $|\downarrow\rangle\rightarrow|e\rangle$ transition with a spatial profile of the form $g_{\mathbf{n},b}=g_c\cos(n_x\varphi)$.
Here, $2g_c$ is the peak single-photon Rabi frequency, and $\varphi=\pi\lambda_l/\lambda_c$ the differential phase between nearest-neighbor sites. 
We consider the parameter regime where the cavity mode $r$ only plays a role in supporting the external laser drives, and interatomic interactions induced by mode $r$ are negligible (see App.~\ref{sec:cavity}-1 and \ref{sec:cavity}-3). 

The central idea is to shape the spatial profile of the atom-light coupling in real space so that the resulting photon-mediated spin exchanges inherit the orbital structure of unconventional $p_x+ip_y$ and $d_{x^2-y^2}+id_{xy}$ pairing symmetries (see Fig.~\ref{fig:cartoon}(a)). 
This spatial pattern naturally emerges when the cavity wavelength $\lambda_c$ is incommensurate with the lattice wavelength $\lambda_l$, a common situation in cavity QED systems \cite{Hu2015,Chu2021}.
As shown in Fig.~\ref{fig:schematic}(a) and its left inset, emulation of the momentum-dependent topological pairings can be accomplished by exciting the $|\uparrow\rangle\rightarrow|e\rangle$ transition by 1) the drive $A$, a running-wave laser beam propagating along the $y$ direction with frequency $\omega_{p,A}$, with a spatial profile of Rabi frequency $\Omega_{\mathbf{n},A}=\Omega_{A}e^{in_y\varphi}/2$, and 2) the drive B, supported by cavity mode $r$ along the $x$ direction with frequency $\omega_{p,B}$ and a standing-wave profile of Rabi frequency $\Omega_{\mathbf{n},B}=\Omega_{B}\cos(n_x\varphi)/2$.

The desired spin exchange interactions between the ground-state levels ($|\uparrow\rangle$ and $|\downarrow\rangle$) are mediated by four-photon exchange processes in which both the excited state $|e\rangle$ and the cavity mode $b$ participate only virtually in the dynamics (see the right inset of Fig.~\ref{fig:schematic}(a)). 
Basically, one atom in $|\uparrow\rangle$ flips to $|\downarrow\rangle$ by absorbing a photon from an external laser drive  (either $A$ or $B$) and emitting a virtual photon into the cavity. Another atom in $|\downarrow\rangle$ flips to $|\uparrow\rangle$  by absorbing the same virtual photon and emitting to the same laser drive (again $A$ or $B$ respectively). 
This is accomplished by making both drives $A$ and $B$ far detuned to the atomic excited state, so to a good approximation they share a similar atomic detuning, $\delta_e=\omega_{p,A}-\omega_e\approx \omega_{p,B}-\omega_e$. 
Moreover, to make the cavity mode $b$ virtual, we set the frequency of the emitted photons highly detuned from the cavity mode $b$, with cavity detunings $\delta_{c,A} = \omega_{p,A}+\tilde{\omega}_0-\tilde{\omega}_{c,b}$, $\delta_{c,B} = \omega_{p,B}+\tilde{\omega}_0-\tilde{\omega}_{c,b}$ (see App.~\ref{sec:cavity}-1 and \ref{sec:cavity}-3 for detailed conditions and derivations). 
Here, $\tilde{\omega}_0=\omega_0 + |\Omega_A|^2/(4\delta_e)$ is the transition frequency between $|\uparrow\rangle$ and $|\downarrow\rangle$ states corrected by AC Stark shifts of drive A, and $\tilde{\omega}_{c,b}=\omega_c+\sum_{\mathbf{n}}g^2_{\mathbf{n},b}/(2\delta_e)$ is the frequency of dressed cavity resonance for mode $b$.
We also assume the frequency scale of $|\omega_{p,A}-\omega_{p,B}|$ is much larger compared to the frequency scale of system dynamics, making their interference effects negligible.
Numerical benchmarking of these approximations is discussed in App.~\ref{sec:numerical}-2.

In this way, external drives $A$ and $B$ can be used to engineer two sets of four-photon couplings which, respectively, implement the desired $p$- and $d$-wave couplings.  
A frequency diagram satisfying all these conditions is shown in Fig.~\ref{fig:schematic}(b).
Under these conditions, the system can be described by the following effective Hamiltonian in the rotating frame of $\tilde{\omega}_0$ (see App.~\ref{sec:cavity}-1 and \ref{sec:cavity}-3),
\begin{equation}
    \begin{aligned}
    \hat{H}_{\rm cav}/\hbar &= \sum_{\mathbf{n}}2J\eta_{\mathbf{n}}^2\hat{S}^z_{\mathbf{n}} - \chi_p \sum_{\mathbf{n}\mathbf{m}}\eta_{\mathbf{n}}\eta_{\mathbf{m}}e^{-i\phi_{\mathbf{n}}}e^{i\phi_{\mathbf{m}}}\hat{S}^{+}_{\mathbf{n}}\hat{S}^{-}_{\mathbf{m}}\\
    & \quad - \chi_d \sum_{\mathbf{n}\mathbf{m}}\eta_{\mathbf{n}}^2\eta_{\mathbf{m}}^2e^{-2i\phi_{\mathbf{n}}}e^{2i\phi_{\mathbf{m}}}\hat{S}^{+}_{\mathbf{n}}\hat{S}^{-}_{\mathbf{m}}.
    \end{aligned}
    \label{eq:caveff}
\end{equation} 
Here, $\hat{S}^z_{\mathbf{n}}$, $\hat{S}^{\pm}_{\mathbf{n}}$ are spin operators for $|\uparrow\rangle$ and $|\downarrow\rangle$ states on lattice site $\mathbf{n}$, and $\eta_{\mathbf{n}}e^{i\phi_{\mathbf{n}}}=\cos(n_x\varphi)e^{-in_y\varphi}$ are dimensionless atom-cavity couplings.  
In Eq.~(\ref{eq:caveff}), $J=|\Omega_B|^2/(8\delta_e)$ is the strength of the inhomogeneous AC Stark shift induced by drive B, which simulates the kinetic energy of electrons, and 
$\chi_p=-|\Omega_A|^2g_c^2/(4\delta_e^2\delta_{c,A})$ and $\chi_d=-|\Omega_B|^2g_c^2/(4\delta_e^2\delta_{c,B})$ are the strengths of the spin exchange interactions involving the external drive A and B respectively. 
We can directly identify the implemented model with effective $p_x+ip_y$ ($\chi_{p}$ term) and $d_{x^2-y^2}+id_{xy}$ ($\chi_{d}$ term) pairing interactions, by mapping the dimensionless atom-cavity couplings $\eta_{\mathbf{n}}e^{i\phi_{\mathbf{n}}}$ to the momentum of pairing electrons $|\mathbf{k}|e^{i\phi_{\mathbf{k}}}$ (see Eq.~(\ref{eq:hamilspinpd})).
Note that we have already applied a local gauge transformation, $\hat{S}^{+}_{\mathbf{n}}\rightarrow\hat{S}^{+}_{\mathbf{n}}e^{-2i\phi_{\mathbf{n}}}$, $\hat{S}^{-}_{\mathbf{n}}\rightarrow\hat{S}^{-}_{\mathbf{n}}e^{2i\phi_{\mathbf{n}}}$, to ensure Eq.~(\ref{eq:caveff}) matches the form of Eq.~(\ref{eq:hamilspinpd}).
We can now define the $p$-wave and $d$-wave order parameters for the implemented model as
\begin{equation}
    \begin{aligned}
    \Delta_p=\chi_p\sum_{\mathbf{n}}\eta_{\mathbf{n}}e^{i\phi_{\mathbf{n}}}\langle\hat{S}^{-}_{\mathbf{n}}\rangle,\quad \Delta_d=\chi_d\sum_{\mathbf{n}}\eta_{\mathbf{n}}^2e^{2i\phi_{\mathbf{n}}}\langle\hat{S}^{-}_{\mathbf{n}}\rangle.
    \end{aligned}
    \label{eq:order}
\end{equation}

A key advantage of our approach is that the BCS pairing problem in momentum space can be 
encoded in real-space internal atomic degrees of freedom, enabling controlled engineering of various pairing channels and avoiding the need of cooling motional levels below quantum degeneracy. 
A further attractive feature of this implementation is that the dynamics of $\Delta_p(t)$ and $\Delta_d(t)$ can be monitored continuously and non-destructively by heterodyne detection of photons leaking out from cavity mode $b$ (see Fig.~\ref{fig:schematic}(a)), since the associated photons are emitted with different frequencies detuned from the resonant frequency of cavity mode $b$ ($\tilde{\omega}_{c,b}$). 
This is achievable in the regime where the cavity mode $b$ adiabatically follows the dynamics of atomic spins, and thus the coherent state amplitude of the cavity mode $b$ is directly determined by the BCS order parameters (see App. \ref{sec:cavity}-2),
\begin{equation}
    \alpha_b(t) \approx \frac{\Delta_p(t)}{\mathcal{G}^{*}_p}e^{-i\delta_{c,A}t} + \frac{\Delta_d(t)}{\mathcal{G}^{*}_d}e^{-i\delta_{c,B}t}.
    \label{eq:readout}
\end{equation}
Above we set  $\mathcal{G}_p=-\Omega_Ag_c/(2\delta_e)$, $\mathcal{G}_d=-\Omega_Bg_c/(2\delta_e)$ and  assume that $\Delta_p(t)$ and $\Delta_d(t)$ are slowly varying compared to the frequency scale of $|\delta_{c,A}-\delta_{c,B}|$.
Measurement induced back-action is negligible in the regime $|\delta_{c,A}|,|\delta_{c,B}|\;\gg \kappa$, with $\kappa$ the cavity intensity decay rate, where the number of photons leaking out of the cavity in every experimental run is small compared to the total number of atoms \cite{Young2024,Young2025}.

\subsection{Preparing mean-field ground states}

The distinct phases of the BCS model can be successfully reproduced by a mean-field (MF) analysis, that ignores quantum fluctuations and approximates $\hat{S}^{+}_{\mathbf{n}}\hat{S}^{-}_{\mathbf{m}}\approx \langle\hat{S}^{+}_{\mathbf{n}}\rangle\hat{S}^{-}_{\mathbf{m}}+\hat{S}^{+}_{\mathbf{n}}\langle\hat{S}^{-}_{\mathbf{m}}\rangle-\langle\hat{S}^{+}_{\mathbf{n}}\rangle\langle\hat{S}^{-}_{\mathbf{m}}\rangle$. 
Here we also introduced a chemical potential $\mu$ to fix the total number of Cooper pairs, $N_{C} = \sum_{\bf{n}}N_{\mathbf{n}}$, where $N_{\mathbf{n}}\equiv \langle\hat{S}^{z}_{\mathbf{n}}\rangle+1/2$ is the momentum distribution of Cooper pairs. Under the MF approximation, the ground state of Eq.~(\ref{eq:caveff}) can be found by minimization of the following MF Hamiltonian,
\begin{equation}
    \hat{H}_{\rm MF}/\hbar = -2\sum_{\mathbf{n}}{\bf B}^{\rm self}_{\mathbf{n}}(\chi_p,\chi_d)\cdot \hat{\mathbf{S}}_{\mathbf{n}},
    \label{eq:meanh}
\end{equation}
where ${\bf B}^{\rm self}_{\mathbf{n}}(\chi_p,\chi_d)$ is an effective magnetic field generated by the average interactions with other atoms in the array,
\begin{equation}
    \begin{aligned}
    (B^{\rm self})^x_{\mathbf{n}}-i(B^{\rm self})^y_{\mathbf{n}}&=\eta_{\mathbf{n}}e^{-i\phi_{\mathbf{n}}}\Delta_p+\eta_{\mathbf{n}}^2e^{-2i\phi_{\mathbf{n}}}\Delta_d,\\
    (B^{\rm self})^z_{\mathbf{n}}&=\mu -J\eta_{\mathbf{n}}^2.
    \end{aligned}
    \label{eq:bself}
\end{equation}

In the following, we discuss a scheme to prepare the MF ground state of Eq.~(\ref{eq:caveff}). 
Notice that the MF Hamiltonian (see Eq.~(\ref{eq:meanh})) becomes an effective single-particle Hamiltonian if one interprets $\Delta_p$ and $\Delta_d$ as free parameters without the need to satisfy self-consistent equations.
This single-particle model can be experimentally realized by adding an external drive $C$ supported by the cavity mode $b$ (see Fig.~\ref{fig:schematic}(a) and more details in App.~\ref{sec:cavity}-4), which drives the $|\downarrow\rangle\rightarrow|e\rangle$ transition with two frequencies $\omega_{p,C}, \omega'_{p,C}$. 
The two tones share the same Rabi frequency spatial profile $\Omega_{\mathbf{n},C}=\Omega_{C}\cos(n_x\varphi)/2$.

We consider the rotating frame of the external laser drives, and impose the two-photon resonance conditions $\omega_{p,A} = \omega_{p,C}-\tilde{\omega}_0-\delta$ and $\omega_{p,B} = \omega'_{p,C}-\tilde{\omega}_0-\delta$, where $\delta$ is the effective two-photon detuning after correcting the AC Stark shift induced by $\Omega_A$. 
One can ignore the interference effects between the two tones when  setting the frequency scale of $|\omega_{p,A}-\omega_{p,B}|$ much faster than system dynamics. 
To suppress cavity-mediated interactions during state preparation, one should  increase the cavity detunings $\delta_{c,A}$ and $\delta_{c,B}$. 
Under these assumptions, the external drive Hamiltonian describing the Raman coupling between $|\uparrow\rangle$ and $|\downarrow\rangle$ states becomes (see App.~\ref{sec:cavity}-4)
\begin{equation}
    \begin{aligned}
    \hat{H}_{\rm drive}/\hbar &= -\sum_{\mathbf{n}}\mathbf{B}^{\rm ext}_{\mathbf{n}}(t')\cdot \hat{\mathbf{S}}_{\mathbf{n}},
     \end{aligned}
\label{eq:drive}
\end{equation}
where $\mathbf{B}^{\rm ext}_{\mathbf{n}}(t')$ can be seen as an effective magnetic field generated by the external drives,
\begin{equation}
    \begin{aligned}
        (\mathbf{B}^{\rm ext})^x_{\mathbf{n}}-i(\mathbf{B}^{\rm ext})^y_{\mathbf{n}}&=\Omega_{p}(t')\eta_{\mathbf{n}}e^{-i\phi_{\mathbf{n}}}+\Omega_{d}(t')\eta_{\mathbf{n}}^2e^{-2i\phi_{\mathbf{n}}},\\
        (\mathbf{B}^{\rm ext})^z_{\mathbf{n}}&= \delta(t') - J'(t')\eta_{\mathbf{n}}^2.
    \end{aligned}
    \label{eq:drive1}
\end{equation}
Here, $\Omega_{p}(t')= -\Omega_A^{*}\Omega_C/(2\delta_e)$ and $\Omega_{d}(t')= -\Omega_B^{*}\Omega_C/(2\delta_e)$ are effective two-photon Rabi frequencies, and $J'(t')=(|\Omega_B|^2-2|\Omega_C|^2)/(4\delta_e)$ is the strength of the inhomogeneous AC Stark shift induced by both external drives $B$ and $C$ along the cavity.
In this way, we engineer a Raman transition with the same spatial dependence as spin-exchange interactions (see MF Hamiltonian in Eq.~(\ref{eq:meanh})).
Note that we have already applied the same local gauge transformation as Eq.~(\ref{eq:caveff}).
For clarity, we will use $t'$ as the label of the evolution time in state preparation, and $t$ for the time during the BCS-model simulation.

The goal, schematically shown in Fig.~\ref{fig:schematic}(c), is to  adiabatically ramp the drive parameters for a time $t'_{\rm prep}$, to prepare the ground state of Eq.~(\ref{eq:drive}), which is equivalent to the MF ground state of the interaction Hamiltonian (see Eq.~(\ref{eq:meanh})). 
This is achieved by satisfying
\begin{equation}
{\bf B}^{\rm self}_{\mathbf{n}}(\chi_{p,i},\chi_{d,i})\propto \mathbf{B}^{\rm ext}_{\mathbf{n}}(t'_{\rm prep}).
\label{eq:magnetic}
\end{equation} 
Here we defined $\chi_{p,i}$ and $\chi_{d,i}$ as the initial $p$-wave and $d$-wave interaction strengths, respectively. 
After preparing the initial state aligned to the self-consistent field $\mathbf{B}^{\rm self}_{\mathbf{n}}(\chi_{p,i},\chi_{d,i})$, quench dynamics can be induced by letting the system evolve under 
self-consistent field $\mathbf{B}^{\rm self}_{\mathbf{n}}(\chi_{p,f},\chi_{d,f})$, where $\chi_{p,f}$ and $\chi_{d,f}$ denote the final interaction strengths (see Fig.~\ref{fig:schematic}(c)).
This corresponds to a sudden quench of interaction strengths from $(\chi_{p,i},\chi_{d,i})$ to $(\chi_{p,f},\chi_{d,f})$. 
Equilibrium physics arises as the special case without a quench, 
i.e., when $(\chi_{p,i},\chi_{d,i})=(\chi_{p,f},\chi_{d,f})$.

As a side note, the topological phase transition discussed in Sec.~\ref{sec:topology}-B imposes a requirement for the adiabatic protocol, i.e. the adiabatic path has to be constrained to the same topological phase. 
This can be achieved for the BCS phase if one starts with all the atoms in $|\uparrow\rangle$ (eigenstate for $\delta\rightarrow+\infty$) and maintains $\delta>0$ along the path.
Similarly, for the BEC phase, one starts with all the atoms in $|\downarrow\rangle$ (eigenstate for $\delta\rightarrow-\infty$) and maintains $\delta<0$ along the path.

\section{Probing dynamical phases and topology}
\label{sec:topology}

\subsection{Dynamical phases after sudden quench}

The non-equilibrium quench dynamics of the BCS model can be characterized by three types of dynamical phases (I, II and III) \cite{Barankov2006,Yuzbashyan2006,Yuzbashyan2006a,Gurarie2009,Foster2013,Foster2014,Yuzbashyan2015,Cui2019,Collado2023}.
They are described using a time-averaged or steady-state order parameter that shows non-analytic change at the dynamical phase boundaries \cite{Marino2022}.
These dynamical phases have been demonstrated experimentally in the $s$-wave BCS model \cite{Mansart2013,Matsunaga2013,Matsunaga2014,Behrle2018,Young2024,Dyke2024,Young2025}, and are also predicted to occur in $p$-wave \cite{Foster2013,Foster2014,Liao2015} and $d$-wave cases \cite{Peronaci2015,Schwarz2020}.

\begin{figure}[t]
    \centering
    \includegraphics[width=1.0\columnwidth]{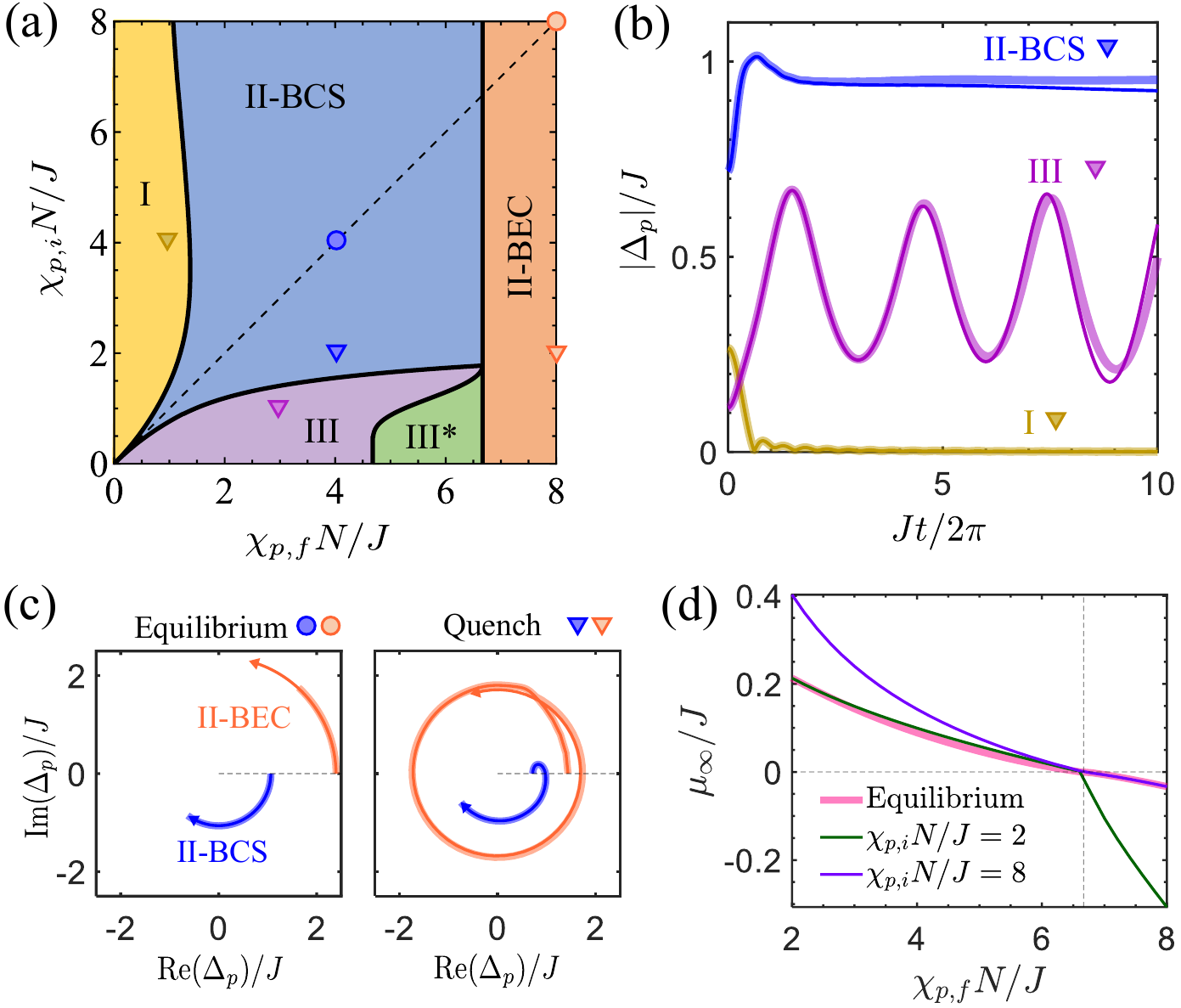}
    \caption{\textbf{Dynamical phases and topological phase transitions for $p$-wave pairing only}. (a) Dynamical phase diagram of suddenly quenching the interaction strength from $\chi_{p,i}$ to $\chi_{p,f}$. We fix the number of Cooper pairs to $N_C/N=0.35$. The solid lines mark the dynamical phase boundary, and the dashed line marks the condition $\chi_{p,i}=\chi_{p,f}$ for equilibrium physics. The circles and triangles mark the position of the curves in (b) and (c) on the phase diagram. (b) Examples of the three dynamical phases (I, II, and III). The yellow lines describe phase I dynamics at $\chi_{p,i}N/J=4$, $\chi_{p,f}N/J=1$, blue lines are for phase II at $\chi_{p,i}N/J=2$, $\chi_{p,f}N/J=4$, and purple lines are for phase III at $\chi_{p,i}N/J=1$, $\chi_{p,f}N/J=3$. In both (b) and (c), the curves with lighter color include Hamiltonian dynamics only (see Eq.~(\ref{eq:caveff})), while the ones with darker color include dissipation due to cavity photon loss with $\kappa/|\delta_{c,A}|\,=2\times 10^{-3}$. (c) Examples of the II-BCS phase and the II-BEC phase. The blue lines are for $\chi_{p,f}N/J=4$ and the orange lines are for $\chi_{p,f}N/J=8$. The left panel shows the equilibrium case with $\chi_{p,i}=\chi_{p,f}$, and the right panel shows quench dynamics from $\chi_{p,i}N/J=2$. In both panels, we show trajectories of $\Delta_p$ with $Jt/2\pi\in[0,2]$. (d) The long-time chemical potential $\mu_{\infty}$ as a function of $\chi_{p,f}$. We only include Hamiltonian dynamics for the evaluation of $\mu_{\infty}$.}
    \label{fig:dpt}
\end{figure}

Here, for clarity, we  will focus on the case with $p$-wave physics only, i.e. $\chi_{d,i},\chi_{d,f}\rightarrow 0$ (achievable by setting $|\delta_{c,B}|\;\gg |\delta_{c,A}|$). 
In Fig.~\ref{fig:dpt}(a) we show the dynamical phase diagram computed both numerically and analytically (via the Lax formalism \cite{Richardson1964,Gaudin1976,Dukelsky2004,Foster2013} in App.~\ref{sec:lax}).
Characteristic traces of the different dynamical phases are shown in Fig.~\ref{fig:dpt}(b).
{\bf Phase I} corresponds to relaxation into a state with a vanishing BCS order parameter at long times (yellow lines). 
{\bf Phase II} exhibits a steady state with a constant non-zero amplitude of the BCS order parameter, $|\Delta_{p}(t)|\,\rightarrow \Delta_{p,\infty}>0$ (blue lines). 
Finally, {\bf Phase III/III*} features oscillations in $|\Delta_{p} (t)|$ that persist for long times (purple lines).

In Fig.~\ref{fig:dpt}(b), we also compare the dynamics of $|\Delta_p(t)|$ with purely  Hamiltonian evolution (see $\hat{H}_{\rm cav}$ in Eq.~(\ref{eq:caveff})) to the one when additional dissipation due to cavity photon loss (see App.~\ref{sec:cavity}-1) is present.
For experimentally realistic conditions ($\kappa/|\delta_{c,A}|\,=2\times 10^{-3}$), dissipation induces only minor quantitative corrections to $|\Delta_p|$, confirming that the key dynamical features are robust.
For all the numerical calculations in this work, we use a $500\times 20$ lattice with $n_x\varphi \,\mathrm{mod}(2\pi)$ and $n_y\varphi \,\mathrm{mod}(2\pi)$ uniformly distributed between $0$ and $2\pi$, where 500 sites are along the $x$ direction and $20$ sites are along the $y$ direction.
Further details of numerical calculations are discussed in App.~\ref{sec:numerical}-1.

In the dynamical phase diagram shown in Fig.~\ref{fig:dpt}(a), we further separate phase II into two different sub-phases (II-BCS and II-BEC) based on a topological phase transition (see Sec.~\ref{sec:topology}-B). The separation between phase III and phase III* is based on the sharp change of oscillation amplitude and frequency (see Sec.~\ref{sec:compete}-D).
It is worth mentioning that our implementation yields similar physical phenomena of standard $p_x+ip_y$ superconductors predicted in Ref.~\cite{Foster2013}. Interestingly, due to different density of states and high-momentum cutoff ($|\eta_{\mathbf{n}}|\;\leq 1$) in our case, we find a modification in the topological phase transition point as well as a new dynamical phase (phase III*).

\subsection{Topological phase transitions in and out of equilibrium}

In equilibrium, the BCS model with $p_x+ip_y$ or $d_{x^2-y^2}+id_{xy}$ pairing symmetry is known for two distinct topological phases depending on the relative strength of the pairing interaction compared to the kinetic energy \cite{Read2000,Gurarie2007}, in contrast to the well-studied and experimentally probed BCS-BEC crossover for $s$-wave pairing \cite{Gurarie2007,Randeria2014}.
At weak pairing strength (BCS phase), the ground state is formed by weakly correlated pairs of fermions. 
In contrast, strong pairing (BEC phase) favors the molecular Bose-Einstein condensate (BEC) formed by strongly coupled pairs of fermions.
The topological BCS-BEC phase transition can be characterized by the Chern number \cite{Read2000,Gurarie2007},
\begin{equation}
    \begin{aligned}
    Q&=\frac{2}{\pi}\int dk_xdk_y \bigg(\frac{\partial\langle {\bf \hat{ S}}_{\bf k}\rangle}{\partial k_x}\times\frac{\partial\langle {\bf \hat{ S}}_{\bf k}\rangle}{\partial k_y}\bigg)\cdot \langle {\bf \hat{ S}}_{\bf k}\rangle\\
    &= \begin{cases}
       \frac{1}{2}(1+\mu/|\mu|) \quad\quad (p_x+ip_y\;\mathrm{pairing})\\
       (1+\mu/|\mu|) \quad\quad (d_{x^2-y^2}+id_{xy}\;\mathrm{pairing})
    \end{cases},
    \end{aligned}
    \label{eq:chern}
\end{equation}
where $\langle {\bf \hat{ S}}_{\bf k}\rangle$ is the pseudospin texture of the ground state in momentum space.
The {\bf BEC phase} always has $Q=0$ and $\mu<0$, while the {\bf BCS phase} has $\mu>0$, with $Q=1$ for $p$-wave ($p$-BCS) and $Q=2$ for $d$-wave ($d$-BCS). Examples of pseudospin textures of all these topological phases are shown in Fig. \ref{fig:cartoon}(b).
Note that our proposed cavity QED simulator is a discrete sampling of the momentum space and we discuss the way to construct Chern number in App.~\ref{sec:meanfield}-2.

To detect the Chern number $Q$ in equilibrium, it is therefore enough to determine the sign of the chemical potential $\mu$, which can be done in our cavity QED simulator by tracking the phase of the BCS order parameter. Given that Eq.~(\ref{eq:caveff}) does not contain chemical potential $\mu$, then if one prepares the MF ground state of Eq.~(\ref{eq:caveff}), the BCS order parameter will evolve as
\begin{equation}
    \Delta_{\alpha}(t) = \Delta_{\alpha,0}e^{-2i\mu t},
    \label{eq:equlibrium}
\end{equation}
where $\alpha$ labels the pairing channel and $\Delta_{\alpha,0}$ is the initial value of the BCS order parameter.
 
This detection technique can be directly generalized to the dynamical phase II where the BCS order parameter reaches a steady-state amplitude at long time,
\begin{equation}
    \Delta_{\alpha}(t\rightarrow\infty) \rightarrow \Delta_{\alpha,\infty}e^{-2i\mu_{\infty}t},
    \label{eq:dynamicmu}
\end{equation}
where $\mu_{\infty}$ is the effective chemical potential for long-time dynamics. 
Based on the analogy between Eq.~(\ref{eq:dynamicmu}) and Eq.~(\ref{eq:equlibrium}), one can define the out-of-equilibrium equivalent of the topological Chern number $W$ to characterize topological transitions for dynamical phases \cite{Foster2013,Shankar2022}, by replacing $\mu\rightarrow\mu_{\infty}$ and $\Delta_{\alpha,0}\rightarrow \Delta_{\alpha,\infty}$ in the definition of $Q$ (see App.~\ref{sec:meanfield}-2). 
Therefore, the {\bf II-BEC phase} always has $W=0$ and $\mu_{\infty}<0$, while the {\bf II-BCS phase} has $\mu_{\infty}>0$, with $W=1$ for $p$-wave and $W=2$ for $d$-wave. 
This allows us to determine the non-equilibrium topological phase transition by tracking the phase of BCS order parameters \cite{Shankar2022}. 
The equilibrium topological phase transition can be considered as a special case of the transition between II-BCS and II-BEC phases, as shown by the dashed line in Fig.~\ref{fig:dpt}(a).

In Fig.~\ref{fig:dpt}(c), for clarity, we focus on the topological phase transition with $p$-wave pairing only. We show the trajectories of the real and imaginary parts of $\Delta_p(t)$ in equilibrium ($\chi_{p,i}=\chi_{p,f}$) and after a quench ($\chi_{p,i}\neq\chi_{p,f}$). The trajectories are clockwise (i.e. $\mu_{\infty}>0$) for II-BCS phase, and counterclockwise (i.e. $\mu_{\infty}<0$) for II-BEC phase.  
To simulate experimental conditions, we also include dissipative effects due to cavity photon loss (see App.~\ref{sec:cavity}-1 and App.~\ref{sec:numerical}-1). The dissipative effects will lead to a negative shift of $\mu_{\infty}$, but they do not alter the physics for short probe times.

In the proposed implementation, due to the high-momentum cutoff $|\eta_{\mathbf{n}}|\;\leq 1$ physically imposed in our scheme (see App.~\ref{sec:meanfield}-2 and App.~\ref{sec:lax}-1), we find the same quantum critical point (QCP) for the $p$-wave topological phase transition both in equilibrium and away from equilibrium (see Fig.~\ref{fig:dpt}(a)), $(\chi_{p,f})_{\rm QCP}N/J = (\frac{1}{2}-\frac{N_C}{N})^{-1}$, contrary to what was predicted in Ref.~\cite{Foster2013}. 
Fig.~\ref{fig:dpt}(d) shows the numerical calculations of $\mu_{\infty}$ performed for a fixed number of Cooper pairs $N_C/N=0.35$, and compares three different cases: 1) equilibrium physics ($\chi_{p,i}=\chi_{p,f}$); 2) quenching from an initial state in $p$-BCS phase ($\chi_{p,i}N/J=2$); 3) quenching from an initial state in BEC phase ($\chi_{p,i}N/J=8$). All these cases share the same zero crossing ($\mu_{\infty}=0$) at $(\chi_{p,f})_{\rm QCP}N/J = 20/3$.
When quenching within the same topological phase, i.e. when  $\chi_{p,i}$ and $\chi_{p,f}$ are on the same side of $(\chi_{p,f})_{\rm QCP}$, we find that $\mu_{\infty}$ is very close to the equilibrium value.
When quenching to a different topological phase, i.e. when  $\chi_{p,i}$ and $\chi_{p,f}$ are on opposite sides of $(\chi_{p,f})_{\rm QCP}$, in contrast we find that $\mu_{\infty}$ differs from the equilibrium value significantly.

\begin{figure*}[t]
    \centering
    \includegraphics[width=0.8\textwidth]{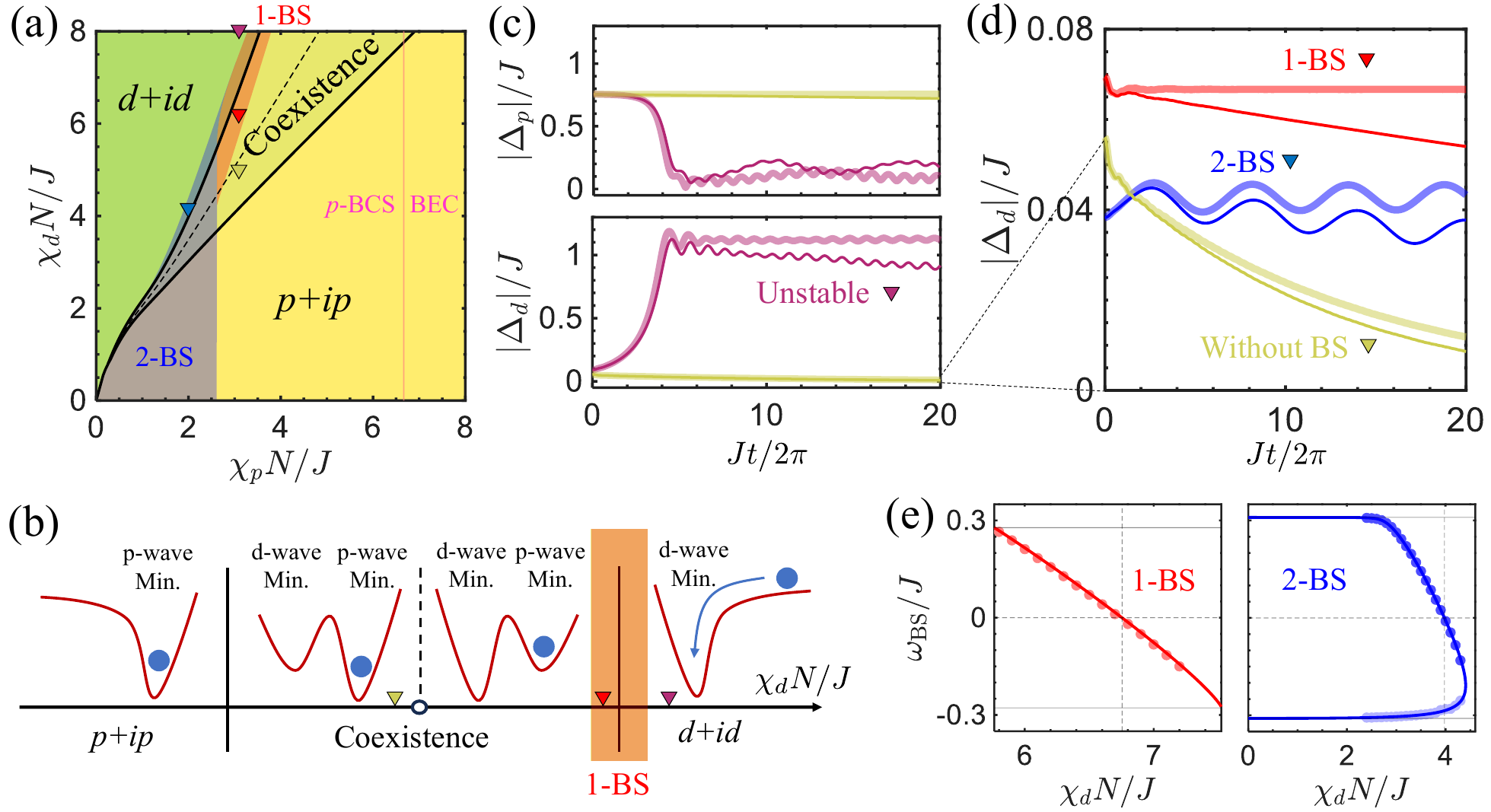}
    \caption{\textbf{Competing $p$-wave and $d$-wave orders and Bardasis-Schrieffer (BS) modes.} (a) Equilibrium phase diagram for Eq.~(\ref{eq:caveff}) with a fixed number of Cooper pairs $N_C/N=0.35$. The black solid lines separate three regimes based on the energetic stability of mean-field eigenstates: $p+ip$ regime (only $p_x+ip_y$ pairing is stable, yellow color), $d+id$ regime (only $d_{x^2-y^2}+id_{xy}$ pairing is stable, green color), coexistence regime (both states are stable). The black dashed line marks the first order phase transition between the $p$-wave and the $d$-wave pairing. The pink line marks the topological transition between $p$-BCS and BEC phases of the $p+ip$ regime discussed in Sec.~\ref{sec:topology}-B. The orange shaded area marks the regimes with a single BS mode (1-BS) in the subdominant $d$-wave channel, and the blue shaded area marks the regimes with two BS modes (2-BS).
    (b) Schematics for the energetic stability of $p_x+ip_y$ pairing with fixed $\chi_p$ (we use $\chi_pN/J=3.1$ as an example). As we increase $\chi_d$, in the $p+ip$ regime the system has a single minimum for $p$-wave pairing; in the coexistence regime the system has two local minima for $p$-wave and $d$-wave pairing respectively. In the $d+id$ regime the system has a single minimum for $d$-wave pairing, and the minimum for $p_x+ip_y$ pairing disappears. A single BS mode (1-BS) with subdominant $d$-wave pairing exists near the boundary between the coexistence and $d+id$ regime, indicated by the orange shaded area.
    (c) We initialize the system in a $p_x+ip_y$ eigenstate at $\chi_p N/J=3.1$, $\chi_d N/J=8.0$ (pink color) and $\chi_p N/J=3.1$, $\chi_d N/J=5.0$ (olive color), with initial $d$-wave perturbation $\varepsilon_d=10^{-1}$. The same initial $d$-wave perturbation on a $p_x+ip_y$ eigenstate is used in (d) and (e). The lines with lighter color include Hamiltonian dynamics only (see Eq.~(\ref{eq:caveff})). The lines with darker color include dissipation due to cavity photon loss ($\kappa/|\delta_{c,A}|\,=2\times 10^{-3}$, $\delta_{c,A}\approx \delta_{c,B}$). The same color convention is used in (d).
    (d) Dynamics of the subdominant $d$-wave order parameter. The red lines (a single BS mode) are at $\chi_pN/J=3.1$, $\chi_dN/J=6.2$, the blue lines (two BS modes) are at $\chi_pN/J=2.0$, $\chi_dN/J=4.2$, and the green lines (without BS modes) are at $\chi_pN/J=3.1$, $\chi_dN/J=5.0$. (e) Comparison of the BS mode frequency $\omega_{\rm BS}$ numerically determined (points) from Hamiltonian dynamics using Eq.~(\ref{eq:1bs}) and Eq.~(\ref{eq:2bs}) and the exact calculation (solid lines) with Eq.~(\ref{eq:bsmode}). The vertical dashed lines mark the stability boundary $\chi_d^{\rm stab}$, and the horizontal solid lines mark the value $\pm\omega_{\rm gap}/J$. The left panel is at $\chi_{p}N/J=3.1$ and the right panel is at $\chi_{p}N/J=2.0$. In the right panel, the darker/lighter points correspond to $\omega_{\rm BS}^{+}$/$\omega_{\rm BS}^{-}$.}
    \label{fig:stability}
\end{figure*}

\section{Probing competing $p$-wave and $d$-wave orders}
\label{sec:compete}  

\subsection{Instability due to competing orders}

We now demonstrate that this platform can access the long-standing problem of 
{\emph{competing unconventional superconducting orders}} by introducing the $d$-wave interaction channel and examining its interplay with the $p$-wave channel.
In this case, the mean-field ground state of Eq.~(\ref{eq:caveff}) is shown to be one of the following two types of solutions (see App.~\ref{sec:meanfield}-1): 1) $p_x+ip_y$ pairing only with $\Delta_p\neq 0$ and $\Delta_d=0$; 2) $d_{x^2-y^2}+id_{xy}$ pairing only with $\Delta_p=0$ and $\Delta_d\neq 0$. 
For the states with $p_x+ip_y$ pairing only, the solution of $\Delta_p$ is independent of $\chi_d$. 
Similarly, for the states with $d_{x^2-y^2}+id_{xy}$ pairing only, the solution of $\Delta_d$ is independent of $\chi_p$. 
However, the energetic stability of these solutions against small perturbations is determined by the interplay between $\chi_p$ and $\chi_d$ (see Fig.~\ref{fig:stability}(a)). 
In fact, by applying a small perturbation, we find that besides the pure $p+ip$ regime and the pure $d+id$ regime, where only the $p$-wave or $d$-wave pairing, respectively, is energetically stable, there is a coexistence regime, where both pairing channels can be energetically stable (see Fig.~\ref{fig:stability}(b)). In the coexistence regime, there are two local minima in the system and the true mean-field ground state abruptly changes from one minimum to another, leading to a first-order phase transition (black dashed line in Fig.~\ref{fig:stability}(a)) between $p_x+ip_y$ and $d_{x^2-y^2}+id_{xy}$ pairings. 

To directly probe the stability of a state with $p_x+ip_y$ pairing only, for a fixed number of Cooper pairs $N_C/N=0.35$, we perturb the system away from its mean-field ground state by adding a small amount of $d_{x^2-y^2}+id_{xy}$ pairing, which is equivalent to adiabatically ramping parameters in Eq.~(\ref{eq:drive1}) to
\begin{equation}
    \Omega_d(t'_{\rm prep}) = \varepsilon_d \Omega_p(t'_{\rm prep}),
    \label{eq:smallperb}
\end{equation}
with $\varepsilon_d\ll 1$ describing the strength of $d$-wave perturbation. 
Note that the small amount of $d$-wave pairing initially ($\Delta_{d,0}$) does not satisfy Eq.~(\ref{eq:magnetic}) since we are perturbing away from the mean-field eigenstate with $\Delta_{d,0}=0$ (see App.~\ref{sec:meanfield}-3).

We understand the $d$-wave perturbation by
expanding $\langle \hat{S}^{-}_{\mathbf{n}}\rangle$ and $\langle \hat{S}^{z}_{\mathbf{n}}\rangle$ in Fourier series,
\begin{equation}
    \langle \hat{S}^{-}_{\mathbf{n}}\rangle = \sum_{m} u_{\mathbf{n},m} e^{-im\phi_{\mathbf{n}}}, \quad \langle \hat{S}^{z}_{\mathbf{n}}\rangle = \sum_{m} w_{\mathbf{n},m} e^{-im\phi_{\mathbf{n}}},
\end{equation}
where $m$ is an integer labeling the angular structure of the pseudospin texture.
Since $\langle \hat{S}^{z}_{\mathbf{n}}\rangle$ is a real number, we have $w_{\mathbf{n},m} = w_{\mathbf{n},-m}^{*}$.
We consider the mean-field equations $\frac{d}{dt}\langle\hat{\mathbf{S}}_{\mathbf{n}}\rangle=-2\mathbf{B}_{\mathbf{n}}^{\rm self}\times \langle\hat{\mathbf{S}}_{\mathbf{n}}\rangle$, where $\mathbf{B}_{\mathbf{n}}^{\rm self}(\chi_p,\chi_d)$ is given by Eq.~(\ref{eq:bself}). Here we choose a rotating frame with frequency $2\mu$ to remove the time dependence in the dominant $p$-wave channel. This leads to $u_{\mathbf{n},1}(t) = \Delta_{p,0}\eta_{\mathbf{n}}/(2E_{\mathbf{n}}^{(0)})$, and $w_{\mathbf{n},0}(t) = \xi_{\mathbf{n}}/(2E_{\mathbf{n}}^{(0)})$, 
where $\xi_{\mathbf{n}}=\mu-J\eta_{\mathbf{n}}^2$, and $E_{\mathbf{n}}^{(0)}=\sqrt{\xi^2_{\mathbf{n}}+\Delta_{p,0}^2\eta_{\mathbf{n}}^2}$.

The small $d$-wave perturbation at the same time excites the terms $u_{\mathbf{n},0}$, $u_{\mathbf{n},2}$ and $w_{\mathbf{n},\pm1}$. 
After linearizing the mean-field equations for these terms, we obtain (see App.~\ref{sec:meanfield}-3)
\begin{equation}
    \begin{aligned}
    \frac{d}{dt}u_{\mathbf{n},0}^{*} &= -2i\xi_{\mathbf{n}}u_{\mathbf{n},0}^{*} +2i\Delta_{p,0}\eta_{\mathbf{n}}w_{\mathbf{n},1},\\
    \frac{d}{dt}u_{\mathbf{n},2} &= 2i\xi_{\mathbf{n}}u_{\mathbf{n},2} - 2i\Big(\Delta_{p,0}\eta_{\mathbf{n}}w_{\mathbf{n,1}} + \Delta_d\eta_{\mathbf{n}}^2w_{\mathbf{n},0}\Big),\\
    \frac{d}{dt}w_{\mathbf{n},1} &= i\Big(\Delta_{p,0}\eta_{\mathbf{n}}u_{\mathbf{n},0}^{*}+\Delta_{d}\eta_{\mathbf{n}}^2u_{\mathbf{n},1}\Big)-i\Delta_{p,0}\eta_{\mathbf{n}}u_{\mathbf{n},2},\\
    \end{aligned}
    \label{eq:linearization}
\end{equation}
which describe the dynamics of the subdominant $d$-wave channel.
Assuming $(u_{\mathbf{n},0}^{*}, u_{\mathbf{n},2}, w_{\mathbf{n},1}, \Delta_d)\propto e^{-i\omega t}$, one can obtain an analytic solution for the excitation frequency $\omega$, which must satisfy  the self-consistent equation, $\Delta_d = \chi_d\sum_{\mathbf{n}}\eta_{\mathbf{n}}^2u_{\mathbf{n},2}$, leading to
\begin{equation}
    \chi_d \Pi_d(\omega) = 1,
    \label{eq:bsmode}
\end{equation}
where
\begin{equation}
    \Pi_d(\omega) = \sum_{\mathbf{n}}\frac{\eta_{\mathbf{n}}^4\Big((E_{\mathbf{n}}^{(0)})^2+\xi_{\mathbf{n}}^2-\xi_{\mathbf{n}}\omega\Big)}{E_{\mathbf{n}}^{(0)}\Big(4(E_{\mathbf{n}}^{(0)})^2-\omega^2\Big)}.
    \label{eq:pid}
\end{equation}

The stability boundary for the state with $p_x+ip_y$ pairing only, shown in Fig.~\ref{fig:stability}(a) with the top  black solid line, is thus given by
\begin{equation}
    \chi_d^{\rm stab} = \frac{1}{\Pi_d(\omega=0)},
    \label{eq:stab_d}
\end{equation}
indicating zero energy cost for the excitation of the subdominant $d$-wave channel. 
Therefore, a $p_x+ip_y$ solution is energetically stable if $\chi_d<\chi_d^{\rm stab}$, but energetically unstable if $\chi_d>\chi_d^{\rm stab}$.
Eq.~(\ref{eq:stab_d}), therefore, marks the boundary between the coexistence regime (two local minima) and the $d+id$ regime (one local minimum) in Fig.~\ref{fig:stability}(b). 
A similar procedure can be used to study the energetic stability of the $d_{x^2-y^2}+id_{xy}$ solution (see App.~\ref{sec:meanfield}-4), leading to the boundary (the bottom black solid line) between the coexistence regime and the $p+ip$ regime.

\begin{figure}[t]
    \centering
    \includegraphics[width=1.0\columnwidth]{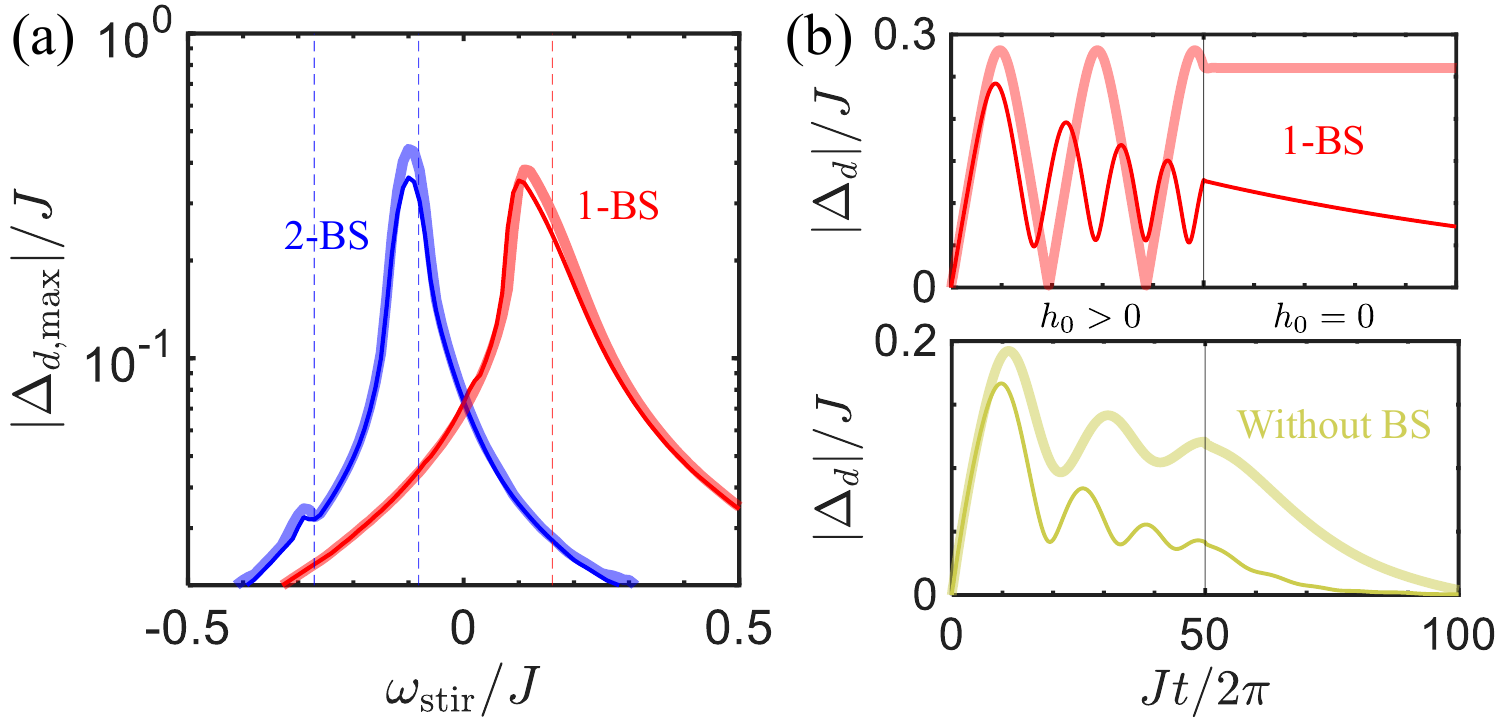}
    \caption{\textbf{Probing Bardasis-Schrieffer (BS) modes via external modulation.} We prepare the $p_x+ip_y$ eigenstate with $N_C/N=0.35$, and apply the external modulation $\hat{H}_{\rm probe}$ (see Eq.~(\ref{eq:probe})) to the system without initial $d$-wave pairing. (a) We let the system evolve under $h_0/J=4\times 10^{-3}$, and show the maximum value of $|\Delta_d|$ within $Jt/2\pi\in [0,12]$ while scanning $\omega_{\rm stir}$. The red lines (a single BS mode) are at $\chi_p N/J=3.1$, $\chi_d N/J=6.2$, and the blue lines (two BS modes) are at $\chi_pN/J=2.0$, $\chi_d N/J=4.2$. The vertical dashed lines are $\omega_{\rm BS}$ evaluated analytically via Eq.~(\ref{eq:bsmode}). (b) We turn on the external modulation $\hat{H}_{\rm probe}$ with $h_0/J=4\times 10^{-3}$ until $Jt/2\pi=50$, and then turn it off. We compare the upper panel where the system has a single BS mode ($\chi_pN/J=3.1$, $\chi_d N/J=6.2$, $\omega_{\rm stir}/J=0.16$), and the lower panel where the system does not contain BS modes ($\chi_pN/J=3.1$, $\chi_d N/J=5.0$, $\omega_{\rm stir}/J=0.45$). In both (a) and (b), the lines with lighter color include Hamiltonian dynamics only, and the lines with darker color include dissipation due to cavity photon loss ($\kappa/|\delta_{c,A}|=2\times 10^{-3}$, $\delta_{c,A}\approx\delta_{c,B}$).  }
    \label{fig:bardasis}
\end{figure}

\subsection{Probing Bardasis-Schrieffer modes via weak quenches in subdominant channel}

Based on the energetic stability analysis above, one might initially expect a small $d$-wave perturbation either to decay or to grow and destabilize the dominant p-wave order. 
Interestingly, the dynamics are richer: the system also supports coherent oscillations in the phase or amplitude of the $d$-wave order parameter, revealing collective modes in the subdominant $d$-wave pairing channel.
Such a collective excitation is protected against decay into broken Cooper pairs when its frequency lies below the pair-breaking threshold,
\begin{equation}
\omega_{\rm gap}=\min_{\mathbf n}2E_{\mathbf n}^{(0)}.
\end{equation}
Below this threshold, it is possible to excite the $d$-wave pairing coherently without destabilizing the dominant $p$-wave order. 
This gapped collective excitation is the Bardasis–Schrieffer (BS) mode~\cite{Bardasis1961,Maiti2015}. 
It reveals the existence of a subdominant competing pairing channel, and its softening signals an approaching instability of the dominant order.
We now show how this mode can be excited and directly measured in our system near the stability boundary $\chi_d=\chi_d^{\rm stab}$.

The BS mode corresponds to a real-frequency solution of
Eq.~(\ref{eq:bsmode}) below the pair-breaking threshold.
For $|\omega|\;<\omega_{\rm gap}$, the denominator in
Eq.~(\ref{eq:pid}) does not vanish, so the response function
$\Pi_d(\omega)$ is real. Such solutions exist for
$\chi_d^L<\chi_d<\chi_d^U$, the shaded orange and blue region in
Fig.~\ref{fig:stability}(a). The determination of these boundaries
is discussed in App.~\ref{sec:meanfield}-3. Within this interval,
a small $d$-wave perturbation undergoes undamped collective
oscillations. 
For weaker coupling, $\chi_d<\chi_d^L$, the
perturbation decays toward $\Delta_d=0$
(see the lower panel of Fig.~\ref{fig:stability}(c) and
Fig.~\ref{fig:stability}(d)). For stronger coupling,
$\chi_d>\chi_d^U$, it grows exponentially, eventually making
$\Delta_d$ the dominant order parameter
(see the lower panel of Fig.~\ref{fig:stability}(c)).
The $p$-wave pairing only decays away from its initial value $\Delta_{p,0}$ in the regime $\chi_d>\chi_d^U$ (see the upper panel of Fig.~\ref{fig:stability}(c)).
In both Fig.~\ref{fig:stability}(c) and (d), we also show that dissipation due to cavity photon loss has negligible effects on the system dynamics. 

Depending on the system parameters, the simulator supports
either one (shaded orange regime in Fig.~\ref{fig:stability}(a)) or two BS modes (shaded blue regime in Fig.~\ref{fig:stability}(a)).
Their frequencies describe
oscillations of the $d$-wave order relative to the $p$-wave
background. 
In Eq.~(\ref{eq:linearization}), we choose a rotating frame in which $\Delta_p$ is stationary, so these
frequencies appear directly in the dynamics of $\Delta_d$.
More generally, this relative dynamics can be measured
through $\Delta_p^{*}(t)\Delta_d(t)$, because the common phase acquired by the two order parameters cancels in this product.

For a single BS mode, the long-time behavior is
\begin{equation}
    \Delta_p^{*}(t)\Delta_d(t)
    \rightarrow
    \Delta_{p,0}\Delta_{d,\infty}e^{-i\omega_{\rm BS}t}.
    \label{eq:1bs}
\end{equation}
Since $|\Delta_p|$ stays at its initial value $\Delta_{p,0}$, the $d$-wave magnitude $|\Delta_d|$ approaches a constant
(see the red lines in Fig.~\ref{fig:stability}(d)), while the relative phase,
$\phi_{pd}(t)=\operatorname{Arg}
[\Delta_p^{*}(t)\Delta_d(t)]$,
winds at a constant rate,
$\phi_{pd}(t)\rightarrow\phi_0-\omega_{\rm BS}t$.
This phase winding provides a direct measurement of
$\omega_{\rm BS}$.

When two BS modes are present, the long-time response is a
superposition,
\begin{equation}
    \Delta_p^{*}(t)\Delta_d(t)
    \rightarrow
    \Delta_{p,0}\Big(
    \Delta_{d,\infty}^{+}e^{-i\omega_{\rm BS}^{+}t}
    +\Delta_{d,\infty}^{-}e^{-i\omega_{\rm BS}^{-}t}
    \Big).
    \label{eq:2bs}
\end{equation}
Interference between the modes produces persistent
oscillations in $|\Delta_d|$ at the difference frequency
$|\omega_{\rm BS}^{+}-\omega_{\rm BS}^{-}|$ (see the blue lines in Fig.~\ref{fig:stability}(d)). 
The two individual frequencies
can be resolved in the Fourier spectrum of
$\Delta_p^{*}(t)\Delta_d(t)$.
It is worth mentioning that when $\chi_d\rightarrow 0$ in this regime, the two frequencies are approaching the pair-breaking threshold, $\omega_{\rm BS}^{+}\rightarrow \omega_{\rm gap}$, $\omega_{\rm BS}^{-}\rightarrow -\omega_{\rm gap}$.

Figure~\ref{fig:stability}(e) shows excellent agreement between the BS mode frequencies extracted numerically from the time-domain dynamics of $\Delta_p^{*}\Delta_d$ and the analytical prediction of Eq.~(\ref{eq:bsmode}). 
The frequency $\omega_{\rm BS}$ passes through zero at $\chi_d=\chi_d^{\rm stab}$, marked by the vertical dashed line, in agreement with Eq.~(\ref{eq:stab_d}). 
This softening signals that a deformation of the pairing order toward the $d$-wave channel costs no energy at the stability boundary.

For Fig.~\ref{fig:stability}(c-e), we choose an initial perturbation
$\varepsilon_d=10^{-1}$ to make the $d$-wave response experimentally
visible. Near the upper boundary $\chi_d^U$, however, this finite
perturbation is large enough to destabilize the $p$-wave order,
preventing extraction of the BS frequency and accounting for
the missing numerical points in Fig.~\ref{fig:stability}(e).
Reducing the perturbation to $\varepsilon_d=10^{-2}$ brings the
numerically observed upper boundary into closer agreement with
the analytical prediction
(see App.~\ref{sec:meanfield}-3).

\subsection{Probing Bardasis-Schrieffer modes via external modulation}
The BS mode can also be excited spectroscopically, without introducing an
initial $d$-wave component, by applying a weak time-dependent external probe.
This protocol is analogous to the use of Raman spectroscopy to detect subdominant pairing channels in solid-state superconductors
\cite{Kretzschmar2013,Boehm2014,Maiti2016}. 
In our system, the BS mode converts pairing
amplitude from the dominant $p$-wave channel, with orbital angular momentum $\ell_p=1$, into the subdominant $d$-wave channel, with $\ell_d=2$. 
The probe must therefore supply the angular-momentum difference $\Delta\ell=\ell_d-\ell_p=1$.

We generate the required spatial structure using two additional laser beams, denoted by $\tilde{A}$ and $\tilde{B}$, in a geometry similar to that shown in Fig.~\ref{fig:schematic}. 
Both beams couple the $|\uparrow\rangle\rightarrow|e\rangle$ transition, with beam $\tilde{A}$ (frequency $\omega_{p,\tilde{A}}$) propagating transverse to the cavity axis and beam $\tilde{B}$ (frequency $\omega_{p,\tilde{B}}$) propagating along it. 
In contrast to the drives used to engineer the interactions, these probe beams are chosen to be far detuned from the cavity resonances, so that their cavity-mediated interactions are negligible. 
Additionally, we choose the frequency difference between the two probe beams, $\omega_{\rm stir}=\omega_{p,\tilde A}-\omega_{p,\tilde B}$, to be comparable to the system dynamical frequency scale, leading to a non-negligible interference term.
This interference term produces a time-dependent AC Stark shift with the desired dipolar angular structure:
\begin{equation}
    \hat{H}_{\rm probe}/\hbar = 4h_0\sum_{\mathbf n}\eta_{\mathbf n}\cos\!\left(\omega_{\rm stir}t+\phi_{\mathbf n}\right)\hat{S}_{\mathbf n}^{z}.
    \label{eq:probe}
\end{equation}
The modulation strength is $h_0=\tilde{\Omega}^{2}/(8\delta_e)$,
where we assume equal Rabi frequencies for the two probe beams,
$\Omega_{\tilde A}=\Omega_{\tilde B}\equiv\tilde{\Omega}$.
Eq.~(\ref{eq:probe}) describes a dipolar spatial profile of AC Stark shift which rotates at frequency $\omega_{\rm stir}$,
selectively coupling the $p$-wave and $d$-wave orders
(see App.~\ref{sec:meanfield}-5).
The individual beams also produce AC Stark shifts that are
either spatially uniform or have the same spatial dependence
as existing single-particle terms. These contributions can
be absorbed into the effective transition frequency
$\tilde{\omega}_0$ and kinetic-energy scale $J$.

In Fig.~\ref{fig:bardasis}(a), we apply a weak probe with $h_0/J=4\times10^{-3}$ to an initially pure $p_x+ip_y$ state ($\Delta_{d,0}=0$). 
For each value of $\omega_{\rm stir}$, we record the maximum of the induced $d$-wave amplitude within $Jt/2\pi\in [0,12]$, denoted by $|\Delta_{d,\rm max}|$. 
The response is resonantly enhanced when $\omega_{\rm stir}\simeq\omega_{\rm BS}$, allowing the BS-mode frequency to be determined spectroscopically.
The spectral resolution of $\omega_{\rm BS}$ also allows us to distinguish between a single BS mode and two BS modes in the system.
The small displacement of the resonance from the linear-response prediction (see Eq.~(\ref{eq:bsmode})) is due to the negative nonlinear frequency shift (softening of BS mode) for a finite $d$-wave excitation. 
Increasing $h_0$ shortens the time required to excite the mode and thereby reduces the influence of dissipation, at the cost of power broadening and a larger nonlinear frequency shift.

A finite $d$-wave response to the external modulation does not
by itself establish the existence of a BS mode, since the
modulation can induce $\Delta_d$ even outside the BS-mode regime.
We therefore use a pump-and-release protocol to test whether
the excitation persists after the modulation is switched off.
We first apply the rotating AC Stark shift described by
$\hat{H}_{\rm probe}$, with $h_0/J=4\times10^{-3}$, to excite
the $d$-wave channel, and then turn off the modulation.
Within the BS-mode regime, $|\Delta_d|$ remains finite and
the response continues as a freely evolving collective excitation
(see the upper panel of Fig.~\ref{fig:bardasis}(b)). Outside this regime,
the induced component decays once the modulation is removed
(see the lower panel of Fig.~\ref{fig:bardasis}(b)).
Note that in Fig.~\ref{fig:bardasis}(b) we keep $h_0>0$ until $Jt/2\pi=50$ to show the coherent oscillations induced by the external probe, while the pump-and-release protocol only requires approaching the first maximum near $Jt/2\pi=10$.
In both Fig.~\ref{fig:bardasis}(a) and (b), we also show that these signatures remain observable with cavity photon loss at experimentally realistic rates
(see App.~\ref{sec:cavity}-1 and App.~\ref{sec:numerical}-1).

\subsection{Dynamically induced competing orders}
The subdominant $d$-wave channel need not be excited by an external probe.
A sudden quench of the dominant $p$-wave interaction can dynamically destabilize the pure $p_x+ip_y$ state, causing an initially weak $d$-wave component to grow and, in turn, reorganize the $p$-wave dynamics. 
We investigate this mechanism by quenching the interaction strength from $\chi_{p,i}$ to $\chi_{p,f}$ in the presence of a finite $d$-wave interaction $\chi_d$, and determine how the emergent competing order modifies the dynamical phases of the system.

\begin{figure}[t]
    \centering
    \includegraphics[width=1.0\columnwidth]{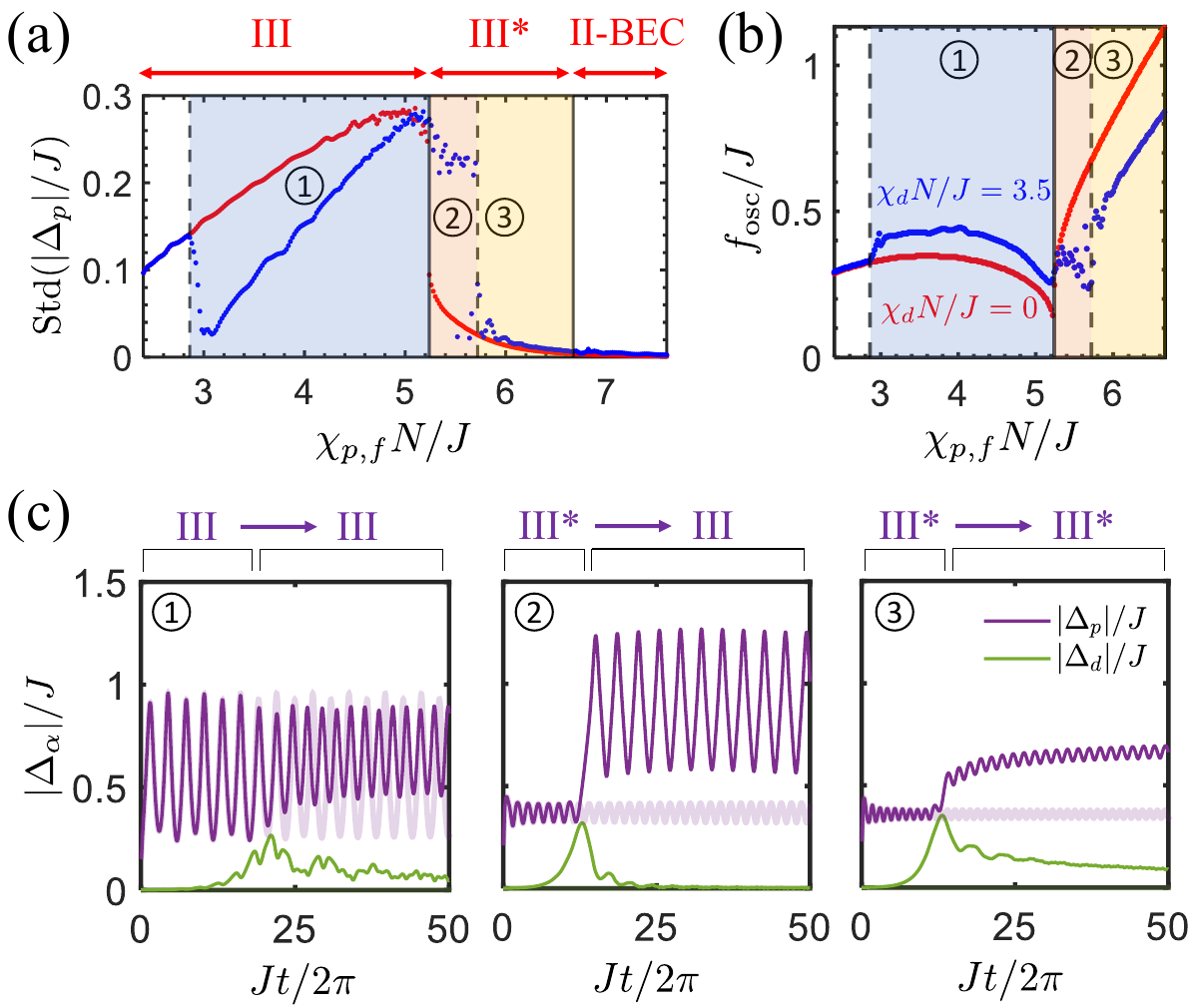}
    \caption{\textbf{Competing $p$-wave and $d$-wave orders away from equilibrium.} (a) We prepare the $p_x+ip_y$ eigenstate at $\chi_{p,i} N/J=1.0$ and $N_C/N=0.35$, and perform a sudden quench of the interaction strength to $\chi_{p,f}N/J$. The dynamical phases of the quench dynamics are characterized by the long-time standard deviation of $|\Delta_p|$ (see text). For the case of $\chi_d=0$ (red), we can identify three dynamical phases (phase III, phase III* and phase II-BEC) shown in Fig.~\ref{fig:dpt}(a). For the case of $\chi_d N/J=3.5$ (blue), a small initial $d$-wave pairing ($\varepsilon_d=10^{-2}$) can lead to significant changes in the quench dynamics, labeled as regime \circled{1}, \circled{2} and \circled{3}. (b) Oscillation frequency $f_{\rm osc}$ of phase III/III* dynamics as a function of $\chi_{p,f}$. The transition between phase III and phase III* is indicated by a frequency dip in the case of $\chi_d=0$. Additional frequency kinks are found in the case of $\chi_dN/J=3.5$. (c) Examples of quench dynamics in regime \circled{1}, \circled{2} and \circled{3} ($\chi_{p,f}N/J=4.0,5.6,6.0$ from left to right). The purple (green) lines describe dynamics of $|\Delta_p|/J$ ($|\Delta_d|/J$) at $\chi_{d,f}N/J=3.5$, and the lines with lighter colors describe dynamics at $\chi_{d,f}N/J=0$. The instability of $|\Delta_d|$ gives rise to an additional quench of the system and can be visualized in $|\Delta_p|$ dynamics. All the numerical results in this figure do not include dissipative effects, and the robustness against dissipation is discussed in App.~\ref{sec:numerical}-3.}
    \label{fig:dynamics}
\end{figure}

If $\chi_d\rightarrow 0$ (pure $p$-wave physics), as shown in Fig.~\ref{fig:dpt}(a), weak-to-strong quenches on $\chi_p$ can lead to either phase II or phase III/III* dynamics.
We further separate phase III and phase III* based on the sharp change of the long-time standard deviation (see red points in Fig.~\ref{fig:dynamics}(a)), $\mathrm{Std}(|\Delta_p|)=\sqrt{\mathrm{Avg}(|\Delta_p|^2)-\mathrm{Avg}(|\Delta_p|)^2}$, where $\mathrm{Avg}(|\Delta_p|)=\lim_{T\rightarrow\infty}
\frac{1}{T}\int_0^T |\Delta_p(t)|\,dt$ is the long-time-averaged order parameter.
This separation is not predicted by Ref.~\cite{Foster2013}, since our implemented model has a different density of states.
The transition between phase III and phase III* is also associated with a dip of oscillation frequency (see Fig.~\ref{fig:dynamics}(b)).
Similar phenomena of frequency dips have been observed in the $s$-wave BCS model \cite{Young2024}, while the analytical properties are distinct from the $s$-wave cases (see calculations via the Lax formalism \cite{Richardson1964,Gaudin1976,Dukelsky2004,Foster2013} in App.~\ref{sec:lax}).

For $\chi_d\neq 0$, we mainly focus on the case that $\chi_{p,f}$ and $\chi_d$ lie in the $p+ip$ regime of Fig.~\ref{fig:stability}(a).
In this regime, the equilibrium physics is dominated by the $p$-wave channel, since only the $p_x+ip_y$ solution is stable in this regime. 
In stark contrast, a finite $d$-wave interaction can play a non-negligible role for phase III/III* quench dynamics, as one can see a significant deviation in $\mathrm{Std}(|\Delta_p|)$ (see Fig.~\ref{fig:dynamics}(a)) and oscillation frequency (see Fig.~\ref{fig:dynamics}(b)) between the $\chi_d=0$ (red points) and  $\chi_d\neq 0$ cases (blue points).
Here we perturb the $p_x+ip_y$ solution of $\chi_{p,i}$ with a small amount of $d_{x^2-y^2}+id_{xy}$ pairing (see Eq.~(\ref{eq:smallperb}) with $\varepsilon_d = 10^{-2}$).
We then evolve the system under interaction strengths $\chi_{p,f}$ and $\chi_d$ (setting $\chi_dN/J=3.5$ in Fig.~\ref{fig:dynamics}).
Note that Ref.~\cite{Zabalo2021} also predicts a significant change of phase III dynamics in the presence of competing $p_x-ip_y$ interactions.

In Fig.~\ref{fig:dynamics}(c), we further analyze the impact of a small perturbation of $d$-wave pairing initially in phase III/III* dynamics.
As we increase $\chi_{p,f}$, we identify three different regimes exhibiting exponential growth of $|\Delta_d|$ (labeled as \circled{1}, \circled{2} and \circled{3}).
When $|\Delta_d|$ reaches its maximum value, the instability of $|\Delta_d|$ gives rise to an additional quench of the system, indicated by the sudden growth of $|\Delta_p|$ seeded by the presence of $|\Delta_d|$.
The three different regimes are separated by the behavior of oscillation amplitude and frequency of $|\Delta_p|$ compared to the $\chi_d=0$ case, after the abrupt change induced by $|\Delta_d|$.
In the left panel (regime \circled{1}), $|\Delta_p|$ transitions to a smaller oscillation amplitude and larger frequency within phase III.
In the middle panel (regime \circled{2}), $|\Delta_p|$ switches from phase III* to phase III, which is shown in the shift of the $\mathrm{Std}(|\Delta_p|)$ jump in Fig.~\ref{fig:dynamics}(a) and the extra frequency dip in Fig.~\ref{fig:dynamics}(b). 
In the right panel (regime \circled{3}), $|\Delta_p|$ transitions to a smaller frequency and a similar oscillation amplitude within phase III*.
We further analyze the existence of these three regimes for different $\chi_d$ values and the robustness against dissipation in App.~\ref{sec:numerical}-3.
These results demonstrate that even weak competing orders, while irrelevant in equilibrium, can provoke dramatic changes in the nonequilibrium phase structure. 

\section{Conclusion and outlook}

We have introduced a cavity QED platform that realizes competing topological $p_x+ip_y$ and $d_{x^2-y^2}+id_{xy}$ superconducting orders by mapping momentum-space pairing physics onto spatially structured light–matter interactions. 
We proposed an efficient and experimentally realistic ground-state preparation protocol and the continuous and non-destructive monitoring of order parameters, enabling direct access to equilibrium and non-equilibrium topological phase transitions, and the competition and coexistence of unconventional pairing orders.

A central result is the controlled excitation and detection of Bardasis-Schrieffer modes in the subdominant pairing channel. 
By tuning the relative interaction strengths, the mode can be followed as it softens toward the stability boundary of the dominant superconducting order. Its frequency can be measured either directly from the relative phase of the two order parameters or spectroscopically using an angular-momentum-selective probe, and remains resolvable in the presence of realistic cavity photon loss. 
When quenching the dominant superconducting order, we find that a weak subdominant channel can grow dynamically and abruptly restructure the amplitude and frequency of the dominant order, even when it has little influence on the equilibrium state.

Several extensions can broaden the scope of the platform. 
Resolving the full pseudospin texture would provide direct access to topological invariants, while an auxiliary atomic level could enable radio-frequency-style spectroscopy of the excitation spectrum~\cite{Young2025}. 
Engineering additional spatial mode structures would allow several pairing channels to be studied simultaneously.
Beyond mean-field physics, it will enable investigation of spectral signatures and quantum chaos in both integrable and non-integrable BCS models \cite{Gaur2024}.
Importantly, by trapping two fermionic atoms per lattice site and mapping $|\uparrow\rangle \rightarrow \hat{c}^{\dagger}_{\mathbf{n},\sigma}\hat{c}^{\dagger}_{\mathbf{n},\sigma'}|0\rangle$, the simulator can directly incorporate fermionic degrees of freedom, which might provide a route to investigate emergent boundary effects and to search for Majorana zero modes in a fully controlled setting.

\begin{acknowledgments}
We thank Dylan J. Young, Allison Carter, Chitose Maruko, Athreya Shankar and Peter Zoller for useful discussions.
This work is supported by the Vannevar-Bush Faculty Fellowship, by the U.S. Department of Energy, Office of Science, National Quantum Information Science Research Centers, Quantum Systems Accelerator, by the National Science Foundation under Grant Numbers 1734006 (Physics Frontier Center), by the Simons Collaboration on Ultra-Quantum Matter (Grant No. 651440), and by NIST.
A. C. acknowledges support from the Simons Foundation (Grant No. 669487) during the completion of this work.
\end{acknowledgments}

\newpage
\appendix

\section{Details of the cavity QED setup}
\label{sec:cavity}

\subsection{Effective Hamiltonian for topological BCS pairing}
As discussed in the main text (see Fig.~\ref{fig:schematic}), we consider an ensemble of $N$ atoms with a Raman coupling between ground state levels trapped in a 2D optical lattice. 
The $x$ direction of the lattice is along a standing-wave optical cavity with an external laser drive, and the $y$ direction of the lattice is along a running-wave laser beam. 
We consider a standing-wave cavity that supports two cavity modes with different polarizations (labelled by $r$ and $b$) and the same frequency $\omega_c$.
The two laser drives $A$ and $B$ have frequencies $\omega_{p,A}$ from the side and $\omega_{p,B}$ along the cavity supported by cavity mode $r$, and are coupled to the $|\uparrow\rangle\rightarrow|e\rangle$ transition (frequency $\omega_e$) with Clebsch-Gordan coefficient $\mathcal{C}_{\uparrow,e}$.
The cavity mode $b$ is coupled to the $|\downarrow\rangle\rightarrow|e\rangle$ transition (frequency $\omega_e+\omega_0$) with Clebsch-Gordan coefficient $\mathcal{C}_{\downarrow,e}$.
For simplicity, we assume there are no other possible excited states that the laser drives and cavity modes can couple to.
We consider $\omega_c,\omega_{p,A},\omega_{p,B}$ have nearly the same wavelength ($\lambda_c=2\pi c/\omega_c$), which is incommensurate with the lattice wavelength ($\lambda_l$), generating inhomogeneities in the atom-light couplings. The atom-cavity couplings can be written as
\begin{equation}
    g_{\mathbf{n},r}=g_c\frac{\mathcal{C}_{\uparrow,e}}{\mathcal{C}_{\downarrow,e}}\cos(n_x\varphi), \quad g_{\mathbf{n},b}=g_c\cos(n_x\varphi),
\end{equation}
where $2g_c$ is the peak single photon Rabi frequency, $\varphi=\pi\lambda_l/\lambda_c$ is the laser phase difference between nearest-neighbor sites, and we use $\mathbf{n}=(n_x,n_y)$ to label the lattice sites. 
The Rabi frequency of the laser drive $A$ can be written as
\begin{equation}
    \Omega_{\mathbf{n},A}=\frac{\Omega_A}{2} e^{in_y\varphi}.
\end{equation}
Laser drive $B$ is generating a coherent pump of the cavity mode $r$ with amplitude $\epsilon$. Therefore, the Hamiltonian for this system takes the following form, $\hat{H}=\hat{H}_{\rm A}+\hat{H}_{\rm L}+\hat{H}_{\rm AL}$, with
\begin{subequations}
    \begin{gather}
        \begin{aligned}
        \hat{H}_{\rm A}/\hbar&=\sum_{\mathbf{n}}\omega_0 |\uparrow_{\mathbf{n}}\rangle\langle\uparrow_{\mathbf{n}}\negmedspace|+\sum_{\mathbf{n}}(\omega_e+\omega_0)|e_{\mathbf{n}}\rangle\langle e_{\mathbf{n}}|\\
        &+\sum_{\mathbf{n}}\Big(\Omega_{\mathbf{n},A}e^{-i\omega_{p,A}t}|e_{\mathbf{n}}\rangle\langle\uparrow_{\mathbf{n}}\negmedspace|+\mathrm{h.c.}\Big),
        \end{aligned}\\
        \hat{H}_{\rm L}/\hbar=\omega_{c} (\hat{a}_r^{\dag}\hat{a}_r+\hat{a}^{\dag}_b\hat{a}_b)+\epsilon e^{-i\omega_{p,B}t}\hat{a}_r^{\dag}+\epsilon^{*}e^{i\omega_{p,B}t} \hat{a}_r,\\
        \begin{aligned}
        \hat{H}_{\rm AL}/\hbar&=\sum_{\mathbf{n}}g_{\mathbf{n},r}\Big(\hat{a}_r|e_{\mathbf{n}}\rangle\langle\uparrow_{\mathbf{n}}\negmedspace|+\mathrm{h.c.}\Big)\\
        &+\sum_{\mathbf{n}}g_{\mathbf{n},b}\Big(\hat{a}_b|e_{\mathbf{n}}\rangle\langle\downarrow_{\mathbf{n}}\negmedspace|+\mathrm{h.c.}\Big).
        \end{aligned}
    \end{gather}
\end{subequations}

We first adiabatically eliminate the excited state $|e\rangle$ using  second-order perturbation theory, justified by assuming that the detuning of the drives to the  atomic transition $\delta_e=\omega_{p,A}-\omega_e$ is the dominant frequency scale (see Eq.~(\ref{eq:adia1}) for the detailed conditions). In particular, we assume $\delta_e\gg\gamma$ with $\gamma$ the spontaneous emission rate of $|e\rangle$, so the excited state decay is suppressed. The effects of residual spontaneous scattering depends on the details of atomic structure, as discussed in App.~\ref{sec:experiment}. The system dynamics can thus be described by the following effective Hamiltonian, $\hat{H}'= \hat{H}'_{\rm A} + \hat{H}'_{\rm L} + \hat{H}'_{\rm AL}$, with
\begin{subequations}
    \begin{gather}
        \hat{H}'_{\rm A}/\hbar = \sum_{\mathbf{n}}\tilde{\omega}_0 \hat{S}^z_{\mathbf{n}},\\
        \hat{H}'_{\rm L}/\hbar = \tilde{\omega}_{c,r} \hat{a}_r^{\dag}\hat{a}_r + \tilde{\omega}_{c,b} \hat{a}^{\dag}_b\hat{a}_b + \epsilon e^{-i\omega_{p,B}t}\hat{a}_r^{\dag}+\epsilon^{*}e^{i\omega_{p,B}t} \hat{a}_r,\\
        \begin{aligned}
        \hat{H}'_{\rm AL}/\hbar &= \frac{1}{\delta_e}\sum_{\mathbf{n}}\Big(g_{\mathbf{n},r}^2\hat{a}^{\dag}_r\hat{a}_r-g_{\mathbf{n},b}^2\hat{a}^{\dag}_b\hat{a}_b\Big)\hat{S}^z_{\mathbf{n}}\\
        &+\frac{1}{\delta_e}\sum_{\mathbf{n}}\Big(\Omega_{\mathbf{n},A}^{*}e^{i\omega_{p,A}t}g_{\mathbf{n},r}\hat{a}_{r}\hat{S}^z_{\mathbf{n}}+\mathrm{h.c.}\Big)\\
        &+ \frac{1}{\delta_e}\sum_{\mathbf{n}}\Big((\Omega_{\mathbf{n},A}^{*}e^{i\omega_{p,A}t}+g_{\mathbf{n},r}\hat{a}_r^{\dag})g_{\mathbf{n},b}\hat{a}_b\hat{S}^{+}_{\mathbf{n}}+\mathrm{h.c.}\Big),
        \end{aligned}
    \end{gather}
    \label{eq:a4}
\end{subequations}
where $\tilde{\omega}_0=\omega_0+|\Omega_A|^2/(4\delta_e)$ is the transition frequency between $|\uparrow\rangle$ and $|\downarrow\rangle$ states corrected by AC Stark shifts of drive $A$, $\tilde{\omega}_{c,r}=\omega_c+\sum_{\mathbf{n}}g_{\mathbf{n},r}^2/(2\delta_e)$ and $\tilde{\omega}_{c,b}=\omega_c+\sum_{\mathbf{n}}g_{\mathbf{n},b}^2/(2\delta_e)$ are the frequencies of the dressed cavity resonances for mode $r$ and $b$ respectively.

Apart from Hamiltonian dynamics, we also consider the cavity photon loss with a rate $\kappa$. The full system dynamics is thus captured by the following Lindblad master equation,
\begin{equation}
    \frac{d}{dt}\hat{\rho}=-\frac{i}{\hbar}[\hat{H}',\hat{\rho}] + \mathcal{D}[\hat{L}_r]\hat{\rho} + \mathcal{D}[\hat{L}_b]\hat{\rho},
    \label{eq:hp}
\end{equation}
where $\mathcal{D}[\hat{L}_{\mu}]\hat{\rho}=\hat{L}_{\mu}\hat{\rho}\hat{L}_{\mu}^{\dag}-(\hat{L}_{\mu}^{\dag}\hat{L}_{\mu}\hat{\rho}+\hat{\rho}\hat{L}_{\mu}^{\dag}\hat{L}_{\mu})/2$, with 
\begin{equation}
    \hat{L}_{r}=\sqrt{\kappa}\hat{a}_r, \quad \hat{L}_{b}=\sqrt{\kappa}\hat{a}_b.
\end{equation}

Here we go to the rotating frame of $\omega_{p,B}$ and $\tilde{\omega}_0$, and then expand the operator $\hat{a}_r$ in the following way,
\begin{equation}
    \hat{a}_r = \alpha_r + \hat{b}_r, \quad \alpha_r=\frac{\epsilon}{(\omega_{p,B}-\tilde{\omega}_{c,r})+i\kappa/2}.
\end{equation}
This allows us to define the Rabi frequency of the laser drive $B$,
\begin{equation}
    \Omega_{\mathbf{n},B} = g_{\mathbf{n},r}\alpha_r = \frac{\Omega_B}{2}\cos(n_x\varphi),
    \label{eq:omegab}
\end{equation}
where $\Omega_B=2g_c\alpha_r\mathcal{C}_{\uparrow,e}/\mathcal{C}_{\downarrow,e}$. Keeping the terms in Eq.~(\ref{eq:hp}) up to linear order of $\hat{b}_r$ and $\hat{a}_b$, we obtain the following Hamiltonian terms,
\begin{subequations}
    \begin{gather}
        \begin{aligned}
        \hat{H}'_{\rm A}/\hbar &= \sum_{\mathbf{n}}\frac{|\Omega_{\mathbf{n},B}|^2}{\delta_e}\hat{S}^z_{\mathbf{n}}\\
        &{ +\sum_{\mathbf{n}}\bigg(\frac{\Omega_{\mathbf{n},A}^{*}\Omega_{\mathbf{n},B}}{\delta_e}e^{i(\omega_{p,A}-\omega_{p,B})t}+\mathrm{h.c.}\bigg)\hat{S}^z_{\mathbf{n}} },
        \end{aligned}\\
        \hat{H}'_{\rm L}/\hbar = -\delta'_{c,B} \hat{b}_r^{\dag}\hat{b}_r -\delta_{c,B} \hat{a}^{\dag}_b\hat{a}_b,\\
        \begin{aligned}
        \hat{H}'_{\rm AL}/\hbar &= \frac{1}{\delta_e}\sum_{\mathbf{n}}(\Omega_{\mathbf{n},A}^{*}e^{i(\omega_{p,A}-\omega_{p,B})t}+\Omega_{\mathbf{n},B}^{*})g_{\mathbf{n},r}\hat{b}_{r}\hat{S}^z_{\mathbf{n}}\\
        &+ \frac{1}{\delta_e}\sum_{\mathbf{n}}(\Omega_{\mathbf{n},A}^{*}e^{i(\omega_{p,A}-\omega_{p,B})t}+\Omega_{\mathbf{n},B}^{*})g_{\mathbf{n},b}\hat{a}_b\hat{S}^{+}_{\mathbf{n}}\\
        &+\mathrm{h.c.},
        \end{aligned}
    \end{gather}
    \label{eq:9}
\end{subequations}
as well as jump operators
\begin{equation}
    \hat{L}_{r}=\sqrt{\kappa}\hat{b}_r, \quad \hat{L}_{b}=\sqrt{\kappa}\hat{a}_b.
    \label{eq:10}
\end{equation}
Here we define the detunings to the dressed cavity resonance of mode $b$, $\delta_{c,A}=\omega_{p,A}+\tilde{\omega}_0-\tilde{\omega}_{c,b}$, $\delta_{c,B}=\omega_{p,B}+\tilde{\omega}_0-\tilde{\omega}_{c,b}$, and the detunings to the dressed cavity resonance of mode $r$, $\delta'_{c,A}=\omega_{p,A}-\tilde{\omega}_{c,r}$, $\delta'_{c,B}=\omega_{p,B}-\tilde{\omega}_{c,r}$.

We then eliminate the cavity modes $r$ and $b$ based on the Reiter-Sorensen approach \cite{Reiter2012}, assuming $\delta_{c,A},\delta_{c,B},\delta'_{c,A},\delta'_{c,B}$ are the dominant frequency scales (see Eq.~(\ref{eq:adia2}) for the detailed conditions). Here we consider the ground manifold is the zero photon subspace for both $\hat{b}_r$ and $\hat{a}_b$ modes, and the excited manifold has a single photon excitation. 
We then drop the fast rotating term at frequency $\omega_{p,A}-\omega_{p,B}$, and get the effective Lindblad master equation as follows,
\begin{equation}
    \frac{d}{dt}\hat{\rho}=-\frac{i}{\hbar}[\hat{H}_{\rm eff},\hat{\rho}] + \mathcal{D}[\hat{L}_{p}]\hat{\rho} + \mathcal{D}[\hat{L}_{d}]\hat{\rho} + \mathcal{D}[\hat{L}'_{p}]\hat{\rho} + \mathcal{D}[\hat{L}'_{d}]\hat{\rho},
\end{equation}
where
\begin{equation}
    \begin{aligned}
    \frac{\hat{H}_{\rm eff}}{\hbar} &= \sum_{\mathbf{n}} 2J\eta_{\mathbf{n}}^2\hat{S}^z_{\mathbf{n}} - \chi_p\sum_{\mathbf{n}\mathbf{m}} \eta_{\mathbf{n}}\eta_{\mathbf{m}}e^{i\phi_{\mathbf{n}}}e^{-i\phi_{\mathbf{m}}}\hat{S}^{+}_{\mathbf{n}}\hat{S}^{-}_{\mathbf{m}}\\
    &-\chi_d\sum_{\mathbf{n}\mathbf{m}}\eta_{\mathbf{n}}^2\eta_{\mathbf{m}}^2\hat{S}^{+}_{\mathbf{n}}\hat{S}^{-}_{\mathbf{m}}-\chi'_p\sum_{\mathbf{n}\mathbf{m}} \eta_{\mathbf{n}}\eta_{\mathbf{m}}e^{i\phi_{\mathbf{n}}}e^{-i\phi_{\mathbf{m}}}\hat{S}^{z}_{\mathbf{n}}\hat{S}^{z}_{\mathbf{m}}\\
    &-\chi'_d\sum_{\mathbf{n}\mathbf{m}}\eta_{\mathbf{n}}^2\eta_{\mathbf{m}}^2\hat{S}^{z}_{\mathbf{n}}\hat{S}^{z}_{\mathbf{m}},
    \end{aligned}
\end{equation}
\begin{equation}
    \hat{L}_{p}= \sqrt{\Gamma_p} \sum_{\mathbf{n}}\eta_{\mathbf{n}}e^{-i\phi_{\mathbf{n}}}\hat{S}^-_{\mathbf{n}}, \quad \hat{L}_{d}= \sqrt{\Gamma_d} \sum_{\mathbf{n}}\eta_{\mathbf{n}}^2\hat{S}^-_{\mathbf{n}},
    \label{eq:jump1}
\end{equation}
\begin{equation}
    \hat{L}'_{p}= \sqrt{\Gamma'_p} \sum_{\mathbf{n}}\eta_{\mathbf{n}}e^{-i\phi_{\mathbf{n}}}\hat{S}^z_{\mathbf{n}}, \quad \hat{L}'_{d}= \sqrt{\Gamma'_d} \sum_{\mathbf{n}}\eta_{\mathbf{n}}^2\hat{S}^z_{\mathbf{n}}.
\end{equation}
Here, the strength of inhomogeneous AC Stark shift due to drive $B$ is given by
\begin{equation}
    J = \frac{|\Omega_B|^2}{8\delta_e},
\end{equation}
and we label the dimensionless atom-light coupling by
\begin{equation}
    \eta_{\mathbf{n}}e^{i\phi_{\mathbf{n}}}=\cos(n_x\varphi)e^{-in_y\varphi}.
\end{equation}
The spin-exchange interaction strengths mimicking $p$-wave (due to drive $A$) and $d$-wave (due to drive $B$) pairing are given by 
\begin{equation}
    \begin{aligned}
    \chi_p&=-\frac{|\Omega_A|^2g_c^2\delta_{c,A}}{4\delta_e^2(\delta_{c,A}^2+\kappa^2/4)}, \\
    \chi_d&=-\frac{|\Omega_B|^2g_c^2\delta_{c,B}}{4\delta_e^2(\delta_{c,B}^2+\kappa^2/4)}.
    \end{aligned}
\end{equation}
Similarly, the Ising-type interaction strengths are given by
\begin{equation}
    \begin{aligned}
    \chi'_p&=-\frac{|\Omega_A|^2g_c^2\delta'_{c,A}}{4\delta_e^2(\delta_{c,A}^{\prime 2}+\kappa^2/4)}\frac{\mathcal{C}_{\uparrow,e}^2}{\mathcal{C}_{\downarrow,e}^2}, \\ \chi'_d&=-\frac{|\Omega_B|^2g_c^2\delta'_{c,B}}{4\delta_e^2(\delta_{c,B}^{\prime 2}+\kappa^2/4)}\frac{\mathcal{C}_{\uparrow,e}^2}{\mathcal{C}_{\downarrow,e}^2}.
    \end{aligned}
\end{equation}
One can also obtain the rates of the dissipative processes by
\begin{equation}
    \begin{aligned}
    \Gamma_p &= \kappa\bigg|\frac{\chi_p}{\delta_{c,A}}\bigg|, \quad \Gamma_d = \kappa\bigg|\frac{\chi_d}{\delta_{c,B}}\bigg|,\\
    \Gamma'_p &= \kappa\bigg|\frac{\chi'_p}{\delta'_{c,A}}\bigg|, \quad \Gamma'_d = \kappa\bigg|\frac{\chi'_d}{\delta'_{c,B}}\bigg|.
    \end{aligned}
\end{equation}

In the regime $|\chi_p|\;\gg |\chi'_p|$, $|\chi_d|\;\gg |\chi'_d|$, we can simply ignore the Ising-type interactions and the corresponding jump operators. This allows us to simplify the effective Lindblad master equation as,
\begin{equation}
    \frac{d}{dt}\hat{\rho}=-\frac{i}{\hbar}[\hat{H}_{\rm eff},\hat{\rho}] + \mathcal{D}[\hat{L}_{p}]\hat{\rho} + \mathcal{D}[\hat{L}_{d}]\hat{\rho},
\end{equation} 
where the Hamiltonian becomes
\begin{equation}
    \begin{aligned}
    \hat{H}_{\rm eff}/\hbar&=\sum_{\mathbf{n}} 2J\eta_{\mathbf{n}}^2\hat{S}^z_{\mathbf{n}} - \chi_p\sum_{\mathbf{n}\mathbf{m}} \eta_{\mathbf{n}}\eta_{\mathbf{m}}e^{i\phi_{\mathbf{n}}}e^{-i\phi_{\mathbf{m}}}\hat{S}^{+}_{\mathbf{n}}\hat{S}^{-}_{\mathbf{m}}\\
    &-\chi_d\sum_{\mathbf{n}\mathbf{m}}\eta_{\mathbf{n}}^2\eta_{\mathbf{m}}^2\hat{S}^{+}_{\mathbf{n}}\hat{S}^{-}_{\mathbf{m}},
    \end{aligned}
\end{equation}
and the jump operators $\hat{L}_p$ and $\hat{L}_d$ are still given by Eq.~(\ref{eq:jump1}).

To match the form of $p$-wave and $d$-wave pairing interactions, we consider the gauge transformation $\hat{S}^{+}_{\mathbf{n}}\rightarrow \hat{S}^{+}_{\mathbf{n}}e^{-2i\phi_{\mathbf{n}}}$, $\hat{S}^{-}_{\mathbf{n}}\rightarrow \hat{S}^{-}_{\mathbf{n}}e^{2i\phi_{\mathbf{n}}}$, leading to
\begin{equation}
    \begin{aligned}
    \hat{H}_{\rm cav}/\hbar&=\sum_{\mathbf{n}} 2J\eta_{\mathbf{n}}^2\hat{S}^z_{\mathbf{n}} - \chi_p\sum_{\mathbf{n}\mathbf{m}} \eta_{\mathbf{n}}\eta_{\mathbf{m}}e^{-i\phi_{\mathbf{n}}}e^{i\phi_{\mathbf{m}}}\hat{S}^{+}_{\mathbf{n}}\hat{S}^{-}_{\mathbf{m}}\\
    &-\chi_d\sum_{\mathbf{n}\mathbf{m}}\eta_{\mathbf{n}}^2\eta_{\mathbf{m}}^2e^{-2i\phi_{\mathbf{n}}}e^{2i\phi_{\mathbf{m}}}\hat{S}^{+}_{\mathbf{n}}\hat{S}^{-}_{\mathbf{m}},
    \label{eq:heff}
    \end{aligned}
\end{equation}
and the jump operators become
\begin{equation}
    \hat{L}_{p}= \sqrt{\Gamma_p} \sum_{\mathbf{n}}\eta_{\mathbf{n}}e^{i\phi_{\mathbf{n}}}\hat{S}^-_{\mathbf{n}}, \quad \hat{L}_{d}= \sqrt{\Gamma_d} \sum_{\mathbf{n}}\eta_{\mathbf{n}}^2e^{2i\phi_{\mathbf{n}}}\hat{S}^-_{\mathbf{n}}.
    \label{eq:jump}
\end{equation}
Note that in real superconducting materials, the kinetic energy of electrons is positive, and the electron-electron interactions mediated by phonons are attractive. So we typically consider the case with $J,\chi_p,\chi_d>0$. 

\subsection{Continuous readout of BCS order parameters}
From the Lindblad master equation with Hamiltonian described by Eq.~(\ref{eq:9}) and jump operators described by Eq.~(\ref{eq:10}), one can derive the mean-field equation for the coherent state amplitude of cavity mode $b$, i.e. $\alpha_b=\langle \hat{a}_b\rangle$. For simplicity, here we go to the rotating frame of $\tilde{\omega}_{c,b}$. The mean-field equation is given by
\begin{equation}
    \frac{d}{dt}\alpha_b = -\frac{\kappa}{2}\alpha_b - i e^{-i\delta_{c,A}t}\frac{\Omega_A g_c}{2\delta_e\chi_p}\Delta_p-ie^{-i\delta_{c,B}t}\frac{\Omega_B g_c}{2\delta_e\chi_d}\Delta_d.
\end{equation}
Here we assume $\Delta_p(t)$ and $\Delta_d(t)$ are slowly varying compared to the frequency scale $\delta_{c,A}$ and $\delta_{c,B}$, so the mean-field equation above leads to
\begin{equation}
    \alpha_b(t) \approx \frac{\Delta_p(t)}{\mathcal{G}_p^{*}}e^{-i\delta_{c,A}t} + \frac{\Delta_d(t)}{\mathcal{G}_d^{*}}e^{-i\delta_{c,B}t},
    \label{eq:cavityoutput}
\end{equation}
where
\begin{equation}
    \mathcal{G}_p=-\frac{\Omega_Ag_c\delta_{c,A}}{2\delta_e(\delta_{c,A}+i\kappa/2)}, \quad \mathcal{G}_d=-\frac{\Omega_Bg_c\delta_{c,B}}{2\delta_e(\delta_{c,B}+i\kappa/2)}.
\end{equation}
We also assume dynamical time scales are longer than $1/\kappa$ so that we can drop the transient response of the cavity field. 
By using heterodyne detection, one can extract $\Delta_p(t)$ by tracking photons emitted at frequency $\omega_{p,A}+\tilde{\omega}_0$, and extract $\Delta_d(t)$ by tracking photons at emitted photon frequency $\omega_{p,B}+\tilde{\omega}_0$.

Let's suppose we choose a time step $t_{\rm step}$ such that $\Delta_p(t)$ and $\Delta_d(t)$ have nearly constant values. Then the number of photons leaking out of the cavity during the time step is given by $\bar{n} = \bar{n}_p+\bar{n}_d$,
\begin{equation}
    \bar{n}_p = \int_{t_0}^{t_0+t_{\rm step}} dt\;\kappa \frac{|\Delta_p(t)|^2}{\mathcal{G}_p^2} \approx 2\pi\xi_p N \frac{\kappa}{|\delta_{c,A}|} \bigg|\frac{\Delta_p(t_0)}{\chi_p N}\bigg|^2,
\end{equation}
where we set $\chi_p Nt_{\rm step}/2\pi=\xi_p$. The approximation above is typically valid for $\xi_p\ll 1$, while $\xi_p$ can have a larger value in the case that the system is dominated by $d$-wave pairing or approaching a steady state. 
One can apply similar calculations to $\bar{n}_d$. 
Note that the signal-to-noise ratio (SNR) of $\Delta_p(t_0)$ is proportional to $(\bar{n}_p)^{1/2}$.
Since we require $|\delta_{c,A}|\;\gg \kappa$ to suppress dissipative effects (Eq.~(\ref{eq:jump})) and measurement back-action, $\bar{n}_p$ is a small number compared to total atom number $N$.
To increase the SNR of $\Delta_p(t)$ and $\Delta_d(t)$, one can average the result over many repetitions.

\subsection{Summary of approximations}
\begin{itemize}
    \item Adiabatic elimination of atomic excited states
    \begin{equation}
        |\delta_e|\;\gg |\omega_{p,A}-\omega_{p,B}|, |\omega_{p,A}-\omega_c|, g_c\sqrt{N}, |\Omega_{A}|, |\Omega_{B}|, \gamma
        \label{eq:adia1}
    \end{equation}
    In the relation above, the condition $|\delta_e|\;\gg |\omega_{p,A}-\omega_{p,B}|$  ensures the drive $A$ and drive $B$ have similar detunings to the atomic excited state (mainly for simplifying notations, does not affect the validity of adiabatic elimination). The inequality $|\delta_e|\;\gg |\omega_{p,A}-\omega_{c}|, g_c\sqrt{N}$   ensures the atomic resonances are not strongly coupled with the cavity resonance. The relation  $|\delta_e|\;\gg |\Omega_{A}|, |\Omega_{B}|$  guarantees  a low population in the atomic excited states. And finally  $|\delta_e|\;\gg \gamma$ enforces that  unitary dynamics is dominant over dissipation effects due to atomic excited states.

    \item Adiabatic elimination of cavity modes
    \begin{equation}
        |\delta_{c,A}|, |\delta_{c,B}|, |\delta'_{c,A}|, |\delta'_{c,B}|\;\gg \bigg|\frac{g_c\Omega_A}{\delta_e}\bigg|\sqrt{N},  \bigg|\frac{g_c\Omega_B}{\delta_e}\bigg|\sqrt{N}, \kappa
        \label{eq:adia2}
    \end{equation}

    The relation  $|\delta_{c,A}|, |\delta_{c,B}|, |\delta'_{c,A}|, |\delta'_{c,B}|\;\gg |\frac{g_c\Omega_A}{\delta_e}|\sqrt{N},  |\frac{g_c\Omega_B}{\delta_e}|\sqrt{N}$  ensures low photon excitations in the cavity modes that couple  the excited state to the atomic ground state levels. The second inequality, $|\delta_{c,A}|, |\delta_{c,B}|, |\delta'_{c,A}|, |\delta'_{c,B}|\;\gg \kappa$  ensures that coherent photon exchange dominates cavity loss.

    \item Ignore quantum fluctuations of cavity mode $r$
    \begin{equation}
        |\chi_p|\;\gg |\chi'_p|, \quad |\chi_d|\;\gg |\chi'_d|
        \label{eq:adia3}
    \end{equation}
    
    This condition is to ensure that photon-mediated interactions generated by cavity mode $r$ are negligible compared to those generated by cavity mode $b$. This is achievable by either $|\delta_{c,A}|\;\ll |\delta'_{c,A}|$, $|\delta_{c,B}|\;\ll |\delta'_{c,B}|$ or $|\mathcal{C}_{\downarrow,e}|\;\gg |\mathcal{C}_{\uparrow,e}|$.

    \item Ignore interference effects between drives $A$ and $B$
    \begin{equation}
        |\omega_{p,A}-\omega_{p,B}|\;\gg J, \chi_p N, \chi_d N
        \label{eq:interference}
    \end{equation}

    This condition  ensures  no interference effects between $p$-wave and $d$-wave interactions, and that the information of $\Delta_p(t)$ and $\Delta_d(t)$ is frequency resolved in the emitted light of cavity mode $b$.

    \item The same laser phase difference $\varphi$ between nearest-neighbor sites
    \begin{equation}
        |\omega_{p,A}-\omega_{p,B}|, |\omega_{p,A}-\omega_{c}|\;\ll \frac{4\pi c}{\lambda_l N_x}, \frac{4\pi c}{\lambda_l N_y}
        \label{eq:a32}
    \end{equation}
    Here $N_x$ is the number of lattice sites in $x$ direction, $N_y$ is the number of lattice sites in $y$ direction. Considering $c/\lambda_l\sim 10^{14}$~Hz, this condition can be easily satisfied for $N_x,N_y <10^4$, which requires that frequency differences between $\omega_{p,A},\omega_{p,B},\omega_c$ are below $10$GHz.

    \item Ignore Gaussian profile of cavity modes and laser drives from the side
    \begin{equation}
        \lambda_l N_x \ll 4z_{\rm cav}, \quad \lambda_l N_y \ll 4w_{\rm cav}
        \label{eq:a33}
    \end{equation}
    \begin{equation}
        \lambda_l N_y \ll 4z_{A}, \quad \lambda_l N_x \ll 4w_{A}
        \label{eq:a34}
    \end{equation}
    Here, $z_{\rm cav}$ and $w_{\rm cav}$ are the Rayleigh length and beam waist of the cavity mode Gaussian profile, respectively.
    Similarly, $z_A$ and $w_A$ are the Rayleigh length and beam waist of laser drive $A$ Gaussian profile, respectively.
    
\end{itemize}

\begin{figure}[t]
    \centering
    \includegraphics[width=1.0\columnwidth]{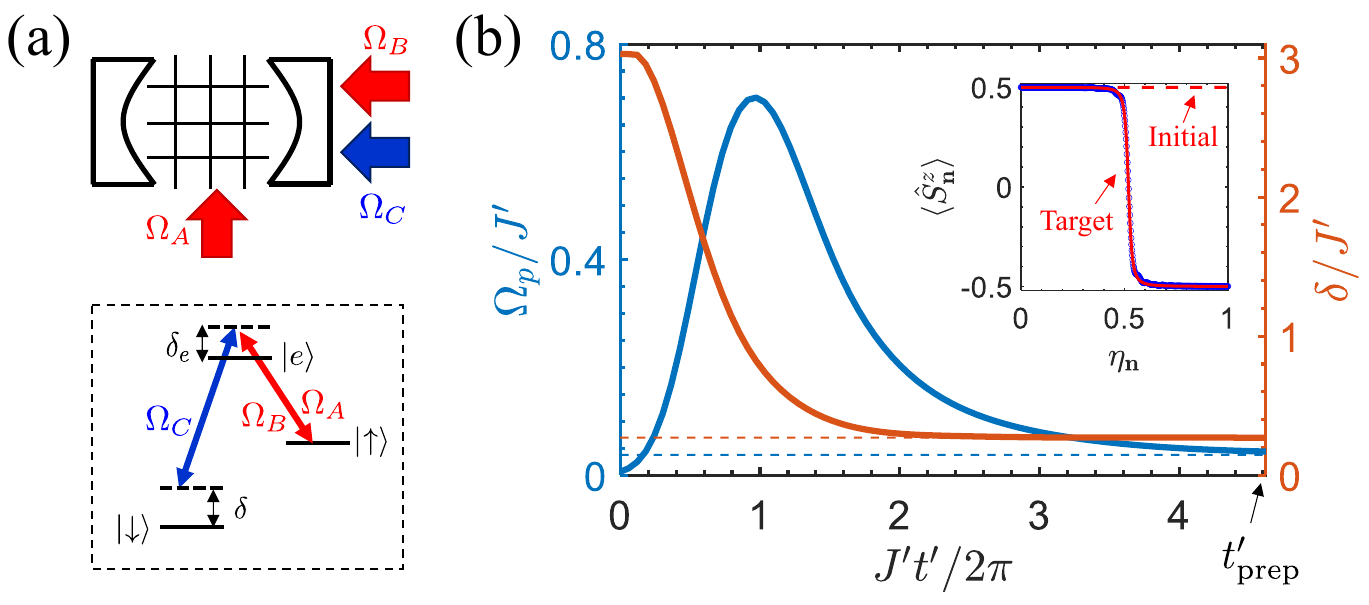}
    \caption{(a) Schematic of laser drives for initial state preparation at the mean-field level. We apply external laser drive $A$ from the side, $B$ and $C$ along the cavity. Drives $A$ and $B$ couple the $|\uparrow\rangle\rightarrow|e\rangle$ transition, and $C$ couples the $|\downarrow\rangle\rightarrow|e\rangle$ transition, forming a Raman coupling between $|\uparrow\rangle$ and $|\downarrow\rangle$ states. (b) The optimized ramp with effective Rabi frequency $\Omega_p(t')$ and detuning $\delta(t')$ for the preparation of $p_x+ip_y$ state with $\chi_{p,i}N/J=1$ and Cooper pair number $N_C/N=0.35$. The inset shows the application of the optimized ramp to the initial state with all spins in the $|\uparrow\rangle$ (red dashed line). The blue points show $\langle\hat{S}^z_{\mathbf{n}}\rangle$ after preparation time $t'_{\rm prep}$, and the red line shows $\langle\hat{S}^z_{\mathbf{n}}\rangle$ for the ideal target state.}
    \label{fig:adiabatic}
\end{figure}

\subsection{Effective Hamiltonian for state preparation}
For state preparation, we apply an additional drive $C$ along the cavity supported by cavity mode $b$ (see Fig.~\ref{fig:adiabatic}(a)), which drives the $|\downarrow\rangle\rightarrow|e\rangle$ transition with two frequencies $\omega_{p,C},\omega'_{p,C}$. 
The choice of two-tone frequency is to satisfy two-photon resonance conditions $\omega_{p,C}-\omega_{p,A}=\omega'_{p,C}-\omega_{p,B}=\tilde{\omega}_0+\delta$, where $\delta$ is the effective two-photon detuning after correcting AC Stark shift induced by $\Omega_A$. 
One can ignore the quantum fluctuations of cavity modes by replacing cavity field operators ($\hat{a}_r$, $\hat{a}_b$) by coherent state amplitudes ($\alpha_r$, $\alpha_b$) in Eq.~(\ref{eq:a4}), which is ensured by setting cavity detunings $\delta_{c,A}$ and $\delta_{c,B}$ to larger values. 
We then define the Rabi frequency of drive $B$ by Eq.~(\ref{eq:omegab}), and the Rabi frequency of drive $C$ (same for the two tones) by
\begin{equation}
    \Omega_{\mathbf{n},C} = g_{\mathbf{n},b}\alpha_b = \Omega_C\cos(n_x\varphi)/2.
\end{equation}
Similar to Eq.~(\ref{eq:interference}), we ignore the interference effects between the two tones.

Using the same local gauge transformation as Eq.~(\ref{eq:caveff}), and in the rotating frame of laser drives, one can obtain the following Hamiltonian,
\begin{equation}
    \begin{aligned}
    \hat{H}_{\rm drive}/\hbar = & \sum_{\mathbf{n}} \bigg(J'\eta_{\mathbf{n}}^2-\delta(t')\bigg)\hat{S}^z_{\mathbf{n}} -\frac{1}{2}\sum_{\mathbf{n}}\bigg[\Big(\Omega_p(t')\eta_{\mathbf{n}}e^{-i\phi_{\mathbf{n}}}\\
    \qquad \qquad &+\Omega_d(t')\eta_{\mathbf{n}}^2e^{-2i\phi_{\mathbf{n}}}\Big)\hat{S}^{+}_{\mathbf{n}}+\mathrm{h.c.}\bigg],\\
    \end{aligned}
    \label{eq:drive2}
\end{equation}
where 
\begin{equation}
    \Omega_p=-\frac{\Omega_A^{*}\Omega_C}{2\delta_e}, \quad \Omega_d=-\frac{\Omega_B^{*}\Omega_C}{2\delta_e},
\end{equation}
are the effective two-photon Rabi frequencies, and
\begin{equation}
    J' = \frac{|\Omega_B|^2-2|\Omega_C|^2}{4\delta_e},
\end{equation}
is the strength of the inhomogeneous AC Stark shift induced by both external drives $B$ and $C$ along the cavity.
Note that we use $t'$ as the label of the evolution time in state preparation, while $t$ is for BCS simulation.

It is worth mentioning that the derivation of $\hat{H}_{\rm cav}$ (see Eq.~(\ref{eq:heff})) and $\hat{H}_{\rm drive}$ (see Eq.~(\ref{eq:drive2})) used  different rotating frames. 
$\hat{H}_{\rm cav}$ is in the rotating frame at frequency $\omega_0+|\Omega'_A|^2/(4\delta_e)$, while $\hat{H}_{\rm drive}$ is in the rotating frame at frequency $\omega_0+|\Omega_A|^2/(4\delta_e)+\delta$, where $\omega_0$ is the bare frequency difference between $|\uparrow\rangle$ and $|\downarrow\rangle$ states, $\Omega'_A$ and $\Omega_A$ are the Rabi frequencies for external drive A in $\hat{H}_{\rm cav}$ and $\hat{H}_{\rm drive}$ respectively, and $\delta$ is the two-photon detuning in $\hat{H}_{\rm drive}$.
This is compatible as long as we never apply these two Hamiltonians simultaneously.
In our case, we apply $\hat{H}_{\rm drive}$ to generate an initial state, and then apply $\hat{H}_{\rm cav}$ for time evolution under BCS Hamiltonian.
Therefore, for converting to the rotating frame of $\hat{H}_{\rm cav}$ we just need to apply the following time-independent unitary transformation to the initial state,
\begin{equation}
    \hat{U} = \exp\bigg(-i\int_0^{t'_{\rm prep}}dt'\Big(\delta(t')+\frac{|\Omega_A(t')|^2-|\Omega'_A|^2}{4\delta_e}\Big)\sum_{\mathbf{n}}\hat{S}^{z}_{\mathbf{n}}\bigg),
\end{equation}
which is a constant rotation determined by the adiabatic ramp sequence. 
Notice that $[\hat{H}_{\rm cav},\hat{U}]=0$, $\hat{U}$ has no effects on the Hamiltonian dynamics, so we can simply drop it for discussions in the main text, which is equivalent to defining the $x$ and $y$ axes of the Bloch sphere for $\hat{H}_{\rm cav}$ based on the adiabatic ramp sequence.

\subsection{Speeding up the initial state preparation}
One caveat for adiabatic state preparation is that the preparation time scale is limited by the energy gap of $\hat{H}_{\rm drive}$, which is typically slow when preparing the mean-field ground state in the vicinity of the topological phase transition, or in the case of a  small BCS order parameter $|\Delta_p|,|\Delta_d|\;\ll J$.
Since we only care about the fidelity of the prepared state, it is unnecessary to ensure the state is always an instantaneous eigenstate of $\hat{H}_{\rm drive}$ during the preparation.
Based on this observation, one can optimize the pulse shape of the external drives \cite{DAlessandro2021} to speed up the preparation time scale and ensure unwanted experimental imperfections do not play a significant role.
Notice that the mean-field ground state takes the form of a product state, $|\psi_{\rm MF}\rangle=\bigotimes_{\mathbf{n}}|\psi_{\mathbf{n},\rm MF}\rangle$, where $|\psi_{\mathbf{n},\rm MF}\rangle$ is the ideal target state for individual atomic spin.
We are seeking the optimal time sequence of $\Omega_p(t')$, $\Omega_d(t')$ and $\delta(t')$ that maximizes the averaged fidelity of individual atomic spin for a given preparation time $t'_{\rm prep}$,
\begin{equation}
    F_{\rm avg} = \frac{1}{N}\sum_{\mathbf{n}}F_{\mathbf{n}}, \quad F_{\mathbf{n}}=|\langle\psi_{\mathbf{n}}(t'=t'_{\rm prep})|\psi_{\mathbf{n},\rm MF}\rangle|^2,
\end{equation}
where $|\psi_{\mathbf{n}}\rangle$ is the actual state of the atomic spin at site $\mathbf{n}$ after preparation time $t'_{\rm prep}$, and $F_{\mathbf{n}}$ is the fidelity with the ideal target state $|\psi_{\mathbf{n},\rm MF}\rangle$ for the atomic spin at site $\mathbf{n}$.
Fig.~\ref{fig:adiabatic}(b) shows an example of an optimized ramp for the preparation of the $p_x+ip_y$ ground state at $\chi_{p,i}N/J=1$ and Cooper pair number $N_C/N=0.35$ (independent of $\chi_{d,i}$), starting with all the atoms in $|\uparrow\rangle$. 
In the inset of Fig.~\ref{fig:adiabatic}(b), we show good agreement between the state after the optimized ramp and the target state.
Numerical calculation further shows that the infidelity $1-F_{\mathbf{n}}$ for each atomic spin can be suppressed to $1-F_{\mathbf{n}} < 10^{-2}$ for all $\mathbf{n}$ using this optimized ramp.

\section{Physical origin for the BCS model with simultaneous $p$-wave and $d$-wave pairing}
\label{sec:origin}

In Eq.~(\ref{eq:hamilspinpd}), we map both the $p_x+ip_y$ and $d_{x^2-y^2}+id_{xy}$ pairing channels using the same pseudospin representation (see Eq.~(\ref{eq:and})).
Here we would like to clarify that this definition is consistent for both $p_x+ip_y$ and $d_{x^2-y^2}+id_{xy}$ pairing interactions, in the context of ultracold contact interaction between fermionic atoms.
This is a useful justification that the Hamiltonian used in our cavity QED simulator is possible to describe some realistic physical system of fermionic superconductors/superfluids.

Considering the scattering of two fermions, the total wave function should be anti-symmetric. 
For even partial waves (e.g. $s$-wave, $d$-wave), the spatial wave function is symmetric, so the spin wave function is anti-symmetric (spin singlet), $(|\Uparrow\Downarrow\rangle - |\Downarrow\Uparrow\rangle)/\sqrt{2}$, which is the only interaction channel.
For odd partial waves (e.g. $p$-wave), the spatial wave function is anti-symmetric, so the spin wave function is symmetric (spin triplet), leading to three different interaction channels, $|\Uparrow\Uparrow\rangle, |\Downarrow\Downarrow\rangle, (|\Uparrow\Downarrow\rangle + |\Downarrow\Uparrow\rangle)/\sqrt{2}$. 
Therefore, if we consider Cooper pairs formed by one fermion in $|\Uparrow\rangle$ state and another fermion in $|\Downarrow\rangle$ state, they can interact via both $p$-wave and $d$-wave channels simultaneously. 

To construct the BCS model, we start from the following Hamiltonian including both $p_x+ip_y$ and $d_{x^2-y^2}+id_{xy}$ contact interactions,
\begin{equation}
    \begin{gathered}
    \hat{H}=\sum_{\alpha=\{\Uparrow,\Downarrow\}}\sum_{\mathbf{k}}\varepsilon_{\mathbf{k}}\hat{c}_{\mathbf{k}\alpha}^{\dag}\hat{c}_{\mathbf{k}\alpha} + \hat{H}_{p+ip} + \hat{H}_{d+id},\\
    \begin{aligned}
\hat{H}_{p+ip}&=\frac{1}{2}\sum_{\alpha\beta=\{\Uparrow,\Downarrow\}}\sum_{\mathbf{k}_1\mathbf{k}_2\mathbf{k}_3\mathbf{k}_4} U_{p,\alpha\beta}\, |\mathbf{k}_1-\mathbf{k}_2|\, |\mathbf{k}_4-\mathbf{k}_3| \\
&e^{-i\phi_{\mathbf{k}_1-\mathbf{k}_2}}e^{i\phi_{\mathbf{k}_4-\mathbf{k}_3}}\hat{c}_{\mathbf{k}_1,\alpha}^{\dag}\hat{c}_{\mathbf{k}_2,\beta}^{\dag}\hat{c}_{\mathbf{k}_3,\beta}\hat{c}_{\mathbf{k}_4,\alpha}\delta_{\mathbf{k}_1+\mathbf{k}_2,\mathbf{k}_3+\mathbf{k}_4},\\
\end{aligned}\\
\begin{aligned}
\hat{H}_{d+id}&= U_{d,\Uparrow\Downarrow}\sum_{\mathbf{k}_1\mathbf{k}_2\mathbf{k}_3\mathbf{k}_4}\, |\mathbf{k}_1-\mathbf{k}_2|^2\, |\mathbf{k}_4-\mathbf{k}_3|^2 \\
&e^{-2i\phi_{\mathbf{k}_1-\mathbf{k}_2}}e^{2i\phi_{\mathbf{k}_4-\mathbf{k}_3}}\hat{c}_{\mathbf{k}_1,\Uparrow}^{\dag}\hat{c}_{\mathbf{k}_2,\Downarrow}^{\dag}\hat{c}_{\mathbf{k}_3,\Downarrow}\hat{c}_{\mathbf{k}_4,\Uparrow}\delta_{\mathbf{k}_1+\mathbf{k}_2,\mathbf{k}_3+\mathbf{k}_4}.
\end{aligned}
\end{gathered}
\end{equation}
Note that $U_{p,\alpha\beta}=U_{p,\beta\alpha}$ is set by the three different $p$-wave scattering volumes (unit: length$^3$), and we only keep the $m_l=+1$ orbital projection for $p_x+ip_y$ interaction.  Similarly,
$U_{d,\Uparrow\Downarrow}$ is determined by the $d$-wave scattering hypervolume (units: length$^5$), and we only keep the $m_l=+2$ orbital projection for $d_{x^2-y^2}+id_{xy}$ interaction. 

Here we consider the case that $(|\Uparrow\Downarrow\rangle + |\Downarrow\Uparrow\rangle)/\sqrt{2}$ is the dominant $p+ip$ interaction channel, i.e. $U_{p,\Uparrow\Downarrow}\gg U_{p,\Uparrow\Uparrow}, U_{p,\Downarrow\Downarrow}$, so we can simply drop the $U_{p,\Uparrow\Uparrow}$ and $U_{p,\Downarrow\Downarrow}$ terms.
We also apply the BCS approximation to select the terms with $\mathbf{k}_1+\mathbf{k}_2=0$ and $\mathbf{k}_3+\mathbf{k}_4=0$, so we can set $\mathbf{k}_1=\mathbf{k}$, $\mathbf{k}_2=-\mathbf{k}$, $\mathbf{k}_3=-\mathbf{q}$ and $\mathbf{k}_4=\mathbf{q}$. All these approximations lead to
\begin{equation}
    \begin{aligned}
    \hat{H}^{pd}_{\rm BCS} &= \sum_{\alpha=\{\Uparrow,\Downarrow\}}\sum_{\mathbf{k}}\varepsilon_{\mathbf{k}}\hat{c}_{\mathbf{k}\alpha}^{\dag}\hat{c}_{\mathbf{k}\alpha}\\
    &+ 4U_{p,\Uparrow\Downarrow}\sum_{\mathbf{k}\mathbf{q}} |\mathbf{k}|\,|\mathbf{q}|\, e^{-i\phi_{\mathbf{k}}}e^{i\phi_{\mathbf{q}}} \hat{c}_{\mathbf{k},\Uparrow}^{\dag}\hat{c}_{-\mathbf{k},\Downarrow}^{\dag}\hat{c}_{-\mathbf{q},\Downarrow}\hat{c}_{\mathbf{q},\Uparrow}\\
    &+ 16U_{d,\Uparrow\Downarrow}\sum_{\mathbf{k}\mathbf{q}} |\mathbf{k}|^2\,|\mathbf{q}|^2\, e^{-2i\phi_{\mathbf{k}}}e^{2i\phi_{\mathbf{q}}} \hat{c}_{\mathbf{k},\Uparrow}^{\dag}\hat{c}_{-\mathbf{k},\Downarrow}^{\dag}\hat{c}_{-\mathbf{q},\Downarrow}\hat{c}_{\mathbf{q},\Uparrow}.
\end{aligned}
\end{equation}
If we define $U_p = -4U_{p,\Uparrow\Downarrow}$, $U_d = -16U_{d,\Uparrow\Downarrow}$, and rewrite the Hamiltonian in terms of pseudospin operators, we can obtain Eq.~(\ref{eq:hamilspinpd}) in the main text,
\begin{equation}
\begin{aligned}
    \hat{H}_{\rm BCS}^{pd}
    &= \sum_{\mathbf{k}} 2\varepsilon_{\mathbf{k}} \hat{S}^{z}_{\mathbf{k}}
    - 
    U_p\sum_{\mathbf{k}\mathbf{q}} 
   |\mathbf{k}|\,|\mathbf{q}|\,e^{-i\phi_{\mathbf{k}}}e^{i\phi_{\mathbf{q}}}
    \hat{S}^{+}_{\mathbf{k}} \hat{S}^{-}_{\mathbf{q}}\\
    &- U_d\sum_{\mathbf{k}\mathbf{q}} 
   |\mathbf{k}|^2\,|\mathbf{q}|^2\,e^{-2i\phi_{\mathbf{k}}}e^{2i\phi_{\mathbf{q}}}
    \hat{S}^{+}_{\mathbf{k}} \hat{S}^{-}_{\mathbf{q}}.
\end{aligned} 
\end{equation}

\section{Mean-field theory for equilibrium physics}
\label{sec:meanfield}

\subsection{Self-consistent mean-field solutions}
Here we would like to discuss the equilibrium phase diagram for the effective Hamiltonian (see Eq.~(\ref{eq:caveff})) we proposed in the cavity QED system.
To minimize the energy of the system with a constraint of conserved number of ``electrons'', we minimize the energy with an additional Lagrangian multiplier, $\hat{H}_{\rm cav}-\hbar\mu\Big(\sum_{\mathbf{n}}(2\hat{S}^z_{\mathbf{n}}+1)-2N_C\Big)$, where $N_C$ is the total number of Cooper pairs, and $\mu$ is the chemical potential.
Considering $\hat{S}^{+}_{\mathbf{n}}\hat{S}^{-}_{\mathbf{m}}\approx \langle\hat{S}^{+}_{\mathbf{n}}\rangle\hat{S}^{-}_{\mathbf{m}}+\hat{S}^{+}_{\mathbf{n}}\langle\hat{S}^{-}_{\mathbf{m}}\rangle-\langle\hat{S}^{+}_{\mathbf{n}}\rangle\langle\hat{S}^{-}_{\mathbf{m}}\rangle$, the mean-field Hamiltonian is given by
\begin{equation}
    \begin{aligned}
    \hat{H}_{\rm MF}/\hbar &= -2\sum_{\mathbf{n}}\hat{\mathbf{S}}_{\mathbf{n}}\cdot \mathbf{B}^{\rm self}_{\mathbf{n}} + \frac{|\Delta_{p}|^2}{\chi_p}+\frac{|\Delta_{d}|^2}{\chi_d}\\
    &-\mu(N-2N_C),
    \end{aligned}
\end{equation}
where $N$ is the total number of atomic spins, and $\mathbf{B}^{\rm self}_{\mathbf{n}}$ is the effective magnetic field,
\begin{equation}
    \begin{aligned}
    (B^{\rm self})^x_{\mathbf{n}}-i(B^{\rm self})^y_{\mathbf{n}} &= \Delta_p \eta_{\mathbf{n}}e^{-i\phi_{\mathbf{n}}}+\Delta_d\eta_{\mathbf{n}}^2e^{-2i\phi_{\mathbf{n}}},\\
    (B^{\rm self})^z_{\mathbf{n}} &= \mu-J\eta_{\mathbf{n}}^2.
    \end{aligned}
\end{equation}

Note that the mean-field ground state is aligned to the effective magnetic field. One can thus obtain a set of self-consistent equations for the $p$-wave and the $d$-wave order parameters,
\begin{equation}
    \Delta_{p} = \chi_p\sum_{\mathbf{n}}\eta_{\mathbf{n}}e^{i\phi_{\mathbf{n}}}\langle\hat{S}^{-}_{\mathbf{n}}\rangle = \chi_p\sum_{\mathbf{n}}\eta_{\mathbf{n}}e^{i\phi_{\mathbf{n}}} \frac{\Delta_{\mathbf{n}}}{2E_{\mathbf{n}}}, 
    \label{eq:1}
\end{equation}
\begin{equation}
    \Delta_{d} = \chi_d\sum_{\mathbf{n}}\eta_{\mathbf{n}}^2e^{2i\phi_{\mathbf{n}}}\langle\hat{S}^{-}_{\mathbf{n}}\rangle = \chi_d\sum_{\mathbf{n}}\eta_{\mathbf{n}}^2e^{2i\phi_{\mathbf{n}}} \frac{\Delta_{\mathbf{n}}}{2E_{\mathbf{n}}},
    \label{eq:2}
\end{equation}
where $E_{\mathbf{n}}=|\mathbf{B}^{\rm self}_{\mathbf{n}}|\;=\sqrt{(J\eta_{\mathbf{n}}^2-\mu)^2+|\Delta_{\mathbf{n}}|^2}$ is the quasi-particle energy, with $\Delta_{\mathbf{n}} = \Delta_p \eta_{\mathbf{n}}e^{-i\phi_{\mathbf{n}}}+\Delta_d\eta_{\mathbf{n}}^2e^{-2i\phi_{\mathbf{n}}}$.
An additional self-consistent equation is given by the conservation of total number of Cooper pairs,
\begin{equation}
    N_{C}=\frac{1}{2}\sum_{\mathbf{n}}\bigg(1-\frac{J\eta_{\mathbf{n}}^2-\mu}{E_{\mathbf{n}}}\bigg).
    \label{eq:3}
\end{equation}

\begin{figure}[t]
    \centering
    \includegraphics[width=1.0\columnwidth]{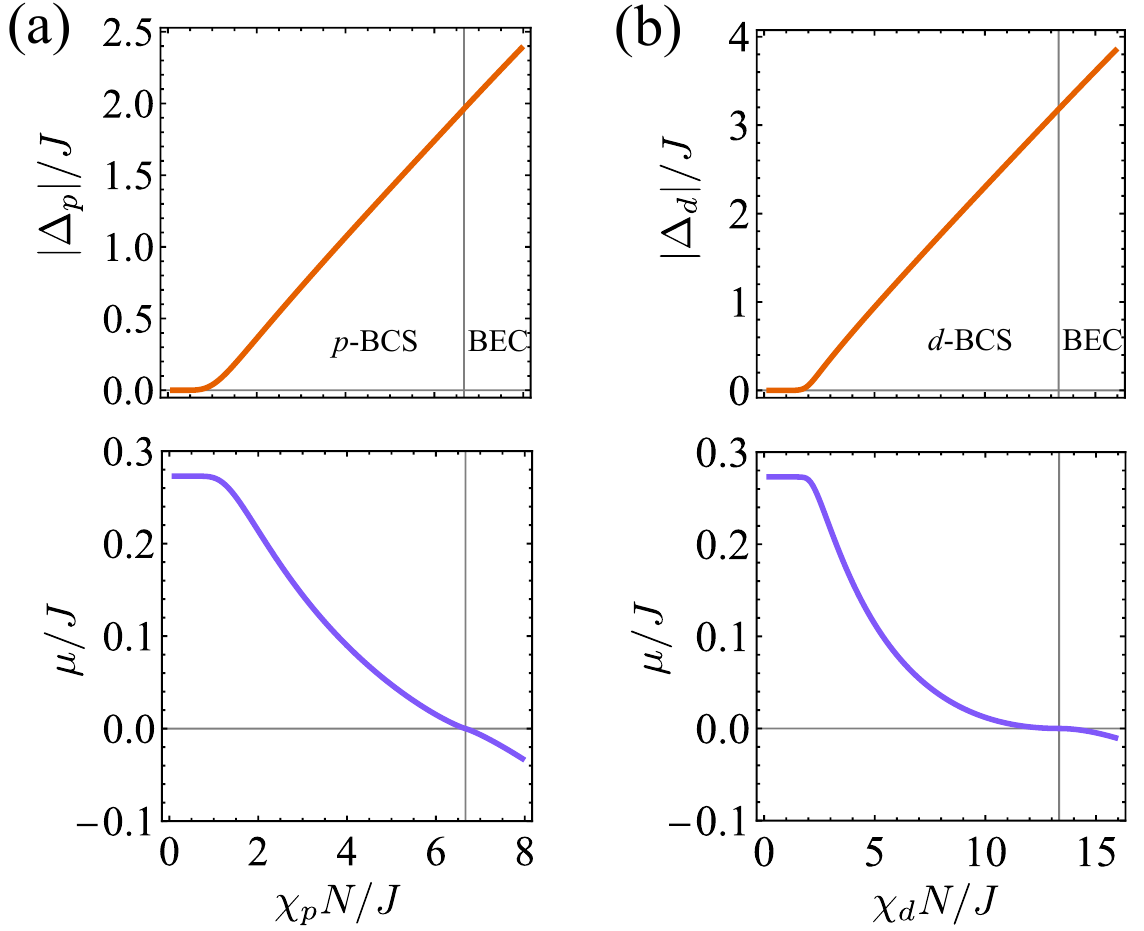}
    \caption{Equilibrium (a) $p_x+ip_y$ solution (b) $d_{x^2-y^2}+id_{xy}$ solution with fixed number of Cooper pairs $N_C/N=0.35$. The top panel shows the BCS order parameters (a) $\Delta_p$ (b) $\Delta_d$, and the bottom panel shows the chemical potential $\mu$. The vertical grid lines mark the critical points of topological phase transition.}
    \label{fig:appen1}
\end{figure}

Now we would like to discuss four different cases of self-consistent solutions:
\begin{itemize}
    \item $\Delta_p=0$, $\Delta_d=0$
    
    In this case, Eq.~(\ref{eq:3}) becomes
    \begin{equation}
        N_C = \frac{1}{2}\sum_{\mathbf{n}}\bigg[1-\mathrm{sgn}(J\eta_{\mathbf{n}}^2-\mu)\bigg].
    \end{equation}
    Using $\eta_{\mathbf{n}}=|\cos(n_x\varphi)|$, we have
    \begin{equation}
        \frac{\mu}{J} = \sin^2\bigg(\frac{\pi N_C}{2N}\bigg).
        \label{eq:munonint}
    \end{equation}
    This is the Fermi distribution without BCS pairings.

    \item $\Delta_p\neq 0$, $\Delta_d=0$ ($p_x+ip_y$ solution)

    In this case, Eq.~(\ref{eq:3}) and Eq.~(\ref{eq:1}) become
    \begin{equation}
    \begin{gathered}
    N_C = \frac{N}{2}-\sum_{\mathbf{n}}\frac{J\eta_{\mathbf{n}}^2-\mu}{2\sqrt{(J\eta_{\mathbf{n}}^2-\mu)^2+|\Delta_p|^2\eta_{\mathbf{n}}^2}}, \\
    1 = \chi_p \sum_{\mathbf{n}}\frac{\eta_{\mathbf{n}}^2}{2\sqrt{(J\eta_{\mathbf{n}}^2-\mu)^2+|\Delta_p|^2\eta_{\mathbf{n}}^2}},
    \end{gathered}
    \label{eq:pconsist}
    \end{equation}
    while Eq.~(\ref{eq:2}) is automatically satisfied. Without loss of generality, we take $\Delta_p$ to be real and positive. One can solve the self-consistent equations above numerically for $\Delta_p$ and $\mu$, which are shown in Fig.~\ref{fig:appen1}(a) with fixed $N_C/N=0.35$.
    
    When $\Delta_p \gg J,|\mu|$, it is possible to simplify the self-consistent solution to 
    \begin{equation}
        \Delta_p\approx \chi_p \sum_{\mathbf{n}}\frac{\eta_{\mathbf{n}}}{2} = \frac{\chi_p N}{\pi}.
        \label{eq:largelimit}
    \end{equation}
    
    When $\Delta_p \ll J,|\mu|$, the chemical potential $\mu$ is almost given by Eq.~(\ref{eq:munonint}) consistently with the first equation of Eq.~(\ref{eq:pconsist}). For the second equation in Eq.~(\ref{eq:pconsist}), we have
    \begin{equation}
        \begin{aligned}
        &\frac{1}{N}\sum_{\mathbf{n}}\frac{\eta_{\mathbf{n}}^2}{2\sqrt{(J\eta_{\mathbf{n}}^2-\mu)^2+|\Delta_p|^2\eta_{\mathbf{n}}^2}}\\
        &\approx\frac{1}{\pi}\int_0^{\pi/2} ds \frac{\cos^2(s)}{\sqrt{(J\cos^2(s)-\mu)^2+\Delta_p^2\cos^2(s)}}\\
        &\approx \frac{1}{2}\frac{\mu}{J}g(\mu/J)\int_0^{1} dy \frac{1}{\sqrt{(Jy-\mu)^2+\Delta_p^2y}}\\
        &\approx \frac{1}{2}\frac{\mu}{J^2}g(\mu/J)\ln\bigg(\frac{4J(J-\mu)}{\Delta_p^2}\bigg),
        \label{eq:psmall}
        \end{aligned}
    \end{equation}
    where $g(y)=1/\Big(\pi\sqrt{y(1-y)}\Big)$ is the normalized density of states. In the third line of Eq.~(\ref{eq:psmall}), we consider the integrand is peaked at a narrow region around $y=\mu/J$ for small $\Delta_p$. One can thus obtain
    \begin{equation}
        \frac{\Delta_p}{J} \propto \exp\bigg(-\frac{J/\mu}{g(\mu/J)}\frac{J}{\chi_p N}\bigg).
        \label{eq:smalllimit}
    \end{equation}
    This result shows that we always have $\Delta_p>0$ for non-zero $\chi_p$, although $\Delta_p$ is exponentially suppressed when $\chi_pN\lesssim J$.

    \item $\Delta_p=0$, $\Delta_d\neq 0$ ($d_{x^2-y^2}+id_{xy}$ solution)

    In this case, Eq.~(\ref{eq:3}) and Eq.~(\ref{eq:2}) become
    \begin{equation}
    \begin{gathered}
    N_C = \frac{N}{2}-\sum_{\mathbf{n}}\frac{J\eta_{\mathbf{n}}^2-\mu}{2\sqrt{(J\eta_{\mathbf{n}}^2-\mu)^2+|\Delta_d|^2\eta_{\mathbf{n}}^4}}, \\
    1 = \chi_d \sum_{\mathbf{n}}\frac{\eta_{\mathbf{n}}^4}{2\sqrt{(J\eta_{\mathbf{n}}^2-\mu)^2+|\Delta_d|^2\eta_{\mathbf{n}}^4}},
    \end{gathered}
    \label{eq:dconsist}
    \end{equation}
    while Eq.~(\ref{eq:1}) is automatically satisfied. Without loss of generality, we take $\Delta_d$ to be real and positive. The solutions are shown in Fig.~\ref{fig:appen1}(b) with fixed $N_C/N=0.35$.
    
    When $\Delta_d\gg J,|\mu|$, it is possible to simplify the self-consistent solution to 
    \begin{equation}
        \Delta_d\approx \chi_d \sum_{\mathbf{n}}\frac{\eta^2_{\mathbf{n}}}{2} = \frac{\chi_d N}{4}.
    \end{equation}
    
    When $\Delta_d\ll J,|\mu|$, the chemical potential $\mu$ is approaching Eq.~(\ref{eq:munonint}). Similar to the discussion of the $p+ip$ solution, we have
    \begin{equation}
        \frac{\Delta_d}{J} \propto \exp\bigg(-\frac{(J/\mu)^2}{g(\mu/J)}\frac{J}{\chi_d N}\bigg).
    \end{equation}
    This result shows that we always have $\Delta_d>0$ for non-zero $\chi_d$, although $\Delta_d$ is exponentially suppressed when $\chi_dN\lesssim J$.

    \item $\Delta_p\neq 0$, $\Delta_d\neq 0$

    In this case, one can solve the self-consistent equations Eq.~(\ref{eq:1}), Eq.~(\ref{eq:2}) and Eq.~(\ref{eq:3}) numerically for $\Delta_p$, $\Delta_d$ and $\mu$. Notice that one can remove the relative phase $\theta$ between $\Delta_p$ and $\Delta_d$ by shifting $\phi_{\mathbf{n}}\rightarrow \phi_{\mathbf{n}}+\theta$.
    Without loss of generality, we assume both $\Delta_p$ and $\Delta_d$ are real and positive numbers.
    
\end{itemize}

The true mean-field ground state of the system is chosen among these cases to be the one with the  minimum mean-field energy. The mean-field energy is given by
\begin{equation}
    E_{\rm MF}/\hbar=-\sum_{\mathbf{n}}E_{\mathbf{n}} + \frac{\Delta_{p}^2}{\chi_p}+\frac{\Delta_{d}^2}{\chi_d}-\mu(N-2N_C).
\end{equation}
Based on numerical calculation, we find that the true ground state is either the $p_x+ip_y$ solution or the $d_{x^2-y^2}+id_{xy}$ solution depending on the choice of system parameters $\chi_p$, $\chi_d$ and $N_C$. 
The first order transition between these two phases is marked by $E_{\mathrm{MF},p}=E_{\mathrm{MF},d}$ (the black dashed line in Fig.~\ref{fig:stability}(a) with fixed $N_C/N=0.35$).

\subsection{Topological phase transition}
In the main text, we define the topological Chern number in terms of an integral assuming  a continuous distribution of the ground-state pseudospin texture (see Eq.~(\ref{eq:chern})). It is more convenient to rewrite the integral  into the following form,
\begin{equation}
    Q=\frac{1}{4\pi}\int dkd\phi \bigg(\frac{\partial\langle\vec{\sigma}_{\mathbf{k}}\rangle}{\partial k}\times\frac{\partial\langle\vec{\sigma}_{\mathbf{k}}\rangle}{\partial \phi}\bigg)\cdot \langle\vec{\sigma}_{\mathbf{k}}\rangle,
    \label{eq:chernapp}
\end{equation}
where $\mathbf{k} = (k_x,k_y)=k(\cos\phi,\sin\phi)$, and $\langle\vec{\sigma}_{\mathbf{k}}\rangle=2\langle \hat{\mathbf{S}}_{\mathbf{k}}\rangle$. 
For a single pairing channel labelled by $\alpha = p,d$ (see Eq.~(\ref{eq:channel})), we have $\langle\vec{\sigma}_{\mathbf{k}}\rangle = \mathbf{B}^{\rm self}_{\mathbf{k},\alpha}/|\mathbf{B}^{\rm self}_{\mathbf{k},\alpha}|$, where $(B^{\rm self})^{x}_{\mathbf{k},\alpha}-i(B^{\rm self})^{y}_{\mathbf{k},\alpha}=f^{\alpha*}_{\mathbf{k}}\Delta_{\alpha}$, $(B^{\rm self})^{z}_{\mathbf{k},\alpha}=\mu-\varepsilon_{\mathbf{k}}/\hbar$.
Notice that one can interpret Eq.~(\ref{eq:chernapp}) as a discrete sum of solid angles on the Bloch sphere formed by the following four unit vectors, $\langle\vec{\sigma}_{k,\phi}\rangle$, $\langle\vec{\sigma}_{k+dk,\phi}\rangle$, $\langle\vec{\sigma}_{k+dk,\phi+d\phi}\rangle$, $\langle\vec{\sigma}_{k,\phi+d\phi}\rangle$.

In our cavity QED simulator, as shown in Fig.~\ref{fig:cartoon}, we can map the 2D lattice sites ($N_x\times N_y$) in the cavity back to the momentum space of a superconductor, forming a discrete sampling of the momentum space.
We understand this discrete sampling as a grid on $k$ and $\phi$, and we label them in ascending order $k_j<k_{j+1}$ ($0\leq k_j\leq k_{\rm max}$, $j=1,2,\cdots,N_x$) and $\phi_l<\phi_{l+1}$ ($0\leq \phi_l< 2\pi$, $l=1,2,\cdots,N_y$). 
Above the cutoff, we assume there are fictitious spins always in the $|\downarrow\rangle$ state, which means adding an extra $k_{N_x+1}$ with $\langle\vec{\sigma}_{k_{N_x+1},\phi_l}\rangle=(0,0,-1)$. 
We also set $\phi_{N_y+1}=\phi_1$ to take account of the periodic boundary of $\phi$.
Similar to Ref.~\cite{Shankar2022}, we have an alternative definition of topological Chern number for discrete sampling of the momentum space,
\begin{equation}
     \begin{aligned}
     Q &= \frac{1}{4\pi}\sum_{j=1}^{N_x}\sum_{l=1}^{N_y} \mathcal{A}\Big(\langle\vec{\sigma}_{k_{j},\phi_l}\rangle, \langle\vec{\sigma}_{k_{j+1},\phi_l}\rangle,\langle\vec{\sigma}_{k_{j+1},\phi_{l+1}}\rangle\Big)\\
     &+ \frac{1}{4\pi}\sum_{j=1}^{N_x}\sum_{l=1}^{N_y} \mathcal{A}\Big(\langle\vec{\sigma}_{k_{j},\phi_l}\rangle, \langle\vec{\sigma}_{k_{j+1},\phi_{l+1}}\rangle,\langle\vec{\sigma}_{k_{j},\phi_{l+1}}\rangle\Big)\\
     &+ \frac{1}{4\pi}\sum_{l=2}^{N_y-1} \mathcal{A}\Big(\langle\vec{\sigma}_{k_{1},\phi_{1}}\rangle, \langle\vec{\sigma}_{k_{1},\phi_l}\rangle,\langle\vec{\sigma}_{k_{1},\phi_{l+1}}\rangle\Big),
     \end{aligned}
     \label{eq:chernapp1}
\end{equation}
where $\mathcal{A}(\vec{a},\vec{b},\vec{c})=2\,\mathrm{atan2}\Big(\vec{a}\cdot(\vec{b}\times\vec{c}),1+\vec{a}\cdot\vec{b}+\vec{b}\cdot\vec{c}+\vec{c}\cdot\vec{a}\Big)$ is the solid angle formed by unit vectors $\vec{a},\vec{b},\vec{c}$. Here, $\mathrm{atan2}(y,x)$ is the 2-argument arctangent calculating the phase of the complex number $x+iy$.
The fictitious spins added above the cutoff ensure that $Q$ takes only integer values.
One can also show that Eq.~(\ref{eq:chernapp1}) reduces to Eq.~(\ref{eq:chernapp}) in the continuum limit.

Here we would like to show that the topological Chern number $Q$ is completely determined by the chemical potential $\mu$ for either $p_x+ip_y$ or $d_{x^2-y^2}+id_{xy}$ solution. In the continuum limit, the general form of $\langle\vec{\sigma}_{k,\phi}\rangle$ can be written as
\begin{equation}
    \langle\vec{\sigma}_{k,\phi}\rangle = \bigg(f_{xy}(k)\cos(n\phi), f_{xy}(k)\sin(n\phi), f_z(k)\bigg),
\end{equation}
where $f_{xy}(k)$ and $f_z(k)$ can be any real-valued functions satisfying $f_{xy}^2+f_z^2=1$, with $f_z(0)=\mathrm{sgn}(\mu)$ and $f_z(\infty)=-1$. We have $n=1$ for $p_x+ip_y$ solution and $n=2$ for $d_{x^2-y^2}+id_{xy}$ solution. Eq.~(\ref{eq:chernapp}) leads to
\begin{equation}
    \begin{aligned}
    Q&=\frac{2\pi n}{4\pi}\int_0^{\infty}dk (-f_{xy})\bigg((\partial_{k}f_z)f_{xy}-(\partial_{k}f_{xy})f_z\bigg)\\
    &= \frac{n}{2}\Big(f_z(0)-f_z(\infty)\Big)\\
    &= \frac{n}{2}\Big(1+\mathrm{sgn}(\mu)\Big).
    \end{aligned}
    \label{eq:qcal}
\end{equation}
In the second line of Eq.~(\ref{eq:qcal}), we use $(\partial_{k}f_{xy})f_{xy}=-(\partial_{k}f_z)f_z$ due to $f_{xy}^2+f_z^2=1$. When $\mu>0$, we have non-trivial Chern number ($Q\neq 0$); when $\mu<0$, we have $Q=0$.

We then discuss the quantum critical point (QCP) of the topological phase transition. 
For the $p_x+ip_y$ solution, plugging in $\mu=0$ to Eq.~(\ref{eq:pconsist}), we have
\begin{equation}
    \begin{gathered}
    N_C = \frac{N}{2}-J\sum_{\mathbf{n}}\frac{\eta_{\mathbf{n}}^2}{2\sqrt{J^2\eta_{\mathbf{n}}^4+|\Delta_p|^2\eta_{\mathbf{n}}^2}},\\
    1 = \chi_{p,\rm QCP} \sum_{\mathbf{n}}\frac{\eta_{\mathbf{n}}^2}{2\sqrt{J^2\eta_{\mathbf{n}}^4+|\Delta_p|^2\eta_{\mathbf{n}}^2}},
    \end{gathered}
\end{equation}
leading to
\begin{equation}
    \frac{\chi_{p,\rm QCP}N}{J} = \bigg(\frac{1}{2}-\frac{N_C}{N}\bigg)^{-1}.
\end{equation}

For the $d_{x^2-y^2}+id_{xy}$ solution, plugging in $\mu=0$ to Eq.~(\ref{eq:dconsist}), we have
\begin{equation}
    N_C = \frac{N}{2}-\frac{NJ}{2\sqrt{J^2+|\Delta_d|^2}}, \quad 1 = \chi_d \frac{N}{4\sqrt{J^2+|\Delta_d|^2}},
\end{equation}
in which we use $\sum_{\mathbf{n}}\eta_{\mathbf{n}}^2=N/2$.
Similarly one can obtain 
\begin{equation}
    \frac{\chi_{d,\rm QCP}N}{J} = \bigg(\frac{1}{4}-\frac{N_C}{2N}\bigg)^{-1}.
\end{equation}

One can further define a long-time dynamical topological Chern number for dynamical phase II \cite{Foster2013,Shankar2022},
\begin{equation}
    W=\frac{1}{4\pi}\int dkd\phi \bigg(\frac{\partial\langle\vec{b}_{\mathbf{k}}\rangle}{\partial k}\times\frac{\partial\langle\vec{b}_{\mathbf{k}}\rangle}{\partial \phi}\bigg)\cdot \langle\vec{b}_{\mathbf{k}}\rangle.
\end{equation}
For a single pairing channel labelled by $\alpha = p,d$, we have $\langle\vec{b}_{\mathbf{k}}\rangle = (\mathbf{B}^{\rm self}_{\mathbf{k},\alpha})_{\infty}/|\mathbf{B}^{\rm self}_{\mathbf{k},\alpha}|_{\infty}$, where $(B^{\rm self}_{\mathbf{k},\alpha})^x_{\infty}-i(B^{\rm self}_{\mathbf{k},\alpha})^y_{\infty}=f^{\alpha*}_{\mathbf{k}}\Delta_{\alpha,\infty}$, $(B^{\rm self}_{\mathbf{k},\alpha})^z_{\infty}=\mu_{\infty}-\varepsilon_{\mathbf{k}}/\hbar$.
Analytical integration shown in Eq.~(\ref{eq:qcal}) can be directly generalized here.
When $\mu_{\infty}>0$, we have non-trivial dynamical Chern number ($W\neq 0$); when $\mu_{\infty}<0$, we have $W=0$.

\subsection{Stability analysis of $p_x+ip_y$ solutions and Bardasis-Schrieffer modes}

We investigate the stability of a stationary $p_x+ip_y$ state
by introducing a small perturbation in the competing
$d_{x^2-y^2}+id_{xy}$ channel. At fixed system parameters
$\chi_p$, $\chi_d$, and $N_C$, we prepare the initial state
\begin{equation}
    \begin{gathered}
    \langle \hat{S}_{\mathbf{n}}^{-}(0)\rangle
    =
    \frac{\Delta_{p,0}
    \left(\eta_{\mathbf{n}}e^{-i\phi_{\mathbf{n}}}
    +\varepsilon_d\eta_{\mathbf{n}}^2
    e^{-2i\phi_{\mathbf{n}}}\right)}
    {2E_{\mathbf{n},p}},
    \\
    \langle \hat{S}_{\mathbf{n}}^{z}(0)\rangle
    =
    \frac{\mu-J\eta_{\mathbf{n}}^2}
    {2E_{\mathbf{n},p}},
    \\
    E_{\mathbf{n},p}
    =
    \sqrt{
    (J\eta_{\mathbf{n}}^2-\mu)^2
    +\Delta_{p,0}^2
    \left(
    \eta_{\mathbf{n}}^2
    +\varepsilon_d^2\eta_{\mathbf{n}}^4
    +2\varepsilon_d\eta_{\mathbf{n}}^3
    \cos\phi_{\mathbf{n}}
    \right)}.
    \end{gathered}
    \label{eq:pinit}
\end{equation}
Here $\varepsilon_d\ll1$ controls the perturbation strength.
We choose $\Delta_{p,0}$ to be real and positive. Its value
and the chemical potential $\mu$ are determined by the
self-consistency equations for the unperturbed $p$-wave state
(see Eq.~(\ref{eq:pconsist})). 
When $\varepsilon_d=0$, the initial state reduces to the
stationary pure $p$-wave solution.

To separate the $p$-wave background from the perturbation,
we expand the pseudospin expectation values in angular harmonics,
\begin{equation}
    \langle \hat{S}^{-}_{\mathbf{n}}\rangle
    =
    \sum_m u_{\mathbf{n},m}e^{-im\phi_{\mathbf{n}}},
    \qquad
    \langle \hat{S}^{z}_{\mathbf{n}}\rangle
    =
    \sum_m w_{\mathbf{n},m}e^{-im\phi_{\mathbf{n}}}.
    \label{eq:mean_fourier}
\end{equation}
The integer $m$ labels the angular harmonics of the pseudospin
texture, with $u_{\mathbf{n},m}=u_m(\eta_{\mathbf{n}},t)$ and
$w_{\mathbf{n},m}=w_m(\eta_{\mathbf{n}},t)$.
Since $\langle\hat{S}^{z}_{\mathbf{n}}\rangle$ is real,
$w_{\mathbf{n},m}=w_{\mathbf{n},-m}^{*}$.
For incommensurate $\varphi/(2\pi)$, distinct angular harmonics
become orthogonal as $N_y\to\infty$; we use this limit in the
following analysis.

At zeroth order in $\varepsilon_d$, the only nonzero
components are those of the stationary $p$-wave background,
\begin{equation}
    u_{\mathbf{n},1}(0)
    =
    \frac{\Delta_{p,0}\eta_{\mathbf{n}}}
    {2E_{\mathbf{n},p}^{(0)}},
    \qquad
    w_{\mathbf{n},0}(0)
    =
    \frac{\xi_{\mathbf{n}}}
    {2E_{\mathbf{n},p}^{(0)}},
\end{equation}
where
$E_{\mathbf{n},p}^{(0)}
=\sqrt{\xi_{\mathbf{n}}^2+\Delta_{p,0}^2\eta_{\mathbf{n}}^2}$
and $\xi_{\mathbf{n}}=\mu-J\eta_{\mathbf{n}}^2$.
Expanding Eq.~(\ref{eq:pinit}) to first order introduces
three additional components,
\begin{equation}
    \begin{gathered}
    u_{\mathbf{n},0}(0)
    =
    -\frac{\Delta_{p,0}^3\eta_{\mathbf{n}}^4}
    {4(E_{\mathbf{n},p}^{(0)})^3}\varepsilon_d,
    \\
    u_{\mathbf{n},2}(0)
    =
    \left(
    \frac{\eta_{\mathbf{n}}^2}{2E_{\mathbf{n},p}^{(0)}}
    -
    \frac{\Delta_{p,0}^2\eta_{\mathbf{n}}^4}
    {4(E_{\mathbf{n},p}^{(0)})^3}
    \right)\Delta_{p,0}\varepsilon_d,
    \\
    w_{\mathbf{n},1}(0)
    =
    w_{\mathbf{n},-1}^{*}(0)
    =
    -\frac{\Delta_{p,0}^2\eta_{\mathbf{n}}^3\xi_{\mathbf{n}}}
    {4(E_{\mathbf{n},p}^{(0)})^3}\varepsilon_d.
    \end{gathered}
\end{equation}
In particular, the $m=2$ component generates a small initial
$d$-wave order parameter,
\begin{equation}
    \Delta_{d,0}
    =
    \chi_d\sum_{\mathbf{n}}
    \eta_{\mathbf{n}}^2u_{\mathbf{n},2}(0).
\end{equation} 

 We evolve the perturbed state in a frame rotating at frequency
$2\mu$, in which the unperturbed $p$-wave order is stationary.
The mean-field equations in this frame are
\begin{equation}
    \begin{aligned}
    \frac{d}{dt}\langle\hat{S}_{\mathbf{n}}^{-}\rangle
    &=
    2i\xi_{\mathbf{n}}\langle\hat{S}_{\mathbf{n}}^{-}\rangle
    -2i\Delta_{\mathbf{n}}\langle\hat{S}_{\mathbf{n}}^{z}\rangle,
    \\
    \frac{d}{dt}\langle\hat{S}_{\mathbf{n}}^{z}\rangle
    &=
    -i\Delta_{\mathbf{n}}^{*}
    \langle\hat{S}_{\mathbf{n}}^{-}\rangle
    +i\Delta_{\mathbf{n}}
    \langle\hat{S}_{\mathbf{n}}^{+}\rangle,
    \end{aligned}
    \label{eq:mean_rotate}
\end{equation}
where
$\Delta_{\mathbf{n}}
=\Delta_p\eta_{\mathbf{n}}e^{-i\phi_{\mathbf{n}}}
+\Delta_d\eta_{\mathbf{n}}^2e^{-2i\phi_{\mathbf{n}}}$.
At zeroth order, the background components remain constant,
\begin{equation}
    u_{\mathbf{n},1}(t)=u_{\mathbf{n},1}(0),
    \qquad
    w_{\mathbf{n},0}(t)=w_{\mathbf{n},0}(0).
\end{equation}
The perturbation therefore evolves on a stationary $p$-wave
background. Keeping only terms linear in $\varepsilon_d$,
we obtain
\begin{equation}
    \begin{aligned}
    \frac{d}{dt}u_{\mathbf{n},0}^{*}
    &=
    -2i\xi_{\mathbf{n}}u_{\mathbf{n},0}^{*}
    +2i\Delta_{p,0}\eta_{\mathbf{n}}w_{\mathbf{n},1},
    \\
    \frac{d}{dt}u_{\mathbf{n},2}
    &=
    2i\xi_{\mathbf{n}}u_{\mathbf{n},2}
    -2i\left(
    \Delta_{p,0}\eta_{\mathbf{n}}w_{\mathbf{n},1}
    +\Delta_d\eta_{\mathbf{n}}^2w_{\mathbf{n},0}
    \right),
    \\
    \frac{d}{dt}w_{\mathbf{n},1}
    &=
    i\left(
    \Delta_{p,0}\eta_{\mathbf{n}}u_{\mathbf{n},0}^{*}
    +\Delta_d\eta_{\mathbf{n}}^2u_{\mathbf{n},1}
    \right)
    -i\Delta_{p,0}\eta_{\mathbf{n}}u_{\mathbf{n},2}.
    \end{aligned}
    \label{eq:un0}
\end{equation}

To determine the collective-mode frequencies, we seek solutions
with a common time dependence $e^{-i\omega t}$,
\begin{equation}
    (u_{\mathbf{n},0}^{*},u_{\mathbf{n},2},
    w_{\mathbf{n},1},\Delta_d)
    =
    (\tilde{u}_{\mathbf{n},0}^{*},
    \tilde{u}_{\mathbf{n},2},
    \tilde{w}_{\mathbf{n},1},\tilde{\Delta}_d)
    e^{-i\omega t}.
\end{equation}
Substitution into Eq.~(\ref{eq:un0}) gives
\begin{equation}
    \begin{gathered}
    \tilde{u}_{\mathbf{n},0}^{*}
    =
    -\frac{\tilde{\Delta}_d\Delta_{p,0}^2\eta_{\mathbf{n}}^4}
    {E_{\mathbf{n},p}^{(0)}
    \left(4(E_{\mathbf{n},p}^{(0)})^2-\omega^2\right)},
    \\
    \tilde{u}_{\mathbf{n},2}
    =
    \frac{\tilde{\Delta}_d\eta_{\mathbf{n}}^2
    \left((E_{\mathbf{n},p}^{(0)})^2
    +\xi_{\mathbf{n}}^2-\xi_{\mathbf{n}}\omega\right)}
    {E_{\mathbf{n},p}^{(0)}
    \left(4(E_{\mathbf{n},p}^{(0)})^2-\omega^2\right)},
    \\
    \tilde{w}_{\mathbf{n},1}
    =
    \frac{\tilde{\Delta}_d\Delta_{p,0}\eta_{\mathbf{n}}^3
    (\omega-2\xi_{\mathbf{n}})}
    {2E_{\mathbf{n},p}^{(0)}
    \left(4(E_{\mathbf{n},p}^{(0)})^2-\omega^2\right)}.
    \end{gathered}
\end{equation} 
These components must also satisfy the self-consistency
condition
$\tilde{\Delta}_d
=\chi_d\sum_{\mathbf{n}}\eta_{\mathbf{n}}^2
\tilde{u}_{\mathbf{n},2}$.
For a nonzero collective-mode amplitude $\tilde{\Delta}_d$,
this condition yields
\begin{equation}
    1=\chi_d\Pi_d(\omega),
    \label{eq:bsfrequency}
\end{equation}
where
\begin{equation}
    \begin{aligned}
    \Pi_d(\omega)
    &=
    \sum_{\mathbf{n}}
    \frac{\eta_{\mathbf{n}}^4
    \left((E_{\mathbf{n},p}^{(0)})^2
    +\xi_{\mathbf{n}}^2-\xi_{\mathbf{n}}\omega\right)}
    {E_{\mathbf{n},p}^{(0)}
    \left(4(E_{\mathbf{n},p}^{(0)})^2-\omega^2\right)}
    \\
    &=
    \sum_{\mathbf{n}}
    \frac{\eta_{\mathbf{n}}^4}{4(E_{\mathbf{n},p}^{(0)})^2}
    \left(
    \frac{(E_{\mathbf{n},p}^{(0)}-\xi_{\mathbf{n}})^2}
    {2E_{\mathbf{n},p}^{(0)}-\omega}
    +
    \frac{(E_{\mathbf{n},p}^{(0)}+\xi_{\mathbf{n}})^2}
    {2E_{\mathbf{n},p}^{(0)}+\omega}
    \right).
    \end{aligned}
    \label{eq:bspid}
\end{equation}

\begin{figure}[t]
    \centering
    \includegraphics[width=1.0\columnwidth]{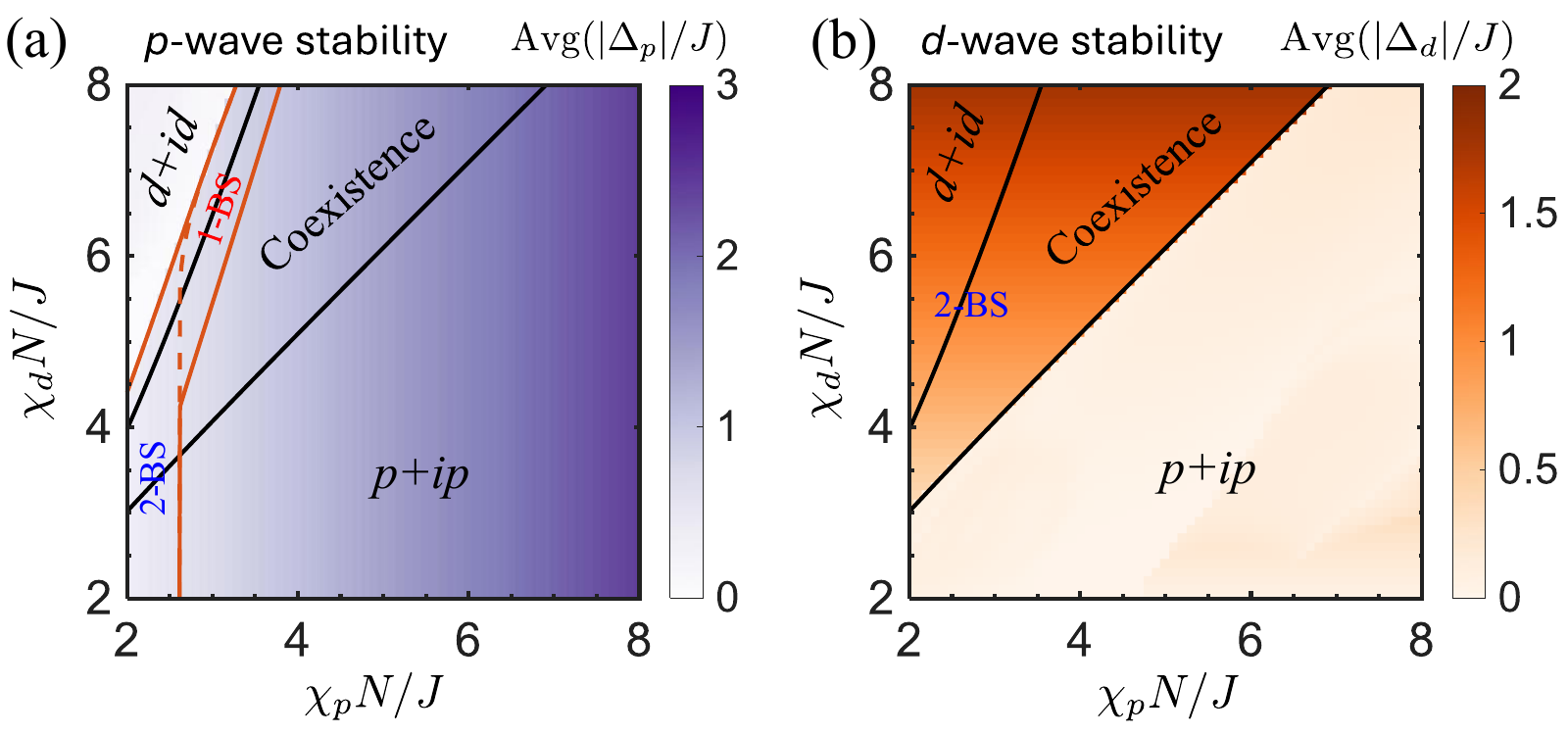}
    \caption{(a) Stability of the $p+ip$ solution. We prepare $p+ip$ initial state based on Eq.~(\ref{eq:pinit}) with $\varepsilon_d=10^{-2}$ and evolve under system parameters $\chi_p$ and $\chi_d$. The long-time averaged $p$-wave order parameter $\mathrm{Avg}(|\Delta_p|/J)$ either remains at its initial value (stable) or decays to $0$ (unstable). The orange solid lines are the boundaries of the regime with Bardasis-Schrieffer (BS) modes. The orange dashed line is separating the regimes with a single BS mode (1-BS) and two BS modes (2-BS). In both (a) and (b), we fix $N_C/N=0.35$. The black solid lines are analytical results of the energetic stability boundaries (see Eq.~(\ref{eq:dstab}) and Eq.~(\ref{eq:pstab})). The heat maps probe dynamical stability. (b) Stability of the $d+id$ solution. We prepare $d+id$ initial state based on Eq.~(\ref{eq:dinit}) with $\varepsilon_p=10^{-2}$ and show the long-time averaged $d$-wave order parameter $\mathrm{Avg}(|\Delta_d|/J)$. The boundary of BS modes in this case is very close to the energetic stability boundary between coexistence regime and $p+ip$ regime (see Eq.~(\ref{eq:pstab})). }
    \label{fig:appen2}
\end{figure}

A Bardasis-Schrieffer (BS) mode is a real-frequency solution
of Eq.~(\ref{eq:bsfrequency}) below the pair-breaking threshold,
$\omega_{\rm gap}=\min_{\mathbf n}2E_{\mathbf n,p}^{(0)}$.
For $|\omega|\;<\omega_{\rm gap}$, the denominators in
Eq.~(\ref{eq:bspid}) do not vanish, and the response function
$\Pi_d(\omega)$ is real. An excitation at such a frequency
cannot decay by breaking Cooper pairs. The range of
couplings $\chi_d^L<\chi_d<\chi_d^U$ that supports these solutions is determined by
\begin{equation}
    \chi_d^L
    =
    \inf_{|\omega|<\omega_{\rm gap}}\frac{1}{\Pi_d(\omega)},
    \quad
    \chi_d^U
    =
    \sup_{|\omega|<\omega_{\rm gap}}\frac{1}{\Pi_d(\omega)}.
\end{equation}
These definitions include limiting values as $\omega$
approaches $\pm\omega_{\rm gap}$ from inside the interval.
At the topological transition,
$\chi_p=\chi_{p,\rm QCP}$, the gap closes, leaving no frequency
interval for an isolated BS mode below the pair-breaking
threshold.

The lower coupling boundary depends on whether the
lowest-energy quasiparticle states couple appreciably to the
competing $d$-wave channel. For $\eta_{\mathbf n}\in[0,1]$,
$J>0$, and $\mu<J$, the energy minimum occurs at
\begin{equation}
    (\eta_{\mathbf n})_{\rm min}
    =
    \begin{cases}
    0,
    & \displaystyle \mu\leq\frac{\Delta_{p,0}^2}{2J},
    \\[6pt]
    \displaystyle
    \frac{\sqrt{2J\mu-\Delta_{p,0}^2}}{\sqrt{2}J},
    & \displaystyle \mu>\frac{\Delta_{p,0}^2}{2J},
    \end{cases}
\end{equation}
giving
\begin{equation}
    \omega_{\rm gap}
    =
    \begin{cases}
    2|\mu|,
    & \displaystyle \mu\leq\frac{\Delta_{p,0}^2}{2J},
    \\[6pt]
    \displaystyle
    \frac{\Delta_{p,0}}{J}\sqrt{4J\mu-\Delta_{p,0}^2},
    & \displaystyle \mu>\frac{\Delta_{p,0}^2}{2J}.
    \end{cases}
\end{equation}

When $(\eta_{\mathbf n})_{\rm min}=0$, the coupling weight
$\eta_{\mathbf n}^4$ in Eq.~(\ref{eq:bspid}) suppresses the contribution of the
lowest-energy states to $\Pi_d(\omega)$. If this suppression
keeps the response finite at the pair-breaking thresholds,
a finite attraction is required to produce a BS mode.
Indeed, a bounded response cannot satisfy
$\Pi_d(\omega)=1/\chi_d$ for arbitrarily small $\chi_d$,
giving $\chi_d^L>0$. In contrast, when
$(\eta_{\mathbf n})_{\rm min}>0$, the states at the energy
minimum retain a nonzero coupling weight. Their contribution
makes the response diverge as $\omega$ approaches either
$-\omega_{\rm gap}$ or $+\omega_{\rm gap}$. An arbitrarily
weak positive attraction can then satisfy the mode equation
at a frequency sufficiently close to one of these thresholds,
giving $\chi_d^L=0$.

To determine whether there are one or two BS modes, we examine
the frequency dependence of $\Pi_d(\omega)$. For a fixed
coupling $\chi_d$, each intersection of this response curve with the
horizontal line at height $1/\chi_d$ gives a BS frequency.
Differentiating Eq.~(\ref{eq:bspid}) gives
\begin{equation}
    \Pi_d'(\omega)
    =
    \sum_{\mathbf n}
    \frac{\eta_{\mathbf n}^4}
    {4(E_{\mathbf n,p}^{(0)})^2}
    \left[
    \frac{(E_{\mathbf n,p}^{(0)}-\xi_{\mathbf n})^2}
    {(2E_{\mathbf n,p}^{(0)}-\omega)^2}
    -
    \frac{(E_{\mathbf n,p}^{(0)}+\xi_{\mathbf n})^2}
    {(2E_{\mathbf n,p}^{(0)}+\omega)^2}
    \right],
\end{equation}
and
\begin{equation}
    \Pi_d''(\omega)
    =
    \sum_{\mathbf n}
    \frac{\eta_{\mathbf n}^4}
    {2(E_{\mathbf n,p}^{(0)})^2}
    \left[
    \frac{(E_{\mathbf n,p}^{(0)}-\xi_{\mathbf n})^2}
    {(2E_{\mathbf n,p}^{(0)}-\omega)^3}
    +
    \frac{(E_{\mathbf n,p}^{(0)}+\xi_{\mathbf n})^2}
    {(2E_{\mathbf n,p}^{(0)}+\omega)^3}
    \right].
\end{equation}
The second derivative $\Pi_d''(\omega)$ is positive for every frequency
satisfying $|\omega|\;<\omega_{\rm gap}$, so the slope of the
response $\Pi_d'(\omega)$ increases continuously across this interval.
The response is therefore either monotonic or decreases
to a single minimum before increasing again. It can
intersect the horizontal line at most twice, giving
at most two distinct BS frequencies.

We define the inverse limiting responses at the two
thresholds as
\begin{equation}
    \chi_d^\pm=\lim_{\substack{\omega\to\pm\omega_{\rm gap}\\|\omega|<\omega_{\rm gap}}}
    \frac{1}{\Pi_d(\omega)}.
\end{equation}
If $\Pi_d'(\omega)$ has the same sign at both ends of the interval,
the response is monotonic and supports one BS mode for
$\min(\chi_d^+,\chi_d^-)<\chi_d<\max(\chi_d^+,\chi_d^-)$.
In this case,
\begin{equation}
    \chi_d^L=\min(\chi_d^+,\chi_d^-),
    \qquad
    \chi_d^U=\max(\chi_d^+,\chi_d^-).
\end{equation}

If instead $\Pi_d'(\omega)$ is negative near $-\omega_{\rm gap}$
and positive near $+\omega_{\rm gap}$, the response $\Pi_d(\omega)$ has a
minimum at a frequency $\omega_*$ satisfying $\Pi_d'(\omega_{*})=0$. Whether there are one
or two modes then depends on the position of $1/\chi_d$
relative to this minimum and the two limiting response
values. If $1/\chi_d$ lies between the two limiting values,
there is only one intersection and hence one BS mode.
If $1/\chi_d$ is smaller than both limiting values but
larger than the minimum, there are two intersections,
one on each side of the minimum, and hence two BS modes.
Thus, there is one BS mode for
\begin{equation}
    \chi_d^L<\chi_d<\max(\chi_d^+,\chi_d^-),
\end{equation}
and two BS modes for
\begin{equation}
    \max(\chi_d^+,\chi_d^-)<\chi_d<\chi_d^U,
\end{equation}
where
\begin{equation}
    \chi_d^L=\min(\chi_d^+,\chi_d^-),
    \qquad
    \chi_d^U=\frac{1}{\Pi_d(\omega_*)}.
\end{equation}
In the regime with two BS modes, increasing $\chi_d$ leads to a decrease of $\Pi_d(\omega)$, bringing the system closer to the minimum of $\Pi_d(\omega)$, so the two BS frequencies become closer together. At $\chi_d=\chi_d^U$, they
merge at $\omega_*$. 
When the response diverges at both thresholds,
$\chi_d^+=\chi_d^-=0$. In that case, there are two BS modes
throughout $0<\chi_d<\chi_d^U$.

The number of excited modes determines the long-time
dynamics. The following expressions describe the surviving
BS-mode contributions after the response from the
quasiparticle continuum has dephased. 
When a single BS mode contributes, the surviving response
in the frame rotating at frequency $2\mu$ is
\begin{equation}
    \Delta_d(t)
    \longrightarrow
    \Delta_{d,\infty}e^{-i\omega_{\rm BS}t}.
\end{equation}
Its magnitude approaches a constant, while the relative
phase between the two orders winds at a constant rate,
\begin{equation}
    \phi_{pd}(t)
    =
    \operatorname{Arg}
    \left[\Delta_p^*(t)\Delta_d(t)\right]
    \longrightarrow
    \phi_0-\omega_{\rm BS}t.
\end{equation}
The slope of the unwrapped relative phase therefore gives
the BS frequency. Because a common phase rotation cancels
in $\Delta_p^*\Delta_d$, this measurement is independent
of the choice of rotating frame.

When both modes are excited, in the frame rotating at frequency $2\mu$ we get
\begin{equation}
    \Delta_d(t)
    \longrightarrow
    \Delta_{d,\infty}^{+}e^{-i\omega_{\rm BS}^{+}t}
    +
    \Delta_{d,\infty}^{-}e^{-i\omega_{\rm BS}^{-}t}.
\end{equation}
Their interference produces persistent oscillations
in $|\Delta_d|$ at the difference frequency
$|\omega_{\rm BS}^{+}-\omega_{\rm BS}^{-}|$.
The individual frequencies can be resolved in the
Fourier spectrum of $\Delta_p^*(t)\Delta_d(t)$.

Finally, the condition that a collective frequency
reaches zero defines the energetic stability boundary,
\begin{equation}
    \chi_d^{\rm stab}
    =
    \frac{1}{\Pi_d(0)}.
    \label{eq:dstab}
\end{equation}
At this boundary, an infinitesimal deformation toward
the competing $d$-wave order costs no energy.
However, energetic instability does not necessarily imply
exponential growth. A collective frequency
can cross zero while remaining real, allowing the
$p$-wave state to remain dynamically stable.
Dynamical instability begins
when $\chi_d>\chi_d^U$ (the frequency solution becomes complex), producing an exponentially growing
perturbation.

In the strong-pairing limit,
$\Delta_{p,0}\gg J,|\mu|$, the approximation used
in Eq.~(\ref{eq:largelimit}) gives
\begin{equation}
    \chi_d^{\rm stab}\approx3\chi_p.
\end{equation}
In the weak-pairing regime,
$\Delta_{p,0}\ll J,|\mu|$, the approximations used
in Eqs.~(\ref{eq:psmall}) and~(\ref{eq:smalllimit}) give
\begin{equation}
    \chi_d^{\rm stab}\approx\frac{J}{\mu}\chi_p.
\end{equation}

\subsection{Stability analysis of $d_{x^2-y^2}+id_{xy}$ solution and Bardasis-Schrieffer modes}

Similarly, we consider a small quench to introduce a small amount of $p_x+ip_y$ pairing ($\varepsilon_p \ll 1$) to the $d_{x^2-y^2}+id_{xy}$ solution at system parameters $\chi_p$, $\chi_d$ and $N_C$,
\begin{equation}
    \begin{gathered}
    \langle \hat{S}_{\mathbf{n}}^{-}(0)\rangle = \frac{\Delta_{d,0} (\varepsilon_p\eta_{\mathbf{n}}e^{-i\phi_{\mathbf{n}}}+\eta_{\mathbf{n}}^2e^{-2i\phi_{\mathbf{n}}})}{2E_{\mathbf{n},d}},\\
    \langle \hat{S}_{\mathbf{n}}^{z}(0)\rangle = \frac{\mu-J\eta_{\mathbf{n}}^2}{2E_{\mathbf{n},d}},\\
    E_{\mathbf{n},d} = \sqrt{(J\eta_{\mathbf{n}}^2-\mu)^2+\Delta_{d,0}^2(\varepsilon_p^2\eta_{\mathbf{n}}^2+\eta_{\mathbf{n}}^4+2\varepsilon_p\eta_{\mathbf{n}}^3\cos(\phi_{\mathbf{n}}))}.
    \end{gathered}
    \label{eq:dinit}
\end{equation}
We choose the initial $d$-wave pairing $\Delta_{d,0}$ to be real and positive. Its value and the chemical potential $\mu$ are determined by the self-consistent equations (see Eq.~(\ref{eq:dconsist})).

Similar to Eq.~(\ref{eq:bsfrequency}) and Eq.~(\ref{eq:bspid}), one can obtain the following equation for BS mode frequency,
\begin{equation}
    1 = \chi_p \Pi_p(\omega),
\end{equation}
where
\begin{equation}
    \begin{aligned}
    \Pi_p(\omega) &= \sum_{\mathbf{n}}\frac{\eta_{\mathbf{n}}^2\Big((E_{\mathbf{n},d}^{(0)})^2+\xi_{\mathbf{n}}^2-\xi_{\mathbf{n}}\omega\Big)}{E_{\mathbf{n},d}^{(0)}\Big(4(E_{\mathbf{n},d}^{(0)})^2-\omega^2\Big)}\\
    &= \sum_{\mathbf{n}}\frac{\eta_{\bf n}^2}{4(E_{\mathbf{n},d}^{(0)})^2}\bigg(\frac{(E_{\mathbf{n},d}^{(0)}-\xi_{\bf n})^2}{2E_{\mathbf{n},d}^{(0)}-\omega}+\frac{(E_{\mathbf{n},d}^{(0)}+\xi_{\bf n})^2}{2E_{\mathbf{n},d}^{(0)}+\omega}\bigg).
    \end{aligned}
    \label{eq:bsdip}
\end{equation}
Here we define $E_{\mathbf{n},d}^{(0)} = \sqrt{\xi_{\mathbf{n}}^2+\Delta_{d,0}^2\eta_{\mathbf{n}}^4}$.
Based on Eq.~(\ref{eq:bsdip}), one can obtain real-frequency solutions if $|\omega|\;<\omega_{\rm gap}$, where $\omega_{\rm gap}= \min_{\mathbf{n}}2E_{\mathbf{n},d}^{(0)}$.
This allows us to determine the regime $\chi_p^L<\chi_p<\chi_p^U$ where BS modes exist, with
\begin{equation}
    \chi_p^L = \inf_{|\omega|<\omega_{\rm gap}}\frac{1}{\Pi_p(\omega)}, \quad \chi_p^U = \sup_{|\omega|<\omega_{\rm gap}}\frac{1}{\Pi_p(\omega)}.
\end{equation} 

In the following, we would like to discuss different cases of BS modes and their connection to the value of $\eta_{\bf n}$ that minimizes $E_{\mathbf{n},d}^{(0)}$. Considering $\eta_{\bf n}\in [0,1]$, $J>0$ and $\mu<J$, we have
\begin{equation}
    (\eta_{\mathbf{n}})_{\rm min} = \begin{cases}
        \displaystyle 0, \quad\quad\quad\quad\quad\quad\quad \mu\leq 0,\\[10pt]
        \displaystyle\sqrt{\frac{J\mu}{J^2+\Delta_{d,0}^2}}, \quad\quad \mu>0,
    \end{cases}
\end{equation}
\begin{equation}
    \omega_{\rm gap} = \begin{cases}
        \displaystyle 2|\mu|, \quad\quad\quad\quad\quad\quad\quad\; \mu\leq 0,\\[10pt]
        \displaystyle\frac{2\mu\Delta_{d,0}}{\sqrt{J^2+\Delta_{d,0}^2}}, \quad\quad\quad\;\; \mu>0.
    \end{cases}    
\end{equation}
In the case of $(\eta_{\bf n})_{\rm min}=0$, the $\eta_{\bf n}^2$ factor in Eq.~(\ref{eq:bsdip}) will remove the poles at $\omega=\pm\omega_{\rm gap}$. This means $\Pi_p(\omega)$ remains finite for all values of $|\omega|\;<\omega_{\rm gap}$, leading to $\chi_p^L>0$. In the case of $(\eta_{\bf n})_{\rm min}>0$, the poles at $\omega=\pm\omega_{\rm gap}$ always lead to $\chi_p^L=0$. 

Following the similar analysis in $\Pi'_p(\omega)$ and $\Pi''_p(\omega)$, one can show that $\Pi''_p(\omega)>0$ for all values of $|\omega|\;<\omega_{\rm gap}$, leading to the monotonic growth of $\Pi'_p(\omega)$.
In the case of $(\eta_{\bf n})_{\rm min}>0$ ($\mu>0$), the poles at $\omega=\pm\omega_{\rm gap}$ always lead to two BS modes.
In the case of $(\eta_{\bf n})_{\rm min}=0$ ($\mu\leq 0$), one can obtain $\xi_{\bf n}<0$, and this leads to $\Pi'_p(\omega)>0$ for all values of $|\omega|\;<\omega_{\rm gap}$, leading to the monotonic growth of $\Pi_p(\omega)$ and a single BS mode in the system.

Similarly, we can also define the stability boundary as
\begin{equation}
    \chi_p^{\rm stab} = \frac{1}{\Pi_p(\omega=0)}.
    \label{eq:pstab}
\end{equation}

When $\Delta_{d,0} \gg J,|\mu|$, similarly one can obtain
\begin{equation}
    \chi_{p,\rm stab}\approx \chi_d.
\end{equation}
When $\Delta_{d,0} \ll J,|\mu|$, similarly one can obtain
\begin{equation}
    \chi_{p,\rm stab}\approx \frac{\mu}{J}\chi_d.
\end{equation}

We benchmark the analytical calculation of the stability boundaries with numerical simulation in Fig.~\ref{fig:appen2}(a) and (b). 
We prepare $p_x+ip_y$ initial state based on Eq.~(\ref{eq:pinit}) with $\varepsilon_d=10^{-2}$ and numerically calculate $\Delta_p(t)$ up to $Jt/2\pi=50$. Based on $\Delta_p(t)$ we then evaluate the long-time average $\mathrm{Avg}(|\Delta_p|/J)$. 
We find that $\mathrm{Avg}(|\Delta_p|/J)$ either remains at its initial value (stable) or decays to $0$ (unstable). Similarly, we prepare $d_{x^2-y^2}+id_{xy}$ initial state based on Eq.~(\ref{eq:dinit}) with $\varepsilon_p=10^{-2}$ and evaluate the long-time averaged $d$-wave order parameter $\mathrm{Avg}(|\Delta_d|/J)$.
We find that the numerical results of instability agree with the analytical calculation of $\chi_d^U$ instead of $\chi_d^{\rm stab}$ in Fig.~\ref{fig:appen2}(a), indicating that the existence of BS modes is modifying the direct picture of energetic instability.   
In Fig.~\ref{fig:appen2}(b), the numerical results of instability agree with $\chi_p^{\rm stab}$, since in this case $\chi_p^U$ is very close to $\chi_p^{\rm stab}$.
Identifying the small difference between $\chi_p^U$ and $\chi_p^{\rm stab}$ is beyond the resolution of our numerical scan.

\subsection{External modulation for Bardasis-Schrieffer modes}
As we discussed in the main text, BS modes can also be excited spectroscopically without introducing an initial $d$-wave component.
This can be achieved by applying  two extra drives (drive $\tilde{A}$ from the side and drive $\tilde{B}$ along the cavity), both of which couple to the $|\uparrow\rangle\rightarrow|e\rangle$ transition.
Unlike the case of engineering cavity-mediated interaction, we consider these two beams to be detuned far from the cavity resonance such that cavity-mediated interactions can be ignored.
We also consider these two beams have similar frequencies $\omega_{p,\tilde{A}}$ and $\omega_{p,\tilde{B}}$, leading to an interference term in AC Stark shifts. So the Hamiltonian is given by 
\begin{equation}
    \begin{aligned}
        \hat{H}_{\rm Stark}/\hbar
        & = \frac{1}{4\delta_e}\sum_{\mathbf{n}}\bigg(\frac{1}{2}+\hat{S}^z_{\mathbf{n}}\bigg)\bigg(|\Omega_{\tilde{A}}|^2+|\Omega_{\tilde{B}}|^2\eta_{\mathbf{n}}^2\\
        &+(\Omega_{\tilde{A}}^{*}\Omega_{\tilde{B}}\eta_{\mathbf{n}}e^{i(\omega_{p,\tilde{A}}-\omega_{p,\tilde{B}})t+i\phi_{\mathbf{n}}}+h.c.)\bigg).
    \end{aligned}
\end{equation}
The first two terms can be absorbed into the original Hamiltonian, so the only new term is the last term describing the interference between the two beams. For simplicity, we assume $\Omega_{\tilde{A}}=\Omega_{\tilde{B}}\equiv\tilde{\Omega}$; the equation above can be rewritten as
\begin{equation}
    \hat{H}_{\rm probe}/\hbar = 4h_0\sum_{\mathbf{n}}\eta_{\mathbf{n}}\cos(\omega_{\rm stir}t+\phi_{\mathbf{n}})\hat{S}^z_{\mathbf{n}},
\end{equation}
where 
\begin{equation}
    h_0 = \frac{\tilde{\Omega}^2}{8\delta_e}, \quad \omega_{\rm stir} = \omega_{p,\tilde{A}}-\omega_{p,\tilde{B}}.
\end{equation}
Adding this term to the mean-field equations, we have
\begin{equation}
    \begin{aligned}
    \frac{d}{dt}\langle \hat{S}^{-}_{\mathbf{n}}\rangle&=2i\Big(\xi_{\mathbf{n}}-2h_0\eta_{\mathbf{n}}\cos(\omega_{\rm stir}t+\phi_{\mathbf{n}})\Big)\langle\hat{S}^{-}_{\mathbf{n}}\rangle\\
    &\quad -2i \Delta_{\mathbf{n}}\langle\hat{S}^z_{\mathbf{n}}\rangle,\\
    \frac{d}{dt}\langle \hat{S}^z_{\mathbf{n}}\rangle&= -i \Delta_{\mathbf{n}}^{*}\langle\hat{S}^{-}_{\mathbf{n}}\rangle+i\Delta_{\mathbf{n}}\langle \hat{S}^{+}_{\mathbf{n}}\rangle.
    \end{aligned}
\end{equation}

In the following, we would like to show the resonant enhancement of $d$-wave pairing when $\omega_{\rm stir}\simeq \omega_{\rm BS}$. We initialize the system into a pure $p_x+ip_y$ eigenstate with no initial $d$-wave pairing. Similar to Eq.~(\ref{eq:mean_fourier}) and Eq.~(\ref{eq:un0}), we expand $\langle\hat{S}^{-}_{\mathbf{n}}\rangle$ and $\langle\hat{S}^{z}_{\mathbf{n}}\rangle$ in Fourier series, and the equations of motion for $u_{\mathbf{n},0}^{*}$, $u_{\mathbf{n},2}$ and $w_{\mathbf{n},1}$ are given by (up to linear order of $h_0$)
\begin{equation}
    \begin{aligned}
    \frac{d}{dt}u_{\mathbf{n},0}^{*} &= -2i\xi_{\mathbf{n}}u_{\mathbf{n},0}^{*} +2i\Delta_{p,0}\eta_{\mathbf{n}}w_{\mathbf{n},1} + 2ih_0\eta_{\mathbf{n}}u_{\mathbf{n},1}^{*}e^{-i\omega_{\rm stir}t},\\
    \frac{d}{dt}u_{\mathbf{n},2} &= 2i\xi_{\mathbf{n}}u_{\mathbf{n},2} - 2i\Big(\Delta_{p,0}\eta_{\mathbf{n}}w_{\mathbf{n,1}} + \Delta_d\eta_{\mathbf{n}}^2w_{\mathbf{n},0}\Big)\\
    &\quad - 2ih_0\eta_{\mathbf{n}}u_{\mathbf{n},1}e^{-i\omega_{\rm stir}t},\\
    \frac{d}{dt}w_{\mathbf{n},1} &= i\Big(\Delta_{p,0}\eta_{\mathbf{n}}u_{\mathbf{n},0}^{*}+\Delta_{d}\eta_{\mathbf{n}}^2u_{\mathbf{n},1}\Big)-i\Delta_{p,0}\eta_{\mathbf{n}}u_{\mathbf{n},2}.\\
    \end{aligned}
\end{equation}

A solution of the form,  $(u_{\mathbf{n},0}^{*}, u_{\mathbf{n},2}, w_{\mathbf{n},1}, \Delta_d)= (\tilde{u}_{\mathbf{n},0}^{*}, \tilde{u}_{\mathbf{n},2}, \tilde{w}_{\mathbf{n},1}, \tilde{\Delta}_d)e^{-i\omega_{\rm stir} t}$, must satisfy the self-consistent equation $\tilde{\Delta}_d=\chi_d\sum_{\mathbf{n}}\eta_{\mathbf{n}}^2\tilde{u}_{\mathbf{n},2}$, which gives
\begin{equation}
    \tilde{\Delta}_d = \chi_d \Pi_d(\omega_{\rm stir})\tilde{\Delta}_d + \chi_d\sigma(\omega_{\rm stir})h_0.
    \label{eq:c61}
\end{equation}
Here, $\Pi_d(\omega)$ is given by Eq.~(\ref{eq:bspid}), and
\begin{equation}
    \sigma(\omega) = \Delta_{p,0}\sum_{\mathbf{n}}\frac{\eta_{\mathbf{n}}^4(2\xi_{\mathbf{n}}-\omega)}{E_{\mathbf{n},p}^{(0)}\Big(4(E_{\mathbf{n},p}^{(0)})^2-\omega^2\Big)}.
\end{equation}
Solving Eq.~(\ref{eq:c61}), we have
\begin{equation}
    \tilde{\Delta}_d = \frac{h_0\chi_d\sigma(\omega_{\rm stir})}{1-\chi_d\Pi_d(\omega_{\rm stir})}.
    \label{eq:tilded}
\end{equation} 

The resonant enhancement can be understood from  the denominator $1-\chi_d\Pi_d(\omega_{\rm stir})$. When $\omega_{\rm stir}\simeq \omega_{\rm BS}$, based on Eq.~(\ref{eq:bsfrequency}), we have $\tilde{\Delta}_d\rightarrow\infty$.
Note that Eq.~(\ref{eq:tilded}) is only valid for the linear regime assuming macroscopic $p$-wave pairing.
When the $d$-wave pairing is comparable with $p$-wave pairing, the linear regime is  no longer a good approximation;  $\tilde{\Delta}_d$ instead reaches a maximum value.

\section{Details of numerical calculations}
\label{sec:numerical}

\begin{figure}[t]
    \centering
    \includegraphics[width=\columnwidth]{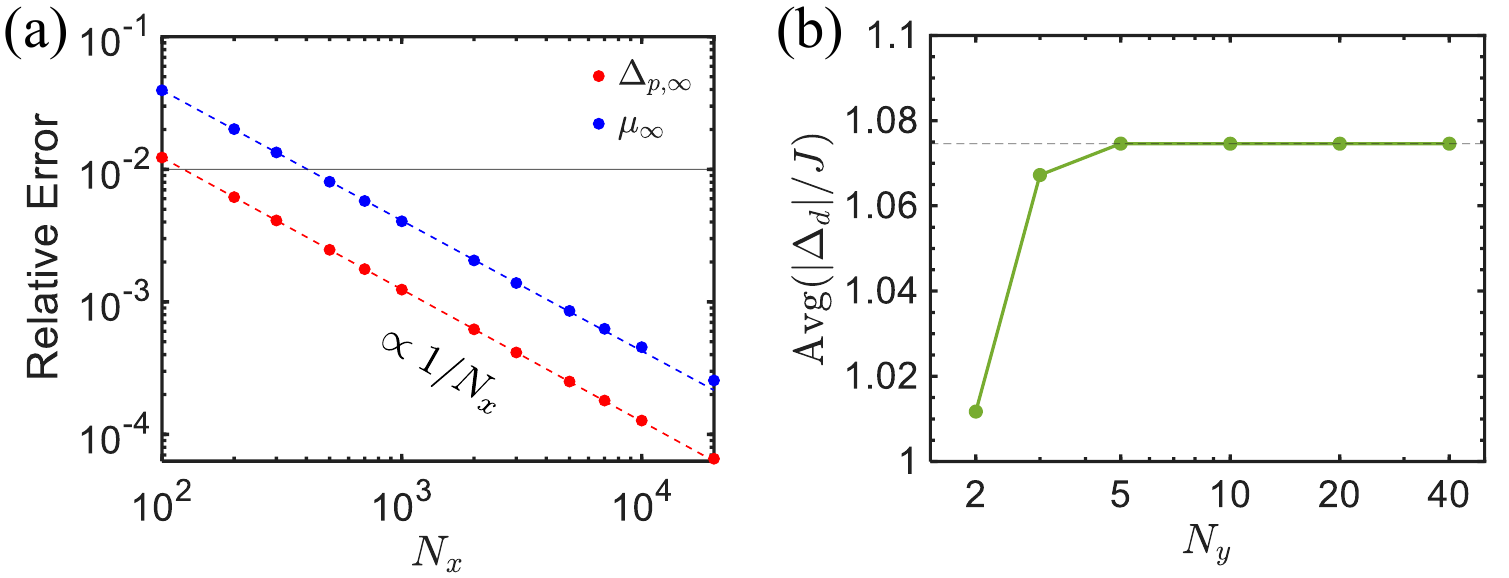}
    \caption{(a) Finite-size scaling analysis of long-time observables as a function of lattice site number $N_x$ in $x$ direction (pure $p$-wave physics is independent of $N_y$). We start from the $p_x+ip_y$ eigenstate at $\chi_{p,i}N/J=3.0$, $N_C/N=0.35$, and focus on the phase II steady state after quenching to $\chi_{p,f}N/J=4.0$, $\chi_dN/J=0$. Both $\Delta_{p,\infty}$ and $\mu_{\infty}$ show $1/N_x$ scaling of relative error compared to the large-$N_x$ limit. (b) Finite-size scaling analysis of long-time observables as a function of lattice site number $N_y$ in $y$ direction (we fixed $N_x=500$). We start from the $p_x+ip_y$ eigenstate with initial $d$-wave perturbation $\varepsilon_d=10^{-2}$ at $\chi_{p,i}N/J=3.0$, $N_C/N=0.35$, and focus on the long-time average of $d$-wave order after exponential growth at $\chi_{p,f}N/J=4.0$, $\chi_dN/J=8.0$. The dashed line marks the large-$N_y$ limit.}
    \label{fig:response5}
\end{figure}

\subsection{Mean-field simulation of system dynamics} 

We obtain the numerical results by solving the mean-field
equations for the atomic pseudospins. Unless otherwise stated, we use a $500\times20$ lattice, with $N_x=500$ sites along the cavity axis and $N_y=20$ sites transverse to it. 
For actual experimental implementation, the distribution of the phases $n_x\varphi\bmod 2\pi$ and $n_y\varphi\bmod 2\pi$ is approaching the uniform distribution over $[0,2\pi)$ at large-$N$ limit, since $\varphi/2\pi$ is incommensurate. 
Since the value of $\varphi$ depends on the choice of cavity wavelength (determined by the choice of atomic transitions) and lattice wavelength, in order to provide a general theoretical description, we instead directly sample these phases from the uniform distribution over $[0,2\pi)$, and ensure the numerical result converges to the large-$N$ limit as we discuss later in this subsection.    

The pseudospin dynamics in our numerical simulation include both the coherent cavity-mediated
interactions and dissipation from cavity photon loss.
The corresponding adjoint Lindblad equation for an operator
$\hat O$ is
\begin{equation}
    \begin{aligned}
    \frac{d}{dt}\hat O
    &=
    \frac{i}{\hbar}[\hat H_{\rm cav},\hat O]
    +\left(
    \hat L_p^\dagger\hat O\hat L_p
    -\frac{1}{2}\{\hat L_p^\dagger\hat L_p,\hat O\}
    \right)
    \\
    &\quad+
    \left(
    \hat L_d^\dagger\hat O\hat L_d
    -\frac{1}{2}\{\hat L_d^\dagger\hat L_d,\hat O\}
    \right),
    \end{aligned}
\end{equation}
where $\hat H_{\rm cav}$ is defined in Eq.~(\ref{eq:heff})
and the jump operators $\hat L_p$ and $\hat L_d$ are defined
in Eq.~(\ref{eq:jump}). We derive the mean-field equations
by taking expectation values and factorizing the operator
products that couple different pseudospins. The resulting
equations at each lattice site are
\begin{equation}
    \begin{aligned}
    \frac{d}{dt}\langle\hat S^x_{\mathbf n}\rangle
    &=
    B^z_{\mathbf n}\langle\hat S^y_{\mathbf n}\rangle
    -B^y_{\mathbf n}\langle\hat S^z_{\mathbf n}\rangle
    +\Gamma^x_{\mathbf n}\langle\hat S^z_{\mathbf n}\rangle,
    \\
    \frac{d}{dt}\langle\hat S^y_{\mathbf n}\rangle
    &=
    B^x_{\mathbf n}\langle\hat S^z_{\mathbf n}\rangle
    -B^z_{\mathbf n}\langle\hat S^x_{\mathbf n}\rangle
    +\Gamma^y_{\mathbf n}\langle\hat S^z_{\mathbf n}\rangle,
    \\
    \frac{d}{dt}\langle\hat S^z_{\mathbf n}\rangle
    &=
    B^y_{\mathbf n}\langle\hat S^x_{\mathbf n}\rangle
    -B^x_{\mathbf n}\langle\hat S^y_{\mathbf n}\rangle-
    \Gamma^x_{\mathbf n}\langle\hat S^x_{\mathbf n}\rangle
    -\Gamma^y_{\mathbf n}\langle\hat S^y_{\mathbf n}\rangle.
    \end{aligned}
\end{equation}
The coherent terms are described by the effective field
\begin{equation}
    \begin{aligned}
    B^x_{\mathbf n}
    &=
    2\mathrm{Re}\left(
    \Delta_p\eta_{\mathbf n}e^{-i\phi_{\mathbf n}}
    +\Delta_d\eta_{\mathbf n}^2e^{-2i\phi_{\mathbf n}}
    \right),
    \\
    B^y_{\mathbf n}
    &=
    -2\mathrm{Im}\left(
    \Delta_p\eta_{\mathbf n}e^{-i\phi_{\mathbf n}}
    +\Delta_d\eta_{\mathbf n}^2e^{-2i\phi_{\mathbf n}}
    \right),
    \\
    B^z_{\mathbf n}
    &=-2J\eta_{\mathbf n}^2,
    \end{aligned}
\end{equation}
while cavity loss contributes
\begin{equation}
    \begin{aligned}
    \Gamma^x_{\mathbf n}
    &=
    \mathrm{Re}\left(
    \frac{\Gamma_p}{\chi_p}
    \Delta_p\eta_{\mathbf n}e^{-i\phi_{\mathbf n}}
    +
    \frac{\Gamma_d}{\chi_d}
    \Delta_d\eta_{\mathbf n}^2e^{-2i\phi_{\mathbf n}}
    \right),
    \\
    \Gamma^y_{\mathbf n}
    &=
    -\mathrm{Im}\left(
    \frac{\Gamma_p}{\chi_p}
    \Delta_p\eta_{\mathbf n}e^{-i\phi_{\mathbf n}}
    +
    \frac{\Gamma_d}{\chi_d}
    \Delta_d\eta_{\mathbf n}^2e^{-2i\phi_{\mathbf n}}
    \right).
    \end{aligned}
\end{equation}
The order parameters $\Delta_p$ and $\Delta_d$, defined in
Eq.~(\ref{eq:order}), are evaluated self-consistently during
the evolution. The dissipative and coherent couplings satisfy
$\Gamma_p/\chi_p=-\kappa/\delta_{c,A}$ and
$\Gamma_d/\chi_d=-\kappa/\delta_{c,B}$.
For purely Hamiltonian dynamics, we set $\kappa=0$.
When including cavity loss, we typically use
$\kappa/|\delta_{c,A}|\;\approx\kappa/|\delta_{c,B}|\;
=2\times10^{-3}$.

We assess convergence with respect to $N_x$ and $N_y$
using the Hamiltonian dynamics shown in
Fig.~\ref{fig:response5}. We first vary $N_x$ in the pure
$p$-wave model (see Fig.~\ref{fig:response5}(a)). In this case,
a gauge transformation of the pseudospins can remove the
$y$-dependent coupling phases, so they do not affect the
dynamics. Starting from the stationary $p_x+ip_y$ state
at $\chi_{p,i}N/J=3.0$, we quench to
$\chi_{p,f}N/J=4.0$. The system approaches a phase-II
state, whose asymptotic pairing amplitude we extract
for each $N_x$. We fit its dependence on $N_x$ to
\begin{equation}
    \left|
    \Delta_{p,\infty}(N_x)
    -\Delta_{p,\infty}(N_x\to\infty)
    \right|
    =
    aN_x^{-b},
\end{equation}
where the extrapolated amplitude
$\Delta_{p,\infty}(N_x\to\infty)$, $a$, and $b$ are
fit parameters. We apply the same procedure to
$\mu_\infty$. Both quantities exhibit relative errors
that decrease approximately as $1/N_x$.
For this benchmark, $N_x=500$ reduces the relative
errors below $10^{-2}$.

We then test convergence with respect to $N_y$ in a
regime where the competing pairing channels make the
angular sampling relevant (see Fig.~\ref{fig:response5}(b)).
With $N_x=500$ fixed, we prepare the stationary
$p_x+ip_y$ state at $\chi_{p,i}N/J=3.0$ and add a small
$d$-wave perturbation, $\varepsilon_d=10^{-2}$.
For the parameters used in this benchmark, the
$d$-wave order initially grows exponentially.
We use its long-time averaged magnitude to assess
convergence and find that it changes little once
$N_y>5$. We therefore use $N_y=20$ in the simulations
unless otherwise stated.

\subsection{Numerical benchmarking for cavity QED implementation}
We have already summarized all the approximations we made in the derivation of the effective model for cavity QED implementation in App.~\ref{sec:cavity}-3.
In the following, we would like to provide further numerical benchmarking of their validity. 

Firstly, we would like to point out that the simultaneous application of adiabatic elimination of excited states (see Eq.~(\ref{eq:adia1})) and adiabatic elimination of cavity modes (see Eq.~(\ref{eq:adia2})) is standard in cavity QED systems and the validity has been tested in various cavity QED setups \cite{luo2025a,luo2025b,Young2024,Young2025}. 
These two approximations lead to the following spin-only Hamiltonian for the specific laser configuration  proposed in our implementation, 
\begin{equation}
    \hat{H}_{\rm eff} = \hat{H}_{\rm target} +  \hat{H}_{\rm corr} + \hat{H}_{\rm RW}.
\end{equation}
Here $\hat{H}_{\rm target}$ is the pseudospin model in Eq.~(\ref{eq:caveff}), which describes the dominant photon-mediated interactions via cavity mode $b$. 
The correction term $\hat{H}_{\rm corr}$ describes photon-mediated interactions via cavity mode $r$,
\begin{equation}
    \begin{aligned}
    \hat{H}_{\rm corr}/\hbar = &- \chi'_p\sum_{\mathbf{n}\mathbf{m}}\eta_{\mathbf{n}}\eta_{\mathbf{m}}e^{i\phi_{\mathbf{n}}}e^{-i\phi_{\mathbf{m}}}\hat{S}^z_{\mathbf{n}}\hat{S}^z_{\mathbf{m}}\\
    &-\chi'_d\sum_{\mathbf{n}\mathbf{m}}\eta_{\mathbf{n}}^2\eta_{\mathbf{m}}^2\hat{S}^z_{\mathbf{n}}\hat{S}^z_{\mathbf{m}},
    \end{aligned}
\end{equation}
and $\hat{H}_{\rm RW}$ describes the fast rotating terms due to different detunings of external drives from the resonance of cavity mode $b$,
\begin{equation}
    \begin{aligned}
    \hat{H}_{\rm RW}&/\hbar = \sum_{\mathbf{n}}2h_0\eta_{\mathbf{n}}e^{i\phi_{\mathbf{n}}}e^{i(\delta_{c,A}-\delta_{c,B})t}\hat{S}^z_{\mathbf{n}}\\
    &-\chi_{pd} \sum_{\mathbf{n}\mathbf{m}}\eta_{\mathbf{n}}\eta_{\mathbf{m}}^2 e^{-i\phi_{\mathbf{n}}}e^{2i\phi_{\mathbf{m}}}\hat{S}^{+}_{\mathbf{n}}\hat{S}^{-}_{\mathbf{m}}e^{i(\delta_{c,A}-\delta_{c,B})t} + \mathrm{h.c.}.
    \end{aligned}
\end{equation}

\begin{figure}[t]
    \centering
    \includegraphics[width=\columnwidth]{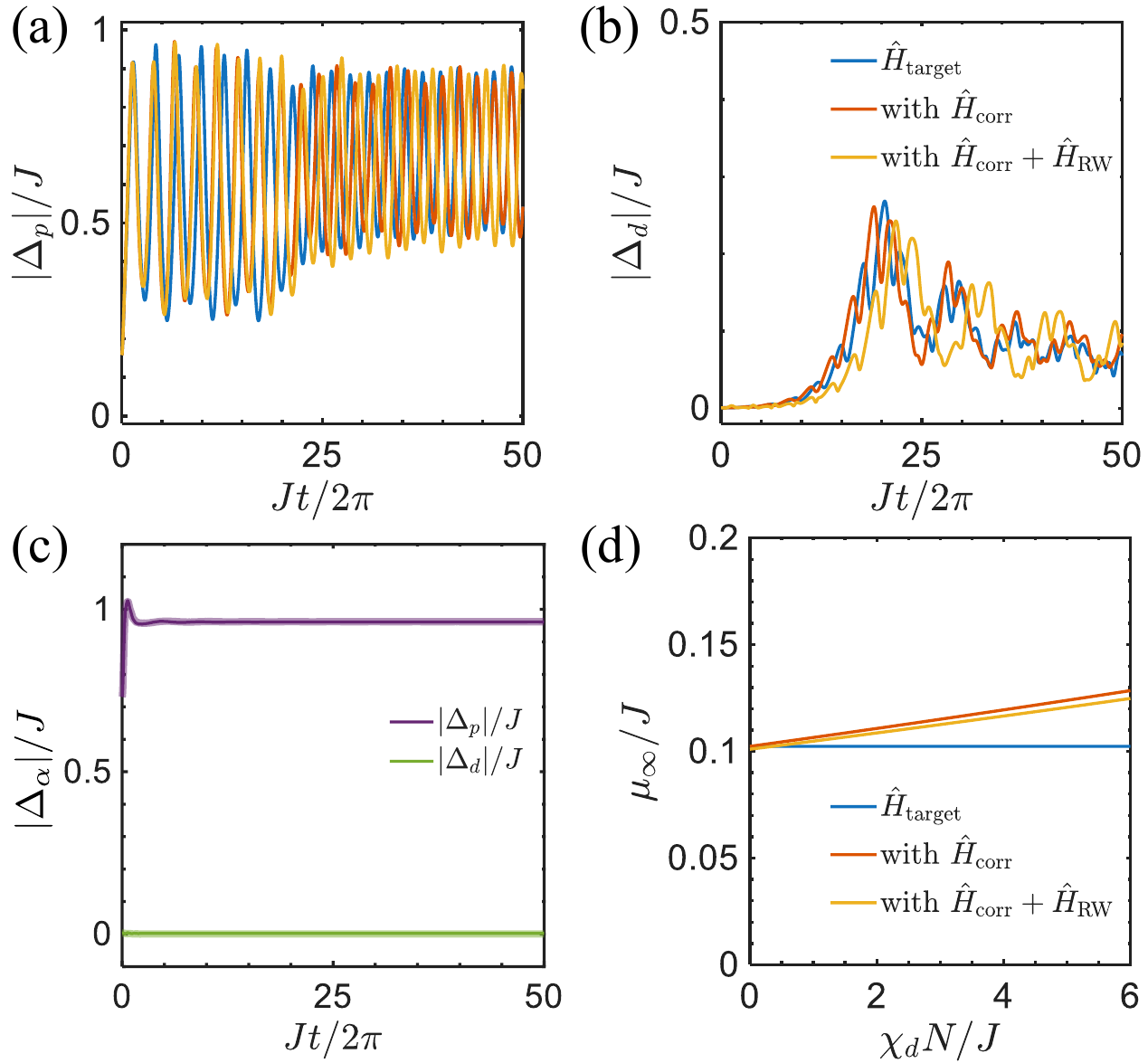}
    \caption{(a,b) Comparison of Hamiltonian dynamics of superconducting order parameters between $\hat{H}_{\rm target}$ (blue color), $\hat{H}_{\rm target}+\hat{H}_{\rm corr}$ (orange color) and $\hat{H}_{\rm target}+\hat{H}_{\rm corr}+\hat{H}_{\rm RW}$ (yellow color). We take the initial state to be a $p_x+ip_y$ eigenstate under $p$-wave interaction strength $\chi_{p,i}N/J=1.0$, with a small initial $d$-wave pairing ($\varepsilon_d=10^{-2}$), and evolve the system under final interaction strengths $\chi_{p,f}N/J=4.0$, $\chi_d N/J=3.5$. (c) Comparison of Hamiltonian dynamics between $\hat{H}_{\rm target}$ (lighter color) and $\hat{H}_{\rm target}+\hat{H}_{\rm corr}+\hat{H}_{\rm RW}$ (darker color) with $\chi_{p,i}N/J=2.0$, $\chi_{p,f}N/J=4.0$, $\chi_d N/J=3.5$. (d) Comparison of long-time chemical potential $\mu_{\infty}$ with $\chi_{p,i}N/J=2.0$ and $\chi_{p,f}N/J=4.0$. The color scheme is the same as (a) and (b).}
    \label{fig:response1}
\end{figure} 

Here we would like to provide further analysis for the validity of ignoring quantum fluctuations of cavity mode $r$ (see Eq.~(\ref{eq:adia3}) for dropping $\hat{H}_{\rm corr}$), as well as the rotating wave approximation for interference effects between different external drives (see Eq.~(\ref{eq:interference}) for dropping $\hat{H}_{\rm RW}$).
We drop $\hat{H}_{\rm corr}$ based on the condition $|\chi_p|\;\gg|\chi'_p|$ and $|\chi_d|\;\gg|\chi'_d|$. This is because the detunings to cavity mode $r$ and $b$ can be controlled separately, so one can suppress the effect of cavity mode $r$ by implementing a much larger detuning for cavity mode $r$. To validate this approximation, here we consider $\chi'_p = \zeta\chi_p$, $\chi'_d=\zeta\chi_d$, and check the effect of a non-zero $\zeta$ on our system.
Note that if $\chi_d=0$, the value $\langle\hat{S}^z_{\mathbf{n}}\rangle$ has no phase dependence, i.e.  $\sum_{\mathbf{n}}\eta_{\mathbf{n}}e^{i\phi_{\mathbf{n}}}\langle\hat{S}^z_{\mathbf{n}}\rangle=0$, thus the contribution of $\chi'_p$ term vanishes at the mean-field level in this case. 
We drop $\hat{H}_{\rm RW}$ based on the condition that the detuning difference between the two external drives, i.e. $\delta_{c,A}-\delta_{c,B}$, is much faster compared to the system dynamics. The parameters in $\hat{H}_{\rm RW}$ are given by $h_0=\Omega_A^{*}\Omega_B/(8\delta_e)$, $\chi_{pd}=-\Omega_A^{*}\Omega_Bg_c^2/(4\delta_e^2\delta_{c,AB})$, with $1/\delta_{c,AB}=(1/\delta_{c,A}+1/\delta_{c,B})/2$.

For both terms $\hat{H}_{\rm corr}$ and $\hat{H}_{\rm RW}$, noticeable differences mainly appear in the case of $\chi_d\neq 0$.
Following the setting in the main text, we initialize the system into a $p_x+ip_y$ eigenstate with a small $d$-wave pairing, leading to rich system dynamics due to competing $p$-wave and $d$-wave orders.
Here we set $\zeta=0.1$ for $\hat{H}_{\rm corr}$, $|\delta_{c,A}-\delta_{c,B}|/J = 200$ for $\hat{H}_{\rm RW}$, and discuss the correction to Hamiltonian dynamics of $\hat{H}_{\rm target}$ due to the term $\hat{H}_{\rm corr}$ only as well as the simultaneous application of $\hat{H}_{\rm corr}$ and $\hat{H}_{\rm RW}$.
In Fig.~\ref{fig:response1}(a) and (b), we use the case $\chi_{p,i}N/J=1.0$, $\chi_{p,f}N/J=4.0$, $\chi_d N/J=3.5$ as an example.
In this case, $\hat{H}_{\rm corr}$ and $\hat{H}_{\rm RW}$ lead to a small modification of $|\Delta_p|$ oscillation, as well as the growth of $d$-wave pairing in very short time.
In Fig.~\ref{fig:response1}(c) and (d), we set $\chi_{p,i}N/J=2.0$, $\chi_{p,f}N/J=4.0$ and scan the value of $\chi_d$. In Fig.~\ref{fig:response1}(c), we use a specific case with $\chi_d N/J=3.5$ to show that the system is dominated by $p$-wave order in phase II, and both $\hat{H}_{\rm corr}$ and $\hat{H}_{\rm RW}$ terms do not modify the value of $|\Delta_p|/J$. Instead, they mainly lead to a small shift of long-time chemical potential $\mu_{\infty}$, as shown in Fig.~\ref{fig:response1}(d).
For all these cases, the qualitative behavior of the dynamical phases remains unchanged, and one can further suppress the differences by choosing system parameters for a smaller $\zeta$ and a larger $|\delta_{c,A}-\delta_{c,B}|/J$. 

\subsection{Further numerical results on non-equilibrium competing orders}

\begin{figure}[t]
    \centering
    \includegraphics[width=\columnwidth]{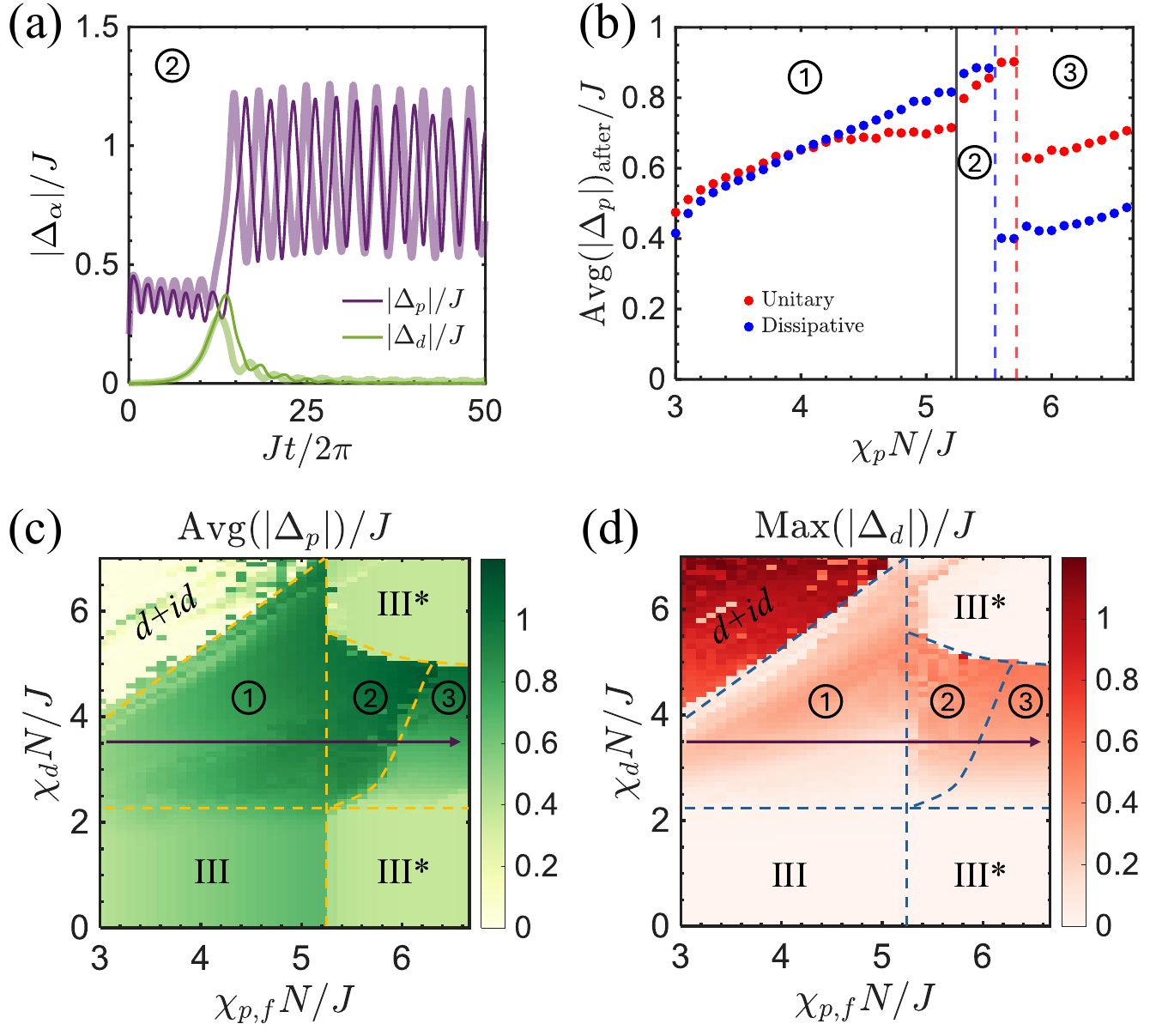}
    \caption{We prepare the $p_x+ip_y$ eigenstate with a small initial $d$-wave perturbation ($\varepsilon_d=10^{-2}$) at $\chi_{p,i}N/J=1.0$ and $N_C/N=0.35$, and perform a sudden quench of the interaction strength to $\chi_{p,f}N/J$ and $\chi_dN/J$. (a) Comparison of $|\Delta_p|$ and $|\Delta_d|$ dynamics in regime \circled{2} with Hamiltonian only (lighter color) and including cavity loss with $\kappa/|\delta_{c,A}|\approx\kappa/|\delta_{c,B}|=2\times 10^{-3}$ (darker color). We set $\chi_{p,f}N/J=5.5$, $\chi_dN/J=3.5$. (b) Average value of $|\Delta_p|$ between $Jt/2\pi\in [25,50]$, labeled as $\mathrm{Avg}(|\Delta_p|)_{\rm after}$. We compare the cases with Hamiltonian dynamics only (red points) and the cases including dissipative effects (blue points). The solid line marks the boundary between regime \circled{1} and \circled{2}, and the dashed lines mark the boundary between regime \circled{2} and \circled{3}. (c,d) Dynamical phase diagram as a function of $\chi_{p,f}$ and $\chi_d$ (Hamiltonian dynamics only). We show the long-time average value of $|\Delta_p|$ in (c), and the maximum value of $|\Delta_d|$ in (d). The dashed lines are a guide to the eye for separating different parameter regimes. The scan of $\chi_{p,f}$ in Fig.~\ref{fig:dynamics} is marked by the position of the arrow.}
    \label{fig:response4}
\end{figure}

Here we would like to provide further numerical results complementing the discussions in Sec.~\ref{sec:compete}-D and Fig.~\ref{fig:dynamics}. 
As discussed in the main text, we prepare the $p_x+ip_y$ eigenstate with a small initial $d$-wave perturbation ($\varepsilon_d=10^{-2}$) at $\chi_{p,i}N/J=1.0$ and $N_C/N=0.35$, and perform a sudden quench of the interaction strength to $\chi_{p,f}N/J$ and $\chi_dN/J$.
In Fig.~\ref{fig:response4}(a) and (b), we show the robustness against dissipative effects due to cavity loss.
When including dissipative effects, we replace the long-time average by the average within $Jt/(2\pi)\in[25,50]$.
This time range is long enough considering the additional quench due to the growth of $|\Delta_d|$, while the effect of dissipation is still typically small. 
We show that the regime \circled{1} (III $\rightarrow$ III) is nearly the same, and parts of the regime \circled{2} (III* $\rightarrow$ III) turn into the regime \circled{3} (III* $\rightarrow$ III*) when including dissipation.

In Fig.~\ref{fig:response4}(c) and (d), we also show the parameter regimes of $\chi_{p,f}$ and $\chi_d$ for observing the phenomena of non-equilibrium competing orders.
Since we have already discussed the dissipative effects in Fig.~\ref{fig:response4}(a) and (b), here we perform the parameter scans with Hamiltonian dynamics only.
When $\chi_d$ is small, the $d$-wave order vanishes and one can only observe the pure $p$-wave dynamical phase III and III* as $\chi_d\rightarrow 0$.
When $\chi_d$ dominates over $\chi_{p,f}$, the dominant $p$-wave order becomes unstable and decays eventually, similar to the physics starting from an equilibrium $p$-wave state.
We label this regime as $d+id$ regime, and the stability boundary has a smaller $\chi_d$ compared to the equilibrium case.
Interestingly, when both $\chi_d$ and $\chi_{p,f}$ are large, the $d$-wave order also vanishes and leads to pure $p$-wave dynamical phase III*.
The regimes \circled{1}, \circled{2} and \circled{3} showing non-equilibrium competing orders can be achieved at intermediate $\chi_d$, leading to the growth of $d$-wave order and a subsequent abrupt change of the dominant $p$-wave order.

In Fig.~\ref{fig:response4}(c), $\mathrm{Avg}(|\Delta_p|)$ in regimes \circled{1}, \circled{2} and \circled{3} shows the $p$-wave order after the abrupt change by the growth of $d$-wave order.
In regimes III and III* with vanishing $d$-wave order, $\mathrm{Avg}(|\Delta_p|)$ agrees with the $p$-wave order before the abrupt change in regimes \circled{1}, \circled{2} and \circled{3}, since the physics before the growth of $d$-wave order is independent of $\chi_d$.   
Within the parameter range studied, regime \circled{2} has the largest abrupt change of the dominant $p$-wave order (without destabilizing the dominant order), and the change is nearly the same across the regime \circled{2}.

\section{Lax formalism for $p_x+ip_y$ dynamical phases}
\label{sec:lax}

Similar to Ref.~\cite{Foster2013}, here we would like to calculate the dynamical phase boundaries for the $p_x+ip_y$ Hamiltonian ($\chi_d=0$) describing our cavity QED system (see Eq.~(\ref{eq:caveff})) using the Lax formalism. 
To simplify the calculation, we perform a gauge transformation to remove all the phases of the $p$-wave pairing,
\begin{equation}
    \hat{H}_{\rm cav}/\hbar = \sum_{\mathbf{n}}2J\eta_{\mathbf{n}}^2\hat{S}^z_{\mathbf{n}} - \chi_p \sum_{\mathbf{n}\mathbf{m}}\eta_{\mathbf{n}}\eta_{\mathbf{m}}\hat{S}^{+}_{\mathbf{n}}\hat{S}^{-}_{\mathbf{m}}.
\end{equation}

We prepare the initial state as the ground state with system parameter $\chi_{p,i}$,
\begin{equation}
    \begin{aligned}
    \langle \hat{S}^{x}_{\mathbf{n}}(0)\rangle&=\frac{1}{2}\frac{\Delta_{p,0}\eta_{\mathbf{n}}}{\sqrt{(J\eta_{\mathbf{n}}^2-\mu)^2+\Delta_{p,0}^2\eta_{\mathbf{n}}^2}}, \\
    \langle \hat{S}^{y}_{\mathbf{n}}(0)\rangle &= 0, \\
    \langle \hat{S}^z_{\mathbf{n}}(0)\rangle&=\frac{1}{2}\frac{\mu-J\eta_{\mathbf{n}}^2}{\sqrt{(J\eta_{\mathbf{n}}^2-\mu)^2+\Delta_{p,0}^2\eta_{\mathbf{n}}^2}},
    \end{aligned}
\end{equation}
where $\Delta_{p,0}$ (assuming real and positive) and $\mu$ are determined by the following self-consistent equations,
\begin{equation}
    \begin{gathered}
    1-\frac{2N_C}{N} = \frac{1}{N}\sum_{\mathbf{n}}\frac{J\eta_{\mathbf{n}}^2-\mu}{\sqrt{(J\eta_{\mathbf{n}}^2-\mu)^2+\Delta_{p,0}^2\eta_{\mathbf{n}}^2}}, \\
    \frac{1}{\chi_{p,i}N} = \frac{1}{N} \sum_{\mathbf{n}}\frac{\eta_{\mathbf{n}}^2}{2\sqrt{(J\eta_{\mathbf{n}}^2-\mu)^2+\Delta_{p,0}^2\eta_{\mathbf{n}}^2}}.
    \end{gathered}
    \label{eq:dyncon}
\end{equation}

We then perform sudden quench and consider mean-field dynamics under system parameter $\chi_{p,f}$. We can define the mean-field Lax vector as follows,
\begin{equation}
    \begin{gathered}
    L^x(u)=2\sqrt{2J}\sum_{\mathbf{n}}\frac{\sqrt{u}\,\eta_{\mathbf{n}}}{u-2J\eta_{\mathbf{n}}^2}\langle\hat{S}_{\mathbf{n}}^x\rangle,\\
    L^y(u)=2\sqrt{2J}\sum_{\mathbf{n}}\frac{\sqrt{u}\,\eta_{\mathbf{n}}}{u-2J\eta_{\mathbf{n}}^2}\langle\hat{S}_{\mathbf{n}}^y\rangle,\\
    L^z(u)=\frac{2J}{\chi_{p,f}}-\sum_{\mathbf{n}}\frac{4J\eta_{\mathbf{n}}^2}{u-2J\eta_{\mathbf{n}}^2}\langle\hat{S}_{\mathbf{n}}^z\rangle.\\
    \end{gathered}
\end{equation}

One can prove that the norm of the Lax vector, $L^2(u)=L^x(u)L^x(u)+L^y(u)L^y(u)+L^z(u)L^z(u)$, is conserved during the mean-field dynamics. The mean-field dynamical phases can be captured by the number of isolated pairs of roots in the equation 
\begin{equation}
    L^2(u)=0.
    \label{eq:root}
\end{equation}
Note that most of the roots of Eq.~(\ref{eq:root}) lie in the positive real axis, so isolated pairs of roots could be complex-conjugated pairs or lie on the negative real axis.
In the following, we list the correspondence between the number of isolated pairs of roots and the dynamical phases for the $p_x+ip_y$ Hamiltonian.
\begin{itemize}
    \item Phase I: $0$ isolated pairs of roots
    
    \item Phase II: $1$ isolated pair of roots

    Topological transition between II-BCS and II-BEC phases is marked by an isolated root $u=0$.

    \item Phase III/III*: $2$ isolated pairs of roots

    All the isolated roots of phase III* are in the negative real axis, while phase III has at least $1$ pair of complex-conjugated roots.
\end{itemize}

Note that the separation of phase III and III* is similar to the separation of phase IIIa and IIIb in the $s$-wave case \cite{Young2024}, but not exactly the same. In the $s$-wave case, both phases IIIa and IIIb have two pairs of complex-conjugated roots. The roots in IIIa have vanishing real parts, while those in IIIb have non-zero real parts. Therefore it is more suitable to use notations distinct from the $s$-wave case.  

\begin{figure}[t]
    \centering
    \includegraphics[width=1.0\columnwidth]{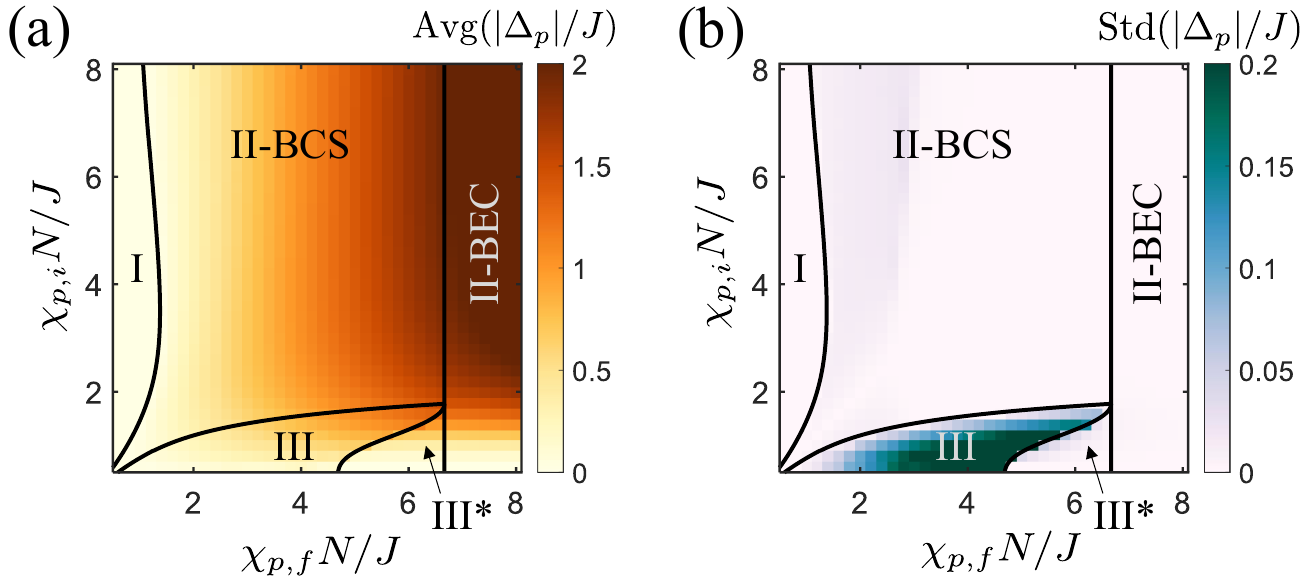}
    \caption{(a) Long-time averaged $p$-wave order parameter $\mathrm{Avg}(|\Delta_p|/J)$ by suddenly quenching $p$-wave interaction strength from $\chi_{p,i}$ to $\chi_{p,f}$. We fix $N_C/N=0.35$ and $\chi_d=0$. The solid lines mark the analytical calculation of dynamical phase boundaries using the Lax formalism. (b) Long-time standard deviation of $p$-wave order parameter $\mathrm{Std}(|\Delta_p|/J)$. The condition is the same as (a).}
    \label{fig:appen3}
\end{figure}

In Fig.~\ref{fig:appen3}, we perform a sudden quench of $p$-wave interaction strength from $\chi_{p,i}$ to $\chi_{p,f}$, and numerically calculate $\Delta_p(t)$ up to $Jt/2\pi=50$. Based on $\Delta_p(t)$, we then evaluate the long-time average $\mathrm{Avg}(|\Delta_p|/J)$ and long-time standard deviation $\mathrm{Std}(|\Delta_p|/J)$. 
We can also characterize the dynamical phases using the numerical results.
\begin{itemize}
    \item Phase I: $\mathrm{Avg}(|\Delta_p|/J)\rightarrow 0$, $\mathrm{Std}(|\Delta_p|/J)\rightarrow 0$
    
    \item Phase II: $\mathrm{Avg}(|\Delta_p|/J)>0$, $\mathrm{Std}(|\Delta_p|/J)\rightarrow 0$
    
    Phase II-BCS has $\mu_{\infty}>0$, and phase II-BEC has $\mu_{\infty}<0$ (see Fig.~\ref{fig:dpt}(c) for numerical results).
    
    \item Phase III/III*: $\mathrm{Avg}(|\Delta_p|/J)>0$, $\mathrm{Std}(|\Delta_p|/J)>0$

    Phase III and phase III* are separated by a sharp change of $\mathrm{Std}(|\Delta_p|/J)$ (see Fig.~\ref{fig:dynamics}(a) for a clearer visualization).
\end{itemize}
We then compare with the analytical calculation using the Lax formalism and show good agreement in Fig.~\ref{fig:appen3}.  

In the following, we would like to explain more details of the analytical calculation using the Lax formalism. If we define
\begin{equation}
    \beta = \frac{2J}{\chi_{p,f}N} - \frac{2J}{\chi_{p,i}N},
\end{equation}
we have
\begin{equation}
    \begin{aligned}
        \frac{L^x(u)}{N} &= \Delta_{p,0}\sqrt{\frac{u}{2J}}f(u),\\
        \frac{L^y(u)}{N} &= 0,\\
        \frac{L^z(u)}{N} &= \beta + \frac{1}{2}(u-2\mu) f(u),
    \end{aligned}
\end{equation}
where
\begin{equation}
    f(u) = \frac{1}{N}\sum_{\mathbf{n}}\frac{2J\eta_{\mathbf{n}}^2}{(u-2J\eta_{\mathbf{n}}^2)\sqrt{(J\eta_{\mathbf{n}}^2-\mu)^2+\Delta_{p,0}^2\eta_{\mathbf{n}}^2}}.
\end{equation}
So Eq.~(\ref{eq:root}) becomes
\begin{equation}
    \frac{L^2(u)}{N^2} = \beta^2 + \beta(u-2\mu) f(u) + [f(u)E(u)]^2 = 0,
    \label{eq:beta}
\end{equation}
with
\begin{equation}
    E(u) = \sqrt{\Big(\frac{u}{2}-\mu\Big)^2+\Delta_{p,0}^2\frac{u}{2J}}.
\end{equation}
Now we can interpret Eq.~(\ref{eq:beta}) as a quadratic equation of $\beta$, leading to
\begin{equation}
    \frac{\beta}{f(u)} = -\Big(\frac{u}{2}-\mu\Big)\pm i\Delta_{p,0}\sqrt{\frac{u}{2J}}.
    \label{eq:betaquad}
\end{equation}

\subsection{Topological transition within phase II}
The first thing we can calculate is the topological transition within phase II. For that, we simply plug in $u=0$. Then  we have
\begin{equation}
    \beta = f(u=0)\mu = -\frac{\mu}{N}\sum_{\mathbf{n}}\frac{1}{\sqrt{(J\eta_{\mathbf{n}}^2-\mu)^2+\Delta_{p,0}^2\eta_{\mathbf{n}}^2}}.
\end{equation}
Based on Eq.~(\ref{eq:dyncon}), we have
\begin{equation}
    \beta = 1-\frac{2N_C}{N} - \frac{2J}{\chi_{p,i}N},
\end{equation}
leading to
\begin{equation}
    \frac{2J}{\chi_{p,f}N} = 1-\frac{2N_C}{N} \quad\Rightarrow\quad \chi_{p,f} = \chi_{p,\rm QCP},
    \label{eq:chif}
\end{equation}
in which $\chi_{p,\rm QCP}$ is the critical point of the topological transition in equilibrium. Note that this result is slightly different from Ref.~\cite{Foster2013}. The difference can be attributed   to the fact that the high-energy cutoff in Ref.~\cite{Foster2013} depends on the chemical potential $\mu$. On the contrary, in our case the high-energy cutoff is independent of $\mu$ and fixed for all system parameters.

\subsection{Dynamical phase boundaries}
Now we analyze the dynamical phase boundaries set by the change of the number of isolated pairs of roots. At dynamical phase boundaries, $u$ should be a real positive number plus infinitesimal imaginary parts ($u\rightarrow u_0\pm i\,\mathrm{sgn}(\beta)\epsilon$). To perform further calculation, it is more convenient to rewrite $f(u)$ into an integral,
\begin{equation}
    \begin{aligned}
    f(u) &= \frac{2}{\pi}\int_0^{\pi/2} dx \frac{2J\cos^2(x)/\Big(u-2J\cos^2(x)\Big)}{\sqrt{(J\cos^2(x)-\mu)^2+\Delta_{p,0}^2\cos^2(x)}}\\
    &= \int_0^1 dy g(y)\frac{2Jy}{(u-2Jy)\sqrt{(Jy-\mu)^2+\Delta_{p,0}^2y}}\\
    \end{aligned}
\end{equation}
where $g(y)=1/\Big(\pi\sqrt{y(1-y)}\Big)$ is the normalized density of states. Using the fact that 
\begin{equation}
    \frac{1}{y-y_0\pm i\epsilon} = P\bigg[\frac{1}{y-y_0}\bigg] \mp i\pi\delta(y-y_0),
\end{equation}
where $P$ denotes the principal value, and $\delta(y-y_0)$ is the Dirac delta function. Here we have $y_0=u_0/2J$, leading to
\begin{equation}
    f(u) = P[f(u_0)] \mp i\pi\,\frac{\mathrm{sgn}(\beta)}{2J}\frac{u_0}{E(u_0)}g(u_0/2J),
\end{equation}
leading to
\begin{equation}
    |\beta|\; = \frac{\pi}{\Delta_{p,0}}\sqrt{u_0/2J}E(u_0)g(u_0/2J),
    \label{eq:ph2}
\end{equation}
\begin{equation}
    P[f(u_0)] = -\mathrm{sgn}(\beta)\frac{\pi}{\Delta_{p,0}}\sqrt{u_0/2J}\frac{u_0/2-\mu}{E(u_0)}g(u_0/2J).
    \label{eq:ph1}
\end{equation}
So the dynamical phase boundary is determined by first solving Eq.~(\ref{eq:ph1}) for $u_0$, and then plugging it in Eq.~(\ref{eq:ph2}). 
Notice that there are two branches of solution for $u_0$ depending on $\mathrm{sgn}(\beta)$.
When $\mathrm{sgn}(\beta)>0$, we have $\chi_{p,i}>\chi_{p,f}$, meaning a strong-to-weak quench, and the solution of $u_0$ marks the dynamical phase boundary between phase I and phase II. 
When $\mathrm{sgn}(\beta)<0$, we have $\chi_{p,i}<\chi_{p,f}$, meaning a weak-to-strong quench, and the solution of $u_0$ marks the dynamical phase boundary between phase II and phase III/III*. 

Here we would like to perform analytical calculation in the limit of $\Delta_{p,0}\ll J,\mu$. In this limit, we can show that $u_0\approx 2\mu$ is a solution. Assuming $|u_0/2-\mu|\;\ll\Delta_{p,0}$, we have
\begin{equation}
    \begin{aligned}
    P[f(u_0)] &\approx -\int_0^1 dy g(y)\frac{1}{\sqrt{(Jy-\mu)^2+\Delta_{p,0}^2y}}\\
    &\approx \frac{g(\mu/J)}{J}\ln\bigg(\frac{\Delta_{p,0}^2}{4J(J-\mu)}\bigg),
    \end{aligned}
\end{equation}
so Eq.~(\ref{eq:ph1}) leads to
\begin{equation}
    u_0 \approx 2\mu -\frac{2\,\mathrm{sgn}(\beta)}{\pi}\frac{\Delta_{p,0}^2}{J}\ln\bigg(\frac{\Delta_{p,0}^2}{4J(J-\mu)}\bigg).
\end{equation}
Based on this result, we have $|u_0/2-\mu|\;\sim O(\Delta_{p,0}^2)$, validating our assumption $|u_0/2-\mu|\;\ll\Delta_{p,0}$. Now we can plug in $u_0 \approx 2\mu$ in Eq.~(\ref{eq:ph2}), leading to
\begin{equation}
    |\beta|\; \approx \pi \frac{\mu}{J} g(\mu/J).
\end{equation}

Note that Ref.~\cite{Foster2013} pointed out that in the limit $\Delta_{p,0}\rightarrow 0$ there is an additional solution $u_0\rightarrow 0^{+}$ for the weak-to-strong quench ($\mathrm{sgn}(\beta)<0$), marking the termination of phase III.
However, since we have a different density of states, the solution $u_0\rightarrow 0^{+}$ instead occurs at a finite $\Delta_{p,0}$. Using Eq.~(\ref{eq:dyncon}) and Eq.~(\ref{eq:betaquad}), we have
\begin{equation}
    P[f(u_0\rightarrow 0^{+})] = \bigg(1-\frac{2N_C}{N}-\frac{2J}{\chi_{p,i}N}\bigg)\frac{1}{\mu},
\end{equation}
and Eq.~(\ref{eq:ph1}) gives
\begin{equation}
    P[f(u_0\rightarrow 0^{+})] = -\frac{1}{\Delta_{p,0}},
\end{equation}
leading to an additional equation
\begin{equation}
    \bigg(1-\frac{2N_C}{N}-\frac{2J}{\chi_{p,i}N}\bigg)\frac{1}{\mu} = -\frac{1}{\Delta_{p,0}}.
    \label{eq:threeter}
\end{equation}
Combining with Eq.~(\ref{eq:dyncon}), one can solve these equations for $\chi_{p,i}$, $\Delta_{p,0}$ and $\mu$ numerically. For $N_C/N=0.35$, we have $\chi_{p,i}N/J=1.77568$. Similar to Eq.~(\ref{eq:chif}), one can obtain $\chi_{p,f}=\chi_{p,\rm QCP}$ for $u_0\rightarrow 0^{+}$.

\begin{figure}[t]
    \centering
    \includegraphics[width=1.0\columnwidth]{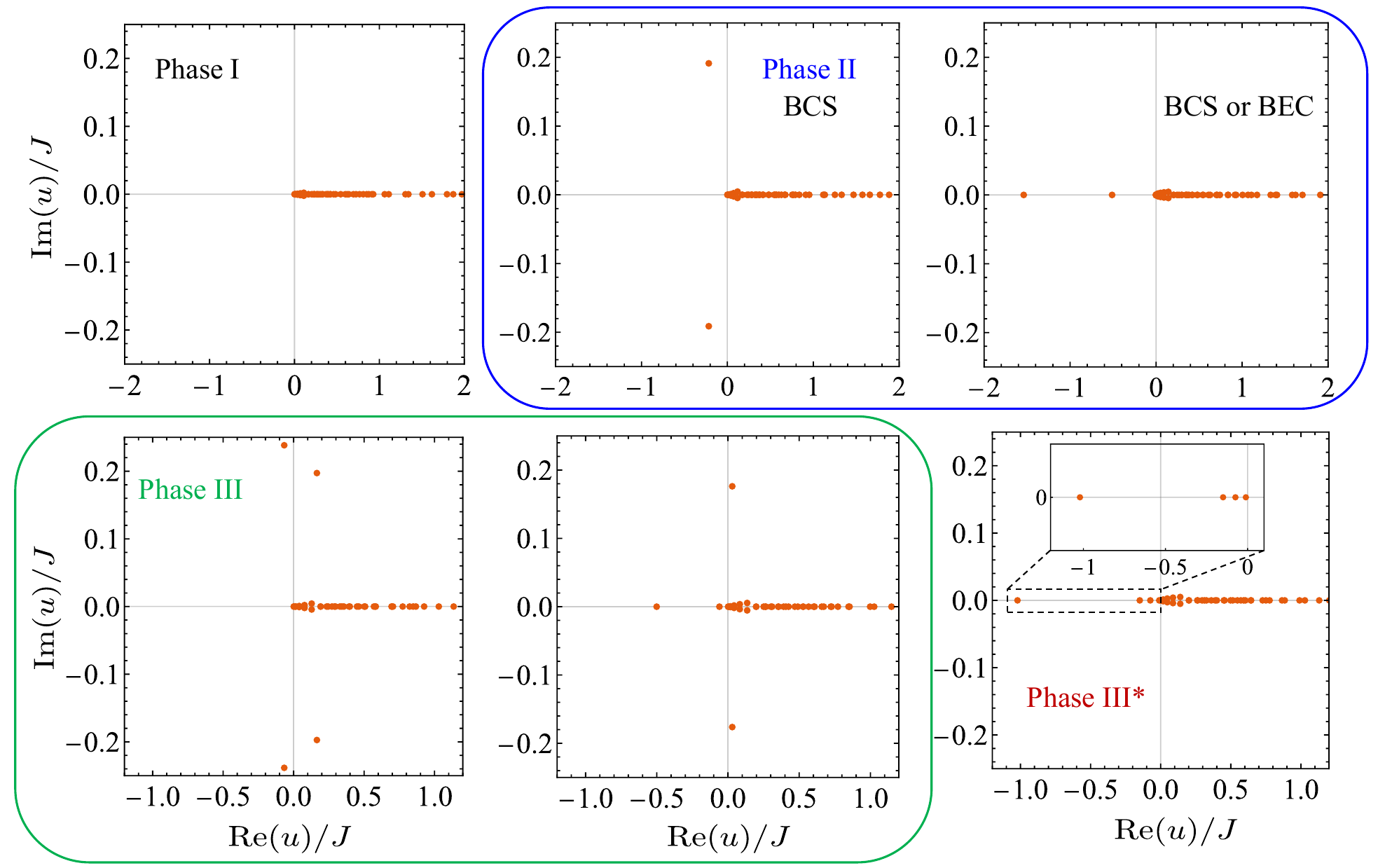}
    \caption{Distribution of roots on the complex plane for equation $L^2(u)=0$. We fix $N=30$, $N_C/N=0.35$ and $\chi_d=0$. A continuum of roots lie in the positive real axis, while isolated pairs of roots are complex-conjugated or negative real. We choose $\chi_{p,i}N/J=4.0$, $\chi_{p,f}N/J=1.0$ for phase I, $\chi_{p,i}N/J=2.0$, $\chi_{p,f}N/J=3.1$ (left) and $\chi_{p,i}N/J=1.0$, $\chi_{p,f}N/J=7.0$ (right) for phase II, $\chi_{p,i}N/J=1.0$, $\chi_{p,f}N/J=4.0$ (left) and $\chi_{p,i}N/J=1.0$, $\chi_{p,f}N/J=4.9$ (right) for phase III, and $\chi_{p,i}N/J=1.0$, $\chi_{p,f}N/J=5.9$ for phase III*.}
    \label{fig:appen5}
\end{figure}

\subsection{Separation between phase III and phase III*}
We would like to go a  step further and discuss the separation between phase III and phase III*. In this case the phase boundary is set by a doubly-degenerate, negative real root pair for $L^2(u)=0$. This is equivalent to an additional condition,
\begin{equation}
    \frac{d}{du}L^2(u)=0.
\end{equation}
Combining with Eq.~(\ref{eq:betaquad}), we have an additional equation for $u$,
\begin{equation}
    \mp \sqrt{-u} = - \frac{\Delta_{p,0}}{\sqrt{2J}}\frac{f(u)+2uf'(u)}{f(u)+(u-2\mu)f'(u)},
    \label{eq:negativereal}
\end{equation}
where $f'(u)=\frac{d}{du}f(u)$, and we require $u<0$. Notice that we have two branches of solution depending on the sign of the LHS of Eq.~(\ref{eq:negativereal}). 

When $\mathrm{LHS}=-\sqrt{-u}$, the solution of $u_0$ lies in both phase II and phase III/III*. In phase II, it marks the change (1 pair of complex-conjugated roots) $\leftrightarrow$ (1 pair of negative real roots). In phase III/III*, it marks the change (2 pairs of complex-conjugated roots) $\leftrightarrow$ (1 pair of negative real roots + 1 pair of complex-conjugated roots). However, we don't find sharp changes in long-time dynamical behavior in numerical calculation. 

When $\mathrm{LHS}=\sqrt{-u}$, the solution of $u_0$ only lies in phase III/III*, and it marks the change (1 pair of negative real roots + 1 pair of complex-conjugated roots) $\leftrightarrow$ (2 pairs of negative real roots). In this case, we find a sharp change in long-time standard deviation in numerical calculation. So this branch of solution separates phase III and phase III*. Note that phase III* is absent in Ref.~\cite{Foster2013} since we have a different density of states.

Here we focus on the $\Delta_{p,0}\rightarrow 0$ limit. To ensure the validity of Eq.~(\ref{eq:negativereal}), in this limit we should have
\begin{equation}
    f(u)+(u-2\mu)f'(u) = 0.
\end{equation}
Notice that 
\begin{equation}
    \begin{aligned}
        &\lim_{\Delta_{p,0}\rightarrow 0}f(u)+(u-2\mu)f'(u)\\
        &= \int_0^{\mu/J}dy g(y) \frac{4Jy}{(u-2Jy)^2}-\int_{\mu/J}^{1}dy g(y) \frac{4Jy}{(u-2Jy)^2},\\
    \end{aligned}
\end{equation}
one can solve for $u$ numerically. We then apply this result to Eq.~(\ref{eq:betaquad}). In the limit of $\Delta_{p,0}\rightarrow 0$, similarly we have
\begin{equation}
    \begin{aligned}
        \frac{2J}{\chi_{p,f}N} &= -\Big(\frac{u}{2}-\mu\Big)f(u) + \frac{2J}{\chi_{p,i}N}\\
        &=\int_0^{\mu/J}dy g(y) \frac{2Jy}{u-2Jy}-\int_{\mu/J}^{1}dy g(y) \frac{2Jy}{u-2Jy}.
    \end{aligned}
\end{equation}
For $N_C/N=0.35$, we have $u=-0.158044$, leading to $\chi_{p,f}N/J=4.68021$.

\section{Detailed experimental implementation schemes}
\label{sec:experiment}

We describe two implementations of the proposed simulator:
one using the ground-state hyperfine levels of ${}^{87}$Rb
(see Fig.~\ref{fig:response2}) and the other using the nuclear-spin states of ${}^{171}$Yb (see Fig.~\ref{fig:response3}).
These examples illustrate implementations with alkali and
alkaline-earth-like atoms, respectively. The same approach
can be adapted to other species with suitable level structures,
although the optical couplings, detunings, and scattering
rates must be evaluated for each case.

\subsection{${}^{87}$Rb atoms}

\begin{figure}[t]
    \centering
    \includegraphics[width=1.0\columnwidth]{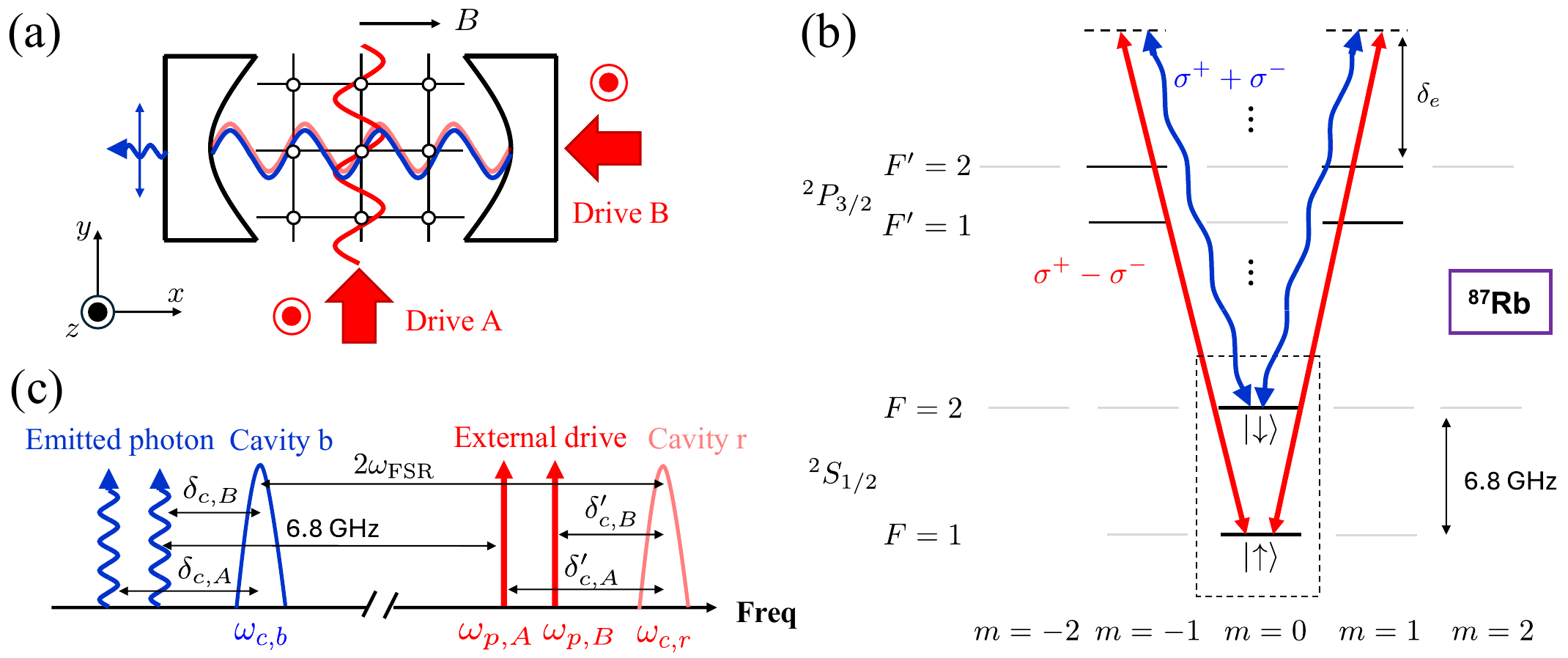}
    \caption{Detailed experimental implementation scheme using ${}^{87}$Rb atoms. (a) We consider atomic spins pinned in 2D lattice sites inside a standing-wave optical cavity, and apply external laser drives (drive A from the side and drive B along the cavity). The polarization of each laser drive and cavity mode is determined based on the quantization axis along the cavity direction. (b) Energy level diagram for ${}^{87}$Rb atoms. We define $|\downarrow\rangle\equiv |F=2,m=0\rangle$ and $|\uparrow\rangle\equiv |F=1,m=0\rangle$ in the ground manifold, with hyperfine splitting $6.8$GHz. The external drives (red color) couple $|\uparrow\rangle$ state to the excited manifold, and the cavity mode $b$ (blue color) couples $|\downarrow\rangle$ state to the excited manifold. The external drives have a linear polarization orthogonal to the linear polarization of cavity mode $b$. (c) Frequency diagram of external drives, cavity modes and emitted photons. All these frequencies are blue detuned to atomic resonances not shown in this diagram.}
    \label{fig:response2}
\end{figure}

For ${}^{87}$Rb, we encode the pseudospin in the ground-state hyperfine levels $|\downarrow\rangle\equiv|F=2,m_F=0\rangle$ and
$|\uparrow\rangle\equiv|F=1,m_F=0\rangle$, whose splitting is $\omega_{\rm hf}/(2\pi) =-\omega_0/(2\pi) = 6.8\,\mathrm{GHz}$.
This transition is insensitive to magnetic fields to first order near zero field. 
The assignment of the two pseudospin states is chosen to give the desired sign of $J$.
We couple the states through off-resonant Raman transitions
on the D$_2$ line at $780\,\mathrm{nm}$.
The Raman fields have orthogonal linear polarizations,
chosen so that the $\sigma^+$ and $\sigma^-$ pathways
contribute constructively to the desired transition.
The level structure and optical geometry are shown in
Fig.~\ref{fig:response2}.
A magnetic field along the cavity axis, taken to be the
$x$ direction, defines the quantization axis.
External drives $A$ and $B$ and cavity mode $r$ are polarized
along $z$, while cavity mode $b$ is polarized along $y$.
These transverse linear polarizations correspond to
different superpositions of $\sigma^+$ and $\sigma^-$
components.
The resonances of modes $r$ and $b$ are separated by two
free spectral ranges, $2\omega_{\rm FSR}$
(see Fig.~\ref{fig:response2}(c)).
With suitable positioning of the atoms near the cavity
center, the two modes have approximately matching
standing-wave profiles over the occupied region.

We estimate the required parameters using an existing
cavity QED setup~\cite{luo2025a,luo2025b}, with cavity linewidth
$\kappa/(2\pi)=50\,\mathrm{kHz}$, peak single-atom coupling
$g_c/(2\pi)=0.5\,\mathrm{MHz}$ on the cycling transition,
and D$_2$ spontaneous-emission rate
$\gamma/(2\pi)=6\,\mathrm{MHz}$.
We similarly define $\Omega_A$ and $\Omega_B$ as the peak
Rabi frequencies referenced to the cycling transition.
Including the Clebsch-Gordan coefficients gives
\begin{equation}
    \begin{gathered}
    J=
    \frac{|\Omega_B|^2}{24}
    \left(
    \frac{1}{\delta_e}
    -\frac{1}{\delta_e+\omega_{\rm hf}}
    \right),
    \\
    \chi_p=
    -\frac{|\Omega_A|^2g_c^2}
    {36\delta_e^2\delta_{c,A}},
    \qquad
    \chi_d=
    -\frac{|\Omega_B|^2g_c^2}
    {36\delta_e^2\delta_{c,B}}.
    \label{eq:st}
    \end{gathered}
\end{equation}
Choosing $\delta_e>0$ and
$\delta_{c,A},\delta_{c,B}<0$ makes
$J$, $\chi_p$, and $\chi_d$ positive.
The free space scattering rate can be estimated by $\gamma_{\rm sc,\mathbf{n}} = \gamma(|\Omega_{A}|^2+|\Omega_B|^2\eta_{\mathbf{n}}^2)/(6\delta_e^2)$.
For example, one can take the detuning to atomic transition $\delta_e/(2\pi)\simeq10\,\mathrm{GHz}$,
and the optical Rabi frequencies $\Omega_A/(2\pi)\simeq\Omega_B/(2\pi)
\simeq50\,\mathrm{MHz}$, which ensures the detuning $\delta_e$ is much larger than the optical Rabi
frequencies $\Omega_A,\Omega_B$ and the excited-state linewidth $\gamma$, supporting
adiabatic elimination of the excited states.
This choice of parameters leads to $J/(2\pi)\simeq4.2\,\mathrm{kHz}$, and the free space scattering rate averaged over lattice sites $\bar{\gamma}_{\rm sc}/(2\pi)\simeq 37.5$Hz.
This result indicates that free space scattering is not expected to alter the underlying coherent mechanism. 
With $N\simeq10^6$ and cavity detunings $\delta_{c,A}/2\pi$, $\delta_{c,B}/2\pi$ between
$-50$ and $-5\,\mathrm{MHz}$,
Eq.~(\ref{eq:st}) gives
$\chi_pN/J$ and $\chi_dN/J$ between approximately $0.8$ and $8$.
Thus, the collective interactions can be tuned from
comparable to $J$ to substantially larger than $J$.

Adiabatic elimination of the cavity modes requires $g_{{\rm eff},A}\sqrt{N}\ll|\delta_{c,A}|$, $g_{{\rm eff},B}\sqrt{N}\ll|\delta_{c,B}|$,
where $g_{{\rm eff},A}=g_c\Omega_A/(6\delta_e)$ and $g_{{\rm eff},B}=g_c\Omega_B/(6\delta_e)$ denote the effective atom-light couplings after eliminating the excited states.
For the parameters above, one can obtain $g_{\mathrm{eff},A}/2\pi\simeq g_{\mathrm{eff},B}/2\pi\simeq 0.4$kHz, satisfying the adiabatic elimination condition.
The ratios
$\kappa/|\delta_{c,A}|$ and $\kappa/|\delta_{c,B}|$
range from $10^{-3}$ to $10^{-2}$, making cavity loss weak compared to the coherent cavity-mediated interactions.

We next estimate corrections from fluctuations of cavity
mode $r$. Its relevant detunings are
$\delta'_{c,A}=\delta_{c,A}+\omega_{\rm hf}-2\omega_{\rm FSR}$, $\delta'_{c,B}
=\delta_{c,B}+\omega_{\rm hf}-2\omega_{\rm FSR}$. We choose their magnitudes to be of order
$2\pi\times100\,\mathrm{MHz}$ by adjusting the difference
between $\omega_{\rm hf}$ and $2\omega_{\rm FSR}$.
The resulting additional couplings are
\begin{equation}
    \begin{gathered}
    \chi'_p=
    -\frac{|\Omega_A|^2g_c^2}{36\delta'_{c,A}}
    \left(
    \frac{1}{\delta_e}
    -\frac{1}{\delta_e+\omega_{\rm hf}}
    \right)^2,
    \\
    \chi'_d=
    -\frac{|\Omega_B|^2g_c^2}{36\delta'_{c,B}}
    \left(
    \frac{1}{\delta_e}
    -\frac{1}{\delta_e+\omega_{\rm hf}}
    \right)^2.
    \end{gathered}
\end{equation}
For these parameters,
$|\chi'_p/\chi_p|$ and $|\chi'_d/\chi_d|$ are below
$10^{-1}$ (see the numerical benchmarking in App.~\ref{sec:numerical}-2).
Additional spin-exchange terms can arise from a cavity
mode lying $2\omega_{\rm FSR}$ above mode $r$.
Its detunings are $\delta''_{c,A}=\delta_{c,A}+2(\omega_{\rm hf}-2\omega_{\rm FSR})$, $   \delta''_{c,B}=\delta_{c,B}+2(\omega_{\rm hf}-2\omega_{\rm FSR})$.
Their magnitudes are of order
$2\pi\times200\,\mathrm{MHz}$, further suppressing these contributions.

To average out interference terms between the external
drives, we choose
$|\delta_{c,A}-\delta_{c,B}|/(2\pi)\simeq1\,\mathrm{MHz}$.
This frequency difference exceeds $J$ by more than
two orders of magnitude, supporting the rotating-wave
approximation discussed in App.~\ref{sec:numerical}-2.

The spatial extent of the atomic sample is limited by
phase mismatch and variations in the optical envelopes.
As discussed by Eq.~(\ref{eq:a32}), frequency differences
below $10\,\mathrm{GHz}$ permit up to approximately $10^4$
sites along either direction under the phase-mismatch
criterion. The Gaussian envelopes impose a more restrictive
practical limit (see Eqs.~(\ref{eq:a33}) and~(\ref{eq:a34})).
In particular, increasing the cavity waist improves spatial
uniformity but reduces the single-atom cooperativity.
For a cavity waist of $70\,\mu\mathrm{m}$ and a lattice
wavelength of $1064\,\mathrm{nm}$, a centered array with
$N_y=50$ has less than $4\%$ variation in the cavity coupling
amplitude between its center and transverse edges.
An array of $10^5$ sites would then require
$N_x=2\times10^3$, corresponding to a longitudinal extent
of approximately $1\,\mathrm{mm}$.
The external drive $A$ and the cavity mode must remain
sufficiently uniform over this length.
The atom number could be increased by replacing each
lattice site with a one-dimensional tube containing
multiple atoms. Ten atoms per tube would give
$N\simeq10^6$ for $10^5$ tubes.
In this configuration, neglecting contact interactions
requires their induced shifts and collision rates to
remain small compared with the cavity-mediated dynamics.

\subsection{${}^{171}$Yb atoms}

\begin{figure}[t]
    \centering
    \includegraphics[width=1.0\columnwidth]{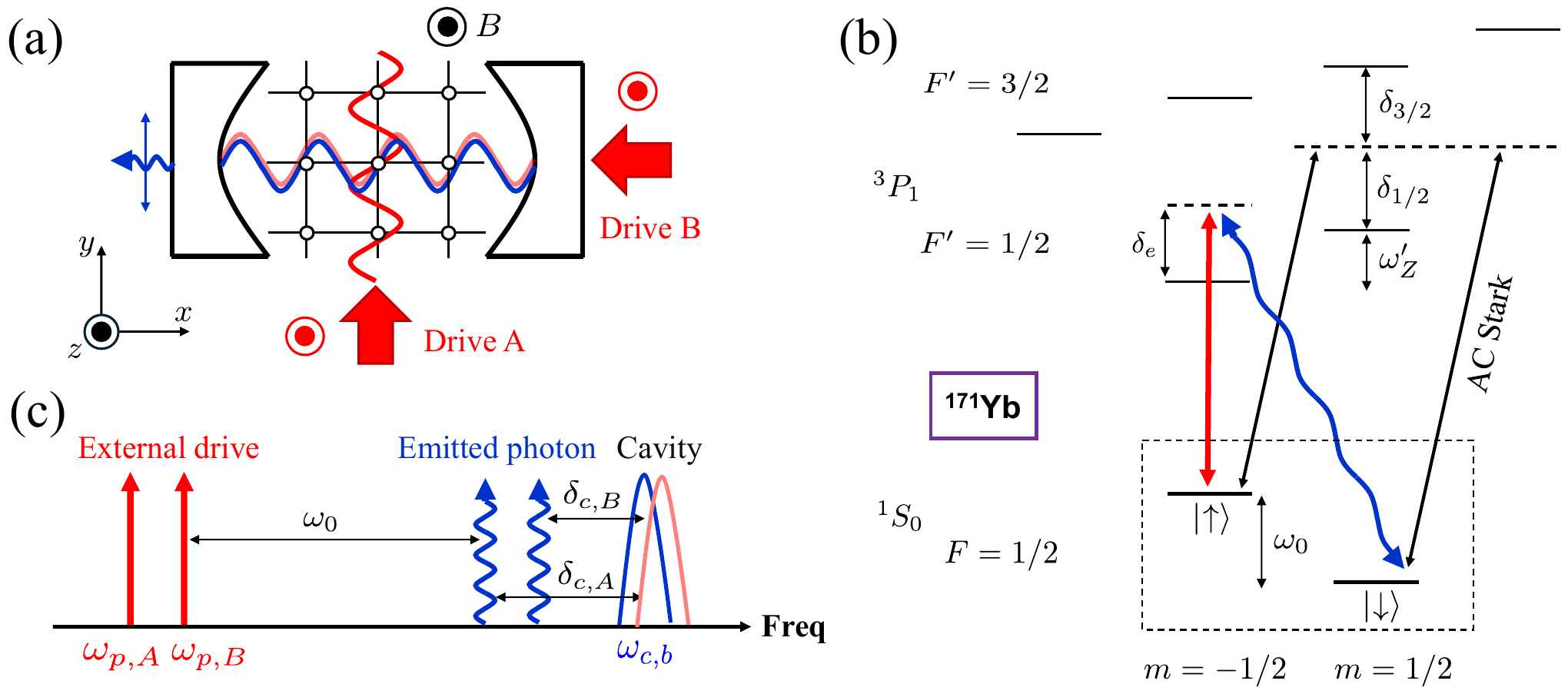}
    \caption{Detailed experimental implementation scheme using ${}^{171}$Yb atoms. (a) We consider atomic spins pinned in 2D lattice sites inside a standing-wave optical cavity, and apply external laser drives (drive A from the side and drive B along the cavity). The polarization of each laser drive and cavity mode is determined based on the quantization axis perpendicular to the 2D lattice plane. (b) Energy level diagram for ${}^{171}$Yb atoms. We define $|\downarrow\rangle\equiv |F=1/2,m=1/2\rangle$ and $|\uparrow\rangle\equiv |F=1/2,m=-1/2\rangle$, with transition frequency $\omega_0$ engineered via differential AC Stark shifts. The external drives (red color) couple $|\uparrow\rangle$ state to the excited manifold, and the cavity mode $b$ (blue color) couples $|\downarrow\rangle$ state to the excited manifold. (c) Frequency diagram of external drives, cavity modes and emitted photons. All these frequencies are blue detuned to atomic resonances not shown in this diagram.}
    \label{fig:response3}
\end{figure}

For ${}^{171}$Yb, we encode the pseudospin in the two
nuclear-spin states of the ${}^{1}S_0$ ground manifold,
$|\downarrow\rangle\equiv|F=1/2,m_F=1/2\rangle$ and
$|\uparrow\rangle\equiv|F=1/2,m_F=-1/2\rangle$.
Their splitting $\omega_0$ is engineered using a
differential AC Stark shift
(see Fig.~\ref{fig:response3}(b)).
A $\sigma^+$-polarized beam couples the ground manifold
off-resonantly to the ${}^{3}P_1$ state.
We denote its detunings from the $F'=3/2$ and $F'=1/2$
manifolds by $\delta_{3/2}$ and $\delta_{1/2}$,
respectively, and its Rabi frequency on the cycling
transition by $\Omega_{\rm Stark}$.
Including the Clebsch-Gordan coefficients gives
$\omega_0= -\omega_{\rm Stark}-\omega_Z$, with
\begin{equation}
    \omega_{\rm Stark}
    =
    \frac{|\Omega_{\rm Stark}|^2}{6}
    \left(
    \frac{1}{\delta_{3/2}}
    -\frac{1}{\delta_{1/2}}
    \right).
\end{equation}
Here $\omega_Z$ is the ground-state Zeeman contribution,
which is assumed small compared with the differential
AC Stark shift. The excited-state hyperfine splitting is
$\delta_{\rm hfs}=\delta_{1/2}-\delta_{3/2}
\simeq2\pi\times5.94\,\mathrm{GHz}$.
Using the ${}^{3}P_1$ spontaneous-emission rate
$\gamma/(2\pi)=182.4\,\mathrm{kHz}$, the Raman scattering rate for $|\uparrow\rangle\to|\downarrow\rangle$ is
\begin{equation}
    \gamma_{\rm Stark}
    =
    \frac{\gamma|\Omega_{\rm Stark}|^2}{18}
    \left(
    \frac{1}{\delta_{3/2}}
    -\frac{1}{\delta_{1/2}}
    \right)^2.
\end{equation}
For this configuration, the Rayleigh-induced coherence decay rate is also given by $\gamma_{\rm Stark}$.

For a fixed differential Stark shift $\omega_{\rm Stark}$ and a beam frequency
restricted to lie between the two hyperfine resonances,
the Raman scattering rate is minimized at their midpoint,
$\delta_{3/2}=-\delta_{\rm hfs}/2$ and
$\delta_{1/2}=\delta_{\rm hfs}/2$.
At this detuning,
\begin{equation}
    \gamma_{\rm Stark}^{\rm opt}
    =
    \frac{4\gamma}{3\delta_{\rm hfs}}
    |\omega_{\rm Stark}^{\rm opt}|,
    \qquad
    \omega_{\rm Stark}^{\rm opt}
    =
    -\frac{2|\Omega_{\rm Stark}|^2}{3\delta_{\rm hfs}}.
\end{equation}
Choosing $\Omega_{\rm Stark}/(2\pi)\simeq100\,\mathrm{MHz}$
gives $|\omega_{\rm Stark}^{\rm opt}|/(2\pi)
\simeq 1.1\,\mathrm{MHz}$ and
$\gamma_{\rm Stark}^{\rm opt}/(2\pi)\simeq46\,\mathrm{Hz}$.
Scattering from this beam is therefore an important
constraint on the protocol duration.
Reducing the engineered splitting lowers the scattering
rate, but also reduces the detunings that suppress
unwanted processes involving mode $r$.

As shown in Fig.~\ref{fig:response3}(a,b), the quantization axis is along $z$. External drives $A$ and $B$ at $556\,\mathrm{nm}$ and
cavity mode $r$ have $\pi$ polarization.
They couple $|\uparrow\rangle$ to
$|e\rangle\equiv|{}^3P_1,F'=1/2,m_F=-1/2\rangle$
with detuning $\delta_e$.
Mode $b$ is linearly polarized along $y$; its
$\sigma^-$ component couples $|\downarrow\rangle$
to $|e\rangle$.
The corresponding frequencies are shown in
Fig.~\ref{fig:response3}(c).
The effective Hamiltonian parameters are thus given by
\begin{equation}
    \begin{gathered}
    J=
    \frac{|\Omega_B|^2}{24}
    \left(
    \frac{1}{\delta_e}
    -\frac{1}{\delta_e-\omega_0-\omega'_Z}
    \right),
    \\
    \chi_p=
    -\frac{|\Omega_A|^2g_c^2}
    {36\delta_e^2\delta_{c,A}},
    \qquad
    \chi_d=
    -\frac{|\Omega_B|^2g_c^2}
    {36\delta_e^2\delta_{c,B}},
    \end{gathered}
    \label{eq:st2}
\end{equation}
where $\omega'_Z = \omega_{F'=1/2,m=1/2}-\omega_{F'=1/2,m=-1/2}$ is the excited-state Zeeman contribution
in the ${}^{3}P_1,F'=1/2$ manifold.
Choosing $\delta_e>0$, $\omega'_Z<-\omega_0$, and
$\delta_{c,A},\delta_{c,B}<0$ gives positive
$J$, $\chi_p$, and $\chi_d$.

For example, we take $\delta_e/(2\pi)\simeq100\,\mathrm{MHz}$
and $\Omega_A/(2\pi)\simeq\Omega_B/(2\pi)
\simeq3\,\mathrm{MHz}$.
The optical Rabi frequencies and excited-state linewidth
are small compared with the detuning, supporting
adiabatic elimination of the excited state.
Scattering from these drives is typically smaller compared to the scattering from the Stark-shift beam.
The excited-state Zeeman shift provides an additional
way to tune $J$. For $\omega'_Z/(2\pi)=-200\,\mathrm{MHz}$
and $\omega_0/(2\pi)\simeq1.1\,\mathrm{MHz}$, Eq.~(\ref{eq:st2}) gives $J/(2\pi)\simeq2.5\,\mathrm{kHz}$.
We take $\kappa/(2\pi)=5\,\mathrm{kHz}$ and
$g_c/(2\pi)\simeq40\,\mathrm{kHz}$, corresponding to
a single-atom cooperativity comparable to the Rb example, but with a longer cavity length.

With $N\simeq3\times 10^5$ and
$\delta_{c,A}/(2\pi)\simeq\delta_{c,B}/(2\pi)
\simeq-0.6\,\mathrm{MHz}$, one can obtain
$\chi_pN/J\simeq\chi_dN/J
\simeq8$.
The Rabi frequencies and cavity detunings allow these
couplings to be tuned.
The effective atom-light couplings $g_{\rm eff,A}/2\pi$, $g_{\rm eff,B}/2\pi$ are of order $10^2$Hz, so the condition for adiabatic elimination of cavity modes is satisfied.
The cavity-loss ratios are
$\kappa/|\delta_{c,A}|\;\simeq
\kappa/|\delta_{c,B}|\;\simeq10^{-2}$.
Unwanted contributions from mode $r$ are suppressed by
the detunings
$\delta'_{c,A}=\delta_{c,A}-\omega_0$ and
$\delta'_{c,B}=\delta_{c,B}-\omega_0$,
together with the cancellation between the relevant
optical pathways. 
One can show that
$|\chi'_p/\chi_p|$ and $|\chi'_d/\chi_d|$
can be suppressed to the order of $10^{-1}$ for the chosen parameters.

For both $^{87}$Rb and $^{171}$Yb atoms, we have illustrated an operating regime with kHz-scale coherent dynamics.
Our estimation on the strength of cavity loss and free space scattering, as well as the extra limitation from the scattering induced by the beam that creates the pseudospin splitting for $^{171}$Yb, in principle supports the proposed implementation.
One can optimize the optical and cavity parameters to further improve the separation between coherent and dissipative time scales in a specific experimental implementation. 

\bibliography{reference}

\end{document}